\documentclass[12pt]{article}

\usepackage{graphics}
\usepackage{feynarts}
\usepackage{amssymb,amsmath}
\usepackage[usenames]{color}
\usepackage{rotating}
\usepackage{cite}
\usepackage{fixltx2e}
\usepackage{afterpage}

\newcommand{\feynhiggs}{{\it FeynHiggs} \cite{Hahn:2009zz,Frank:2006yh,Degrassi:2002fi,hep-ph/9812472,Heinemeyer:1998yj}}
\newcommand{\cpsuperh}{{\it CPsuperH} \cite{Lee:2003nta,Lee:2007gn}}
\newcommand{\higgsbounds}{{\it HiggsBounds}
\cite{Bechtle:2008jh,Bechtle:2011sb}}
\newcommand{\hb}{{\it HiggsBounds}}
\newcommand{\looptools}{{\it LoopTools} \cite{Hahn:1998yk}}
\newcommand{\feynarts}{{\it FeynArts} \cite{Kublbeck:1990xc,Hahn:2000kx,Hahn:2001rv}}

\newcommand{\formcalc}{{\it FormCalc} \cite{Hahn:2001rv,Hahn:1998yk}}
\newcommand{\cph}{{\it CPH}}
\newcommand{\hdecay}{{\it HDECAY} \cite{Djouadi:1997yw}}

\newcommand{\CPX}{\rm{CPX}} 
\newcommand{\CPXdrbar}{\overline{\rm{CPX}}}

\newcommand{\hZtobbZ}{h_1 Z\to b \bar{b} Z \, ({\color{red}\blacksquare})}
\newcommand{\HZtobbZ}{h_2 Z\to b \bar{b} Z \, ({\color{blue}\blacksquare})}
\newcommand{\HZtohhZtobbbbZ}{h_2 Z\to h_1 h_1 Z \to b \bar{b} b \bar{b} Z \, (\square)}
\newcommand{\Hhtobbbb}{h_2 h_1\to b \bar{b} b \bar{b} \, ({\color{cyan}\blacksquare})}
\newcommand{\Hhtohhhtobbbbbb}{ h_2 h_1 \to h_1 h_1 h_1 \to b \bar{b} b \bar{b} b \bar{b} \, ({\color{yellow}\blacksquare})}
\newcommand{\Ahtobbbb}{ h_3 h_1 \to b \bar{b} b \bar{b} \, ({\color{green}\blacksquare}) }

\newcommand{\muXt}{\mu X_t}

\newcommand{\AtXt}{A_tX_t}

\newcommand{\Cfuncshort}{\mathcal{C}}
\newcommand{\Dfuncshort}{\mathcal{D}}
\newcommand{\Efuncshort}{\mathcal{E}}

\input paperdef

\newcommand{\SLASH}[2]{\makebox[#2ex][l]{$#1$}/}
\newcommand{\pslash}{\SLASH{p}{.2}}

\newcommand{\RemuXt}{\re{\left[\mu X_t\right]}}
\newcommand{\ResqmuXt}{\left(\re{\left[\mu X_t\right]}\right)^2}
\newcommand{\ImmusqXtsq}{\im{\left[\mu^2 X_t^2\right]}}
\newcommand{\ImmuXt}{\im{\left[\mu X_t\right]}}
\newcommand{\ImsqmuXt}{\left(\im{\left[\mu X_t\right]}\right)^2}
\newcommand{\ReAtCXt}{\re{\left[A_t^*X_t\right]}}
\newcommand{\RemuAt}{\re{\left[\mu  A_t\right]}}
\newcommand{\Cfunc}{\mathcal{C}_{112-122}}

\newcommand{\Dfunc}{\mathcal{D}_{1112-1222}}
\newcommand{\Efunc}{\mathcal{E}_{11122-11222}}
\newcommand{\logMStebyMStz}{\log\left(\frac{m_{\tilde{t}_1}}{m_{\tilde{t}_2}}\right)}
\newcommand{\logMSteMStzbyMTsq}{\log\left(\frac{m_{\tilde{t}_1}m_{\tilde{t}_2}}{m_t^2}\right)}

\newcommand{\CfuncL}{\mathcal{C}_{12L}}
\newcommand{\CfuncLeq}{\mathcal{C}_{11L}}

\graphicspath{{plots/}}

\evensidemargin \oddsidemargin
\allowdisplaybreaks
\begin{document}
\thispagestyle{empty}

\def\thefootnote{\fnsymbol{footnote}}

\begin{flushright}
BONN-TH-2011-01\\
DESY 10-168\\
FR-PHENO-2011-001
\end{flushright}

\vspace{1cm}

\begin{center}

{\Large\sc {\bf Higher order corrections to Higgs boson decays\\[1em]
in the MSSM with complex parameters}}

\vspace{1cm}

{\sc
Karina E.~Williams$^1$%
\footnote{Email: williams@th.physik.uni-bonn.de},%
~Heidi~Rzehak$^2$%
\footnote{Email: hr@particle.uni-karlsruhe.de}
~and ~Georg~Weiglein$^3$%
\footnote{Email: Georg.Weiglein@desy.de}%
}

\vspace*{.7cm}

{\sl
$^{1}$ Bethe Center for Theoretical Physics, 
Physikalisches Institut der
Universit\"at Bonn
Nussallee 12, D--53115 Bonn, Germany \\
$^{2}$ Physikalisches Institut
Albert-Ludwigs-Universit\"at Freiburg,\\
Hermann-Herder-Str. 3, D-79104 Freiburg im Breisgau, Germany \\
$^{3}$ DESY, Notkestr.~85, D--22607 Hamburg, Germany
}

\end{center}

\vspace*{0.1cm}

\begin{abstract}
\noindent
We discuss Higgs boson decays in the CP-violating
MSSM, and examine their phenomenological impact using cross section
limits from the LEP Higgs searches. This includes a discussion of the full 1-loop
results for the partial decay widths of neutral Higgs bosons into
lighter neutral Higgs bosons ($h_a \to h_b h_c$) and of neutral Higgs
bosons into fermions ($h_a \to f \bar{f}$).  In calculating the genuine
vertex corrections, we take into account the full spectrum of
supersymmetric particles and all complex phases of the supersymmetric
parameters. These genuine vertex corrections are supplemented with Higgs
propagator corrections incorporating the full one-loop and the dominant
two-loop contributions, and we illustrate a method of consistently
treating diagrams involving mixing with Goldstone and Z bosons. In
particular, the genuine vertex corrections to the process $h_a \to h_b
h_c$ are found to be very large and, where 
this process is kinematically allowed, can have a significant effect on
the regions of the CPX benchmark scenario which can be excluded by the results of the Higgs searches at LEP. However, there remains an unexcluded region of CPX parameter space at a lightest neutral Higgs boson mass of $ \sim 45 \gev $. In the analysis, we pay particular attention to the conversion between parameters defined in different renormalisation schemes and are therefore able to make a comparison to the results found using renormalisation group improved/effective potential calculations.
\end{abstract}
\def\thefootnote{\arabic{footnote}}
\setcounter{page}{0}
\setcounter{footnote}{0}

\newpage

\section{Introduction}
\label{intro}

High energy colliders, at present and in the past,
have given the search for Higgs bosons a high priority. The LEP and Tevatron 
experiments, in particular, have been able to turn the non-observation of Higgs bosons into constraints on the Higgs
sector, which have been very useful in reducing the available parameter space of some of the most popular particle physics models, such as the Standard Model
(SM)~\cite{Barate:2003sz} and the 
Minimal Supersymmetric Standard Model (MSSM)~\cite{Schael:2006cr}. For
first results on the Higgs searches at the LHC, see
\citeres{Atlas:2011.005,arXiv:1102.5429}.

However, MSSM scenarios involving CP violation in the Higgs sector,
which induces a mixing of all three neutral Higgs bosons, can prove
particularly difficult to restrict using the Higgs search data. This is
due to the fact that the CP violation can result in suppressed couplings
of the lightest Higgs boson to two gauge bosons and to the non-standard decay
mode of a heavier SM-like Higgs boson into a pair of light Higgs bosons, 
resulting in an experimentally rather challenging final state. The
$\CPX$ benchmark scenario~\cite{Carena:2000ks} is an example of such a
situation in the MSSM. In the original combined LEP analysis by the LEP
Higgs Working group and the LEP collaborations (LHWG), it was found that
substantial regions of the $\CPX$ parameter space could not be
excluded~\cite{Schael:2006cr} where the lightest Higgs mass is
substantially below the limit on the Standard Model Higgs mass~\cite{Barate:2003sz} of $M_H=114.4\gev$.

In this paper, we will present complete one-loop results for the decay
widths of neutral Higgs bosons into lighter neutral Higgs bosons (Higgs
cascade decays) and the decay widths of neutral Higgs bosons into fermions in the CP-violating MSSM. The results are obtained in the 
Feynman-diagrammatic approach, taking into account the full dependence on the spectrum of
supersymmetric particles and all complex phases of the supersymmetric
parameters. The genuine  
vertex contributions are supplemented with two-loop propagator-type corrections, yielding the currently most precise prediction for this class 
of processes. One-loop propagator-type mixing between neutral Higgs bosons and Goldstone and Z bosons is also consistently taken into account. 

Both of these calculations require loop corrections to the neutral Higgs
mass matrix $\matr{M}$, which are well known for the real and complex
MSSM and are frequently used to add propagator corrections to processes
involving external neutral Higgs particles. These corrections are
incorporated in the two main public codes for calculating the complex MSSM Higgs sector, {\feynhiggs} and {\cpsuperh}. 
{\fh} is based on the Feynman-diagrammatic approach and on-shell mass renormalisation while {\cpsh} is based on a renormalisation group improved effective potential calculation and $\drbarm$ renormalisation. Therefore, to compare between these results it is necessary to perform a parameter conversion. We shall discuss this issue in \refse{sec:convren}. We also investigate the numerical impact of parametrising the neutral Higgs self-energies (in the Feynman-diagrammatic approach) in terms of the $\msbarm$ top mass, rather than the on-shell top mass, which is formally a 3-loop effect.

The Higgs cascade decays often dominate the Higgs decay width where they
are kinematically allowed. They directly involve the Higgs
self-couplings, the observation and measurement of which is a crucial
goal for the experimental confirmation of the Higgs mechanism. We will
present two momentum-dependent approximations for the loop-corrected 
triple Higgs couplings, which can be used, for instance,
for predictions of the Higgs production process $e^+e^-\to Zh_ah_a$ at 
the ILC \cite{Djouadi:2007ik} or CLIC \cite{Accomando:2004sz}. 

The genuine vertex corrections to the triple Higgs decay can be very
large. In the MSSM with real parameters, the leading Yukawa vertex
corrections and the complete 1-loop vertex corrections have been
calculated
\cite{Barger:1991ed,Chankowski:1992es,Heinemeyer:1996tg,Osland:1998hv,Hollik:2001px,Dolgopolov:2003kv,Philippov:2006th}.
However, for the complex MSSM, previous to our result, first described in 
\citere{Williams:2007dc}, only effective coupling approximations were
available~\cite{Choi:1999uk,Carena:2002bb}, as provided by the program
{\cpsuperh}. The genuine vertex corrections we present will be
incorporated into the code {\fh}. As we will demonstrate, the $h_2\to h_1 h_1 $ decay width has a critical influence on the size and shape of one of the regions of $\CPX$ parameter space which the LEP Higgs search results are unable to exclude. 

The fermionic decay modes of the neutral Higgs bosons are crucially
important to collider phenomenology. These modes have been used when
obtaining a lower bound on the Standard Model Higgs mass
\cite{Barate:2003sz} and to exclude significant regions of the MSSM
parameter space \cite{Schael:2006cr,Aglietti:2006ne,Duperrin:2008in}. In
particular, an accurate prediction for the Higgs decay to b-quarks has
been vital for these analyses, since, for Standard Model Higgs bosons
with mass less than about $130 \gev$ and for most SUSY scenarios, $h_a\to b\bar{b}$ is the dominant decay mode. 
The decay to $\tau$-leptons can also be very important for Higgs
searches, as demonstrated for instance for various benchmark MSSM 
scenarios in the high $\tb$ region at the Tevatron \cite{arxiv:1003.3363}.

In the Standard Model, the fermionic decay width is extremely well known
(for a review, see e.g.\ 
\citere{Djouadi:2005gi} and references therein), and the treatment of
higher-order QCD (gluon-exchange) and QED corrections can be taken over
to the MSSM case.
The SUSY QCD corrections can be sizable for the $h_a\to b\bar{b}$ decay and should be resummed (see, for example, \citere{Carena:1999py}, for an investigation into these effects). Results supplemented with leading 2-loop propagator corrections \cite{Heinemeyer:2000fa} and full electroweak contributions \cite{hep-ph/9503443} are also available in the MSSM with real parameters. 

Predictions for the $h_a\to f\bar{f}$ decay widths for the Standard 
Model and the MSSM with real parameters can be obtained from the
programs {\hdecay} and {\it HFOLD} \cite{Frisch:2010gw}. For the complex MSSM, the program {\cpsuperh} 
is available. It is based on calculations involving 
effective $h_af\bar{f}$ couplings, as described in \citere{Carena:2002bb}.

The program {\feynhiggs} calculates the $h_a\to f\bar{f}$ decay width using the Feynman-diagrammatic approach, including the most significant QCD corrections, resummed SUSY QCD corrections and propagator corrections incorporating the full neutral Higgs self-energies. This calculation is valid in the real and complex MSSM.  The full 1-loop electroweak vertex corrections presented here have recently been incorporated into {\fh}. 

The corrections to the Higgsstrahlung and Higgs pair production processes 
at LEP in the MSSM with real parameters have been studied in \citeres{hep-ph/9303309,Driesen:1995ew,Driesen:1995ib,Akeroyd:2001aka,Heinemeyer:2001iy,Beccaria:2005un}
and the CP-violating MSSM in
\citeres{Demir:1998dp,Carena:2000yi,Akeroyd:2001kt,Arhrib:2002ti,Ham:2007gw}.
In the present paper, we will investigate the $t,\tilde{t},b,\tilde{b}$ corrections to these production processes in the Feynman-diagrammatic approach in the CP-violating MSSM, and supplement these with full propagator-type corrections. This type of corrections were not included in the Feynman-diagrammatic analysis of the $\CPX$ scenario in \citere{Schael:2006cr}. 

The parameter region in the MSSM with complex parameters that could not
be excluded with the Higgs searches at LEP,
characterised by a rather light Higgs boson with a mass of
about $45 \gev$ and moderate values of $\tb$, persists also in view of
the present search limits from the Tevatron~\cite{arxiv:1003.3363}.
This parameter region will be difficult to
cover also with the standard Higgs search channels at the
LHC~\cite{Buescher:2005re,Schumacher:2004da,Accomando:2006ga}, 
while it can be thoroughly
investigated at the ILC \cite{Djouadi:2007ik}. The
phenomenology of scenarios with such a light Higgs boson has recently
found considerable interest in the literature, see 
\citeres{Akeroyd:2003jp,Ghosh:2004cc,Cheung:2007sva,Bandyopadhyay:2007cp,
arXiv:0712.2466, arXiv:0909.5165,
arXiv:0911.0034,arXiv:1008.3339,arXiv:1010.3701} for
discussions
of other (non-standard) possible LHC search channels to access this 
parameter region.

In the present paper we make use of our improved theoretical predictions
for the Higgs branching ratios into a pair of lighter Higgs bosons and
into a fermion pair to examine their impact on the parameter region with
a light Higgs boson left unexcluded by the LEP Higgs searches. For this
purpose we employ the topological cross section limits obtained at LEP,
as implemented in the program {\higgsbounds}. We investigate the
sensitivity of the excluded parameter region with respect to variations
in the parameters of the $\CPX$ scenario. 
This analysis updates and considerably extends our previous results
reported in \citere{Williams:2007dc}.
We then compare our results to 
the results obtained with the code {\cpsh}, using various ways of 
performing the parameter conversion.

The paper is organized as follows: After introducing complex parameters in
\refse{section:complex} and the CPX scenario in \refse{section:cpx}  we discuss
contributions to the Higgs masses and mixings including also resummed SUSY QCD
corrections in \refse{section:Higgsmasses}. In \refse{sec:convren} we focus on
 the conversion between different renormalization schemes as well as on the
effect of a different parameterization of the top quark mass. In
\refse{sec:hihjhk} and in \refse{chapter:Hiff} we discuss the Higgs cascade
decay  and the Higgs decay into SM fermions, respectively, and the different
contributions to their partial decay widths and possible approximations.
After the investigation of the partial decay widths we turn our focus
particularly on the branching ratios of the Higgs cascade decay
processes in \refse{section:higgsBR}. In \refse{sec:Higgsprod} Higgs
production channels which were relevant at LEP are investigated.
Finally, in \refse{sec:LHWGresults} we discuss the phenomenological
impact of the improved predictions obtained in this paper. We
investigate in particular 
the parameter dependence of the CPX scenario and we perform a thorough 
comparison with the results obtained with the program {\cpsh}.
\refse{sec:concl} contains our conclusions.

\section{The MSSM with complex parameters at tree level}
\label{section:complex} 

In its general form, the MSSM allows various parameters to be complex. 
This includes the trilinear couplings $A_f$, the Higgsino mass parameter $\mu$, the gluino mass parameter $M_3$ and the soft SUSY breaking parameters $M_1$ and $M_2$ from the neutralino/chargino sector. These complex parameters can induce CP violation. 
Below we list the relevant quantities to fix our notation, which
closely follows that in \citere{Frank:2006yh}.

We write the two MSSM Higgs doublets as
\begin{align}
\label{eq:higgsdoublets}
\cHe=\begin{pmatrix} H_{11} \\ H_{12} \end{pmatrix} &=
\begin{pmatrix} v_1 + \tfrac{1}{\sqrt{2}} (\phi_1-i \chi_1) \\
  -\phi^-_1 \end{pmatrix} \notag ,\\ 
\cHz= \begin{pmatrix} H_{21} \\ H_{22} \end{pmatrix} &=
\begin{pmatrix} \phi^+_2 \\ v_2 + \tfrac{1}{\sqrt{2}} (\phi_2+i
  \chi_2) \end{pmatrix},
\end{align}
where $v_1$ and $v_2$ are the vacuum expectation values, and 
$\tb \equiv v_2/v_1$. Here we have made use of the fact that the MSSM
Higgs sector is CP-conserving at lowest order, i.e.\ complex phases
occurring in the Higgs potential can be rotated away (or vanish via the minimisation of the Higgs potential). 

The tree level neutral mass eigenstates $h,H,A,G$ are related to the tree level neutral fields $\phi_1,\phi_2,\chi_1,\chi_2$ through a unitary matrix, 
\begin{align}
\begin{pmatrix} h \\ H \\ A \\ G \end{pmatrix} = \begin{pmatrix}
        - \sina & \cosa &                 0 &           0 \\
    \quad \cosa & \sina &                 0 &           0 \\
              0 &     0 &     - \sin \betan & \cos \betan \\
              0 &     0 & \quad \cos \betan & \sin \betan
  \end{pmatrix} \cdot
\begin{pmatrix} \phi_1 \\ \phi_2 \\ \chi_1 \\ \chi_2 \end{pmatrix},
\label{basischange}
\end{align}
in which the CP-even eigenstates $\phi_1,\phi_2$ do not mix with the 
CP-odd eigenstates $\chi_1,\chi_2$. Unless otherwise stated, 
$h,H,A,G$ will always represent tree level neutral (mass eigenstate)
fields throughout 
this paper. At tree level, the off-diagonal mass terms must vanish, 
leading to the condition $\betan=\beta$. 

The Higgs sector at lowest order is given in terms of two independent
parameters (besides the gauge couplings), conventionally chosen as $\tb$
and either $m_A$ or $m_{H^{\pm}}$. Since CP violation can be induced via
potentially large higher-order corrections, in general all three neutral
Higgs bosons will mix once higher-order corrections are included, so
that the CP-odd $A$~boson is no longer a mass eigenstate. For the
general case of the MSSM with complex parameters it is therefore
convenient to use $m_{H^{\pm}}$ as input parameter. In our notation
lower-case Higgs masses indicate tree-level masses, while upper case 
masses refer to loop-corrected masses.

We write the squark mass matrices as
\begin{align}
M_{\sq} =
\begin{pmatrix}
        M_L^2 + \mq^2 + \MZ^2 \CZb (I_3^q - Q_q \sw^2) & \mq \; \Xq^* \\
        \mq \; \Xq    & M_{\tilde{q}_R}^2 + \mq^2 + \MZ^2 \CZb Q_q \sw^2
\end{pmatrix}, 
\label{eq:squarkmassmatrix}
\end{align}
where
\BEA
\Xq &=& A_q - \mu^* \{\CTb, \tb\}, 
\EEA
and $\CTb$ or $\tb$ applies to u-type or d-type quarks,
respectively. The eigenvalues of \refeq{eq:squarkmassmatrix} are
\begin{align}
m_{\tilde q_{1,2}}^2 = \mq^2
  + \edz 
&\Bigl[
M_L^2 + M_{\tilde{q}_R}^2 + I_3^q \MZ^2 \CZb \non\\&
           \mp \sqrt{[M_L^2 - M_{\tilde{q}_R}^2 + \MZ^2 \CZb(I_3^q -2 Q_q
  \sw^2)]^2 + 4 \mq^2 |\Xq|^2}~\Bigr].
\label{eq:stopmasses}
\end{align}

In the complex MSSM, the trilinear coupling $A_q$ and the higgsino mass parameter $\mu$ can have non-zero complex phases. The mass matrix $M_{\sq}$ can be diagonalised by the matrix $\matr{U}_{\sq}$. 
Here
\BEA
\VL \sqe \\ \sqz \VR = \matr{U}_{\sq} \VL \sql \\ \sqr \VR,&{\rm where}&
\matr{U}_{\sq} = \ML \ctq & \stq \\ -\stq^* & \ctq \MR,
\label{eq:squarkmixmatrix}
\EEA
and $\ctq$ is real, $\stq$ is complex,
and $\ctq^2+|\stq|^2=1$.

The coefficient of the gluino mass term in the Lagrangian, $M_3$, 
is in general complex. The gluino mass is given by
$m_{\tilde{g}}=|M_3|$, while the phase $\phi_{M_3}$ can be absorbed into
the gluino fields~\cite{Feng:2004me}. The phase
of $M_3$ thus appears in the quark--squark--gluino couplings.

For the chargino mass matrix we use
\begin{align}
  \matr{M_{\rm chargino}} =
  \begin{pmatrix}
    \MTwo & \sqrt{2} \sinb \MW \\
    \sqrt{2} \cosb \MW & \mu
  \end{pmatrix},
\end{align}
which includes the soft SUSY-breaking term $M_2$, which can be complex.
For the neutralino mass matrix we use
\begin{align}
  \matr{M_{\rm neutralino}} =
  \begin{pmatrix}
    \MOne                  & 0                & -\MZ \, \sw \cosb
    & \MZ \, \sw \sinb \\ 
    0                      & \MTwo            & \quad \MZ \, \cw \cosb
    & \MZ \, \cw \sinb \\ 
    -\MZ \, \sw \cosb      & \MZ \, \cw \cosb & 0
    & -\mu             \\ 
    \quad \MZ \, \sw \sinb & \MZ \, \cw \sinb & -\mu                    & 0
  \end{pmatrix},
\end{align}
which includes furthermore the soft SUSY-breaking term $M_1$, which can
also be complex.

It should be noted that not all phases mentioned above are physical, but
only certain combinations. In particular, the phase of the parameter
$M_2$ (chosen by convention) and, as mentioned above, the phase appearing in
the Higgs sector at lowest order can be rotated away.

\section{Phenomenology and the $\CPX$ scenario}
\label{section:cpx}

CP-violating effects, which can enter the Higgs sector via potentially large
higher-order corrections, can give rise to important
phenomenological consequences. CP phases in the loop corrections to the
Higgs particles can have a large impact on the predictions for the
masses (all three neutral Higgs bosons mix in the CP-violating case)
and the Higgs
couplings~\cite{Choi:2000wz,Pilaftsis:1998dd,Pilaftsis:1999qt,Carena:2000yi,Frank:2006yh}.

Studies of the possible impact of CP-violating effects on the MSSM Higgs
sector have often been carried out in the CPX benchmark 
scenario~\cite{Carena:2000ks}. As input values for the CPX scenario we use in this paper
\begin{itemize}
\item $\mt = 173.1 \gev $
\item $\msusy = 500 \gev(=M_{L}^{\rm on-shell}=M_{\tilde{q}_R}^{\rm on-shell})$
\item $\mu = 2000 \gev $
\item $|M_3| = 1000 \gev$
\item $M_2 = 200 \gev$, $M_1=\frac{5}{3}\frac{s_W^2}{c_W^2}M_2$ (see
e.g.~\citere{HiggsHunter}).
\item $|A_{t}^{\rm on-shell}| =|A_{b}| = 900 \gev $
\item $\phi_{A_{t}^{\rm on-shell}}=\phi_{A_{b}} = \phi_{M_3} = \frac{\pi}{2}$
\item $\MHpm\le 1000 \gev$
\end{itemize}

With the phases of the parameters $A_{t,b}$ and $M_3$ set to the maximal
value of $\pi/2$ and the relatively large value of $\mu$, this scenario 
has been devised to illustrate the possible importance of CP-violating
effects.

The above values differ from the ones defined in \citere{Carena:2000ks}
in the following ways: Firstly, we use an
on-shell value for the absolute value of the trilinear coupling $A_{t}$ and the soft SUSY breaking mass parameters $M_{L}$ and $M_{\tilde{q}_R}$, rather than $\drbarm$ values, and we therefore use a numerical value of $|A_{t}|$ that is somewhat
shifted compared to that specified in \citere{Carena:2000ks} in order to remain in an area of parameter space with similar phenomenology (the value specified in \citere{Carena:2000ks} is
$|A_{t}^{\drbarm}| = 1000 \gev$). Secondly, we use $\mt = 173.1 \gev$, which was the world 
average top-quark mass in March 2009~\cite{:2009ec}.

We use an on-shell definition of $A_t$, $M_{L}$ and $M_{\tilde{q}_R}$
since this is the natural choice for a Feynman-diagrammatic calculation. We will discuss how to convert between the different
parameter definitions in 
\refse{sec:convren}. For the purposes of this discussion, we use a second scenario using the parameter values given above, except with $A_t$, $M_{L}$, $M_{\tilde{q}_R}$ defined according to the $\drbarm$ scheme at the scale $M_S := \sqrt{\msusy^2+\mt^2}$ and with $|A_{t}^{\drbarm}(M_S)| = 1000 \gev(=|A_{b}|)$, $\phi_{A_t^{\drbarm}}(M_S) = \frac{\pi}{2}$, $M_{L}^{\drbarm}(M_S)= 500 \gev$ and $M_{\tilde{q}_R}^{\drbarm}(M_S)= 500 \gev$, which we will call the $\CPXdrbar$ scenario (i.e. this scenario is more similar to that in \citere{Carena:2000ks}). 

The LEP Higgs Working Group study\cite{Schael:2006cr} of the CPX scenario also used $|A_{t}^{\drbarm}| = 1000 \gev$. The majority of its analyses were
performed using $m_t=174.3 \gev$. We will investigate the dependence of
our results on $m_t$ in \refse{sec:exclplots}.

It should be noted that there are constraints on the CP phases in the
complex MSSM from experimentally measured upper limits on electric dipole
moments, such as those of the electron and neutron (for a recent discussion, 
see e.g.\ \citere{arXiv:1006.1440}). These provide particularly significant constraints on the CP phases in the first two generations. The constraints on the phases of the third generation are less
restrictive. In the definition of the CPX benchmark scenario, existing
bounds on CP phases were taken into account, see \citere{Carena:2000yi}
for more details.

\section{Loop corrections to the Higgs masses and Higgs mixing matrices}
\label{section:Higgsmasses}

Higher-order corrections to Higgs masses and mixing properties are
known to be very important for the phenomenology of the MSSM Higgs
sector, see \citeres{Djouadi:2005gj,Heinemeyer:2004ms,Heinemeyer:2004gx} for reviews.

In the MSSM with real parameters, the full 1-loop result~\cite{PHLTA.B257.83,PTPKA.85.1,PRLTA.66.1815,Brignole:1992uf,PHLTA.B286.307,hep-ph/9303309,hep-ph/9503443,hep-ph/9409375} and the dominant 2-loop corrections~\cite{hep-ph/9812472,hep-ph/9903404,hep-ph/0411114,hep-ph/0001002,hep-ph/9808299,hep-ph/9912236,hep-ph/0105096,hep-ph/9401219,Brignole:2001jy,hep-ph/0003246,hep-ph/0104047,hep-ph/0206101,hep-ph/0305127,hep-ph/9504316,hep-ph/9508343,hep-ph/9407389}
 have been calculated, and the $\tb$-enhanced terms \order{\alb(\als\tb)^n} have been resummed~\cite{PHRVA.D49.6168,hep-ph/9306309,hep-ph/9402253,Carena:1999py,hep-ph/9912463,hep-ph/0305101}. A full 2-loop effective potential calculation is known~\cite{hep-ph/0111209,hep-ph/0206136,hep-ph/0211366,hep-ph/0307101,hep-ph/0312092,hep-ph/0405022,hep-ph/0502168,hep-ph/0501132}\footnote{In principle, the effective potential calculation is also applicable to the complex MSSM.}. In addition, some dominant 3-loop contributions have been calculated \cite{arXiv:0803.0672,hep-ph/0701051,arXiv:1005.5709}.
 
In the complex MSSM, 1-loop corrections from the fermion/sfermion sector and some leading logarithmic corrections from the gaugino sector and the dominant 2-loop results have been calculated in the renormalisation group improved effective potential approach~\cite{Demir:1999hj,Pilaftsis:1999qt,Choi:2000wz,Carena:2000yi,Ibrahim:2000qj,Ibrahim:2002zk,hep-ph/0211366,hep-ph/0405022,hep-ph/0701051}. In the Feynman-diagrammatic approach, leading 1-loop contributions have been obtained in~\citere{Pilaftsis:1998dd,Heinemeyer:2001qd}, and the full 1-loop result has been calculated in \citere{Frank:2006yh}. At 2-loop order, the $\mathcal{O}(\alpha_t\alpha_s)$ corrections are available~\cite{Heinemeyer:2007aq}. 

Most of these results for the complex MSSM have been incorporated either into the public code {\fh}~\cite{Frank:2006yh,Degrassi:2002fi,hep-ph/9812472,Heinemeyer:1998yj,Heinemeyer:1998jw,Heinemeyer:1998kz,Frank:2002qa}, which uses the Feynman-diagrammatic approach, or the public code {\cpsuperh}, which uses the renormalisation group improved effective potential approach\footnote{Unless explicitly stated otherwise, `{\fh}' will refer to {\fh} version 2.6.5 and `{\cpsh}' to {\cpsh} version 2.2 throughout this paper.}.

In this paper, when calculating the Higgs masses and mixings, we will use renormalised neutral Higgs self-energies
calculated by {\fh}, to take advantage of the fact that it includes the complete 1-loop result and \order{\alpha_t\alpha_s} corrections of \citere{Heinemeyer:2007aq} with full phase dependence. {\fh} additionally allows the option of including sub-leading 2-loop corrections which are known so far only for the MSSM with real parameters~\cite{Brignole:2001jy,hep-ph/0206101,hep-ph/0003246,hep-ph/0305127,hep-ph/0411114}. If the user wishes to apply these corrections in an MSSM calculation with complex phases, {\fh} evaluates these corrections at a phase of $0$ and $\pi$ for each complex parameter, then an interpolation is performed to arrive at an approximation to these corrections for arbitrary complex phases. 
However, this prescription can be problematic in a rather `extreme'
scenario like the CPX scenario. In fact, it can happen in this case that 
one of the combinations of real parameters needed as input for the
interpolation turns out to be in an unstable region of the parameter
space where the reliability of the perturbative predictions is
questionable. This would skew the interpolation towards the unstable values. 
Therefore, unless otherwise stated, we will use in the present paper 
the leading \order{\alpha_t\alpha_s} corrections to the Higgs
self-energies from {\fh}, but not the sub-leading 2-loop corrections. A
discussion of the incorporation of the subleading 2-loop contributions via 
the interpolation from the results for real parameters is given in
\refse{sec:comp_cpsh}. As discussed in more detail below, besides the 
irreducible 2-loop contributions of \order{\alpha_t\alpha_s} we do
incorporate into our results higher order $\tb$--enhanced terms (for
arbitrary complex parameters), which we
take into account by introducing an effective b-quark mass.

\subsection{Determination of neutral Higgs masses}

In general, the neutral Higgs masses are obtained from the real parts of the complex poles of the propagator matrix. In the determination of the Higgs masses, we neglect mixing with the Goldstone and Z bosons as these are sub-leading 2-loop contributions to the Higgs masses. We therefore use   
a $3\times3$ propagator matrix $\matr{\Delta}(p^2)$ in the $(h,H,A)$ basis. 

In order to determine
the neutral Higgs masses we must first find the three solutions to 
\begin{equation}
\left| \matr{\Delta}^{-1}(p^2)\right| = 0 
\label{eq:findpoles},
\end{equation}
which, in the case with non-zero mixing between all three neutral Higgs bosons, is equivalent to solving 
\begin{equation}
\frac{1}{\Delta_{ii}(p^2)}=0
\label{eq:findpoles2},
\end{equation}
where $i=h,H$ or $A$. The propagator matrix is related to the $3\times3$ matrix of the irreducible 2-point vertex-functions $ \matr{\hat{\Gamma}}_2(p^2)$ through the equation
\begin{equation}
\left[-\matr{\Delta}(p^2)\right]^{-1} = \matr{\hat{\Gamma}}_2(p^2)=  i \left[p^2 \id - \matr{M}(p^2) \right]
\label{eq:deltagammaM},
\end{equation}
where
\BEA
  \matr{M}(p^2) &=
  \begin{pmatrix}
    \mh^2 - \ser{hh}(p^2) & - \ser{hH}(p^2) & - \ser{hA}(p^2) \\
    - \ser{hH}(p^2) & \mH^2 - \ser{HH}(p^2) & - \ser{HA}(p^2) \\
    - \ser{hA}(p^2) & - \ser{HA}(p^2) & \mA^2 - \ser{AA}(p^2)
  \end{pmatrix}. 
\EEA
As before, $m_h$,$m_H$,$m_A$ refer to the tree level masses. $\ser{ij}(p^2)$ are renormalised Higgs self-energies. 
The explicit form of the counterterms used in this paper is given in \refapp{sec:apprencon}. For the majority of these renormalisation conditions, we use the on-shell scheme. 
However, we shall use $\drbarm$ renormalisation for the Higgs fields
(see \citere{Frank:2006yh}). If there is CP conservation,
$\ser{hA}(p^2)=\ser{HA}(p^2)=0$, and the CP-even Higgs bosons $h,H$ do not mix with the CP-odd Higgs boson $A$.

In general, the renormalised Higgs self-energies can be complex, due to absorptive parts. Therefore, the three poles of the propagator matrix ${\cal M}_a^2$ can be written as
\begin{equation}
{\cal M}_{h_a}^2 = M_{h_a}^2 - i M_{h_a} W_{h_a}, 
\end{equation}
where $M_{h_a}$ is real and is interpreted as the loop-corrected 
(i.e.\ physical) mass, $W_{h_a}$ is the Higgs width, and $a=1,2,3$.

In the MSSM with complex parameters, the loop-corrected masses are labelled 
in size order such that
\begin{equation}
M_{h_1} \leq M_{h_2} \leq M_{h_3}.
\end{equation}
In the CP-conserving case, the masses are labelled such that the CP-even loop-corrected Higgs bosons have masses $M_{h}$ and $M_{H}$ with $M_{h}\leq M_{H}$ and the CP-odd loop-corrected Higgs boson has mass $M_{A}$ (the numerical value of $M_{A}$ is affected by loop corrections only if $M_{H^{\pm}}$ is chosen as an independent input parameter). 

Solving \refeq{eq:findpoles} with full momentum dependence involves an 
iterative procedure, since the self-energies also depend on the momentum. 
In order to deal with the complex momentum argument it is convenient to use
an expansion about the real part of the pole, $M^2_{h_a}$, such that 
\BEA
\ser{jk}({\cal M}_{h_a}^2)&\simeq&\ser{jk}(M_{h_a}^2)+i\im \left[ {\cal M}_{h_a}^2 \right] \ser{jk}^{\prime}(M_{h_a}^2),
\label{se1order}
\EEA
with $j=h,H,A$ and $k=h,H,A$. We obtain $\ser{jk}(M_{h_a}^2)$ and 
$\ser{jk}^{\prime}(M_{h_a}^2)$ (where the prime indicates the derivative
w.r.t.\ the external momentum squared) from {\feynhiggs}.
For each $h_a$, we use a momentum-independent approximation to obtain an
initial value for the iteration. This solution is then refined using 
\BEA
{\cal M}_{h_a}^{2,\left[n+1\right]}&=& \mbox{ $a$th eigenvalue of } \matr{M}({\cal M}_{h_a}^{2,\left[n\right]}) ,
\EEA
where the eigenvalues have been sorted into ascending value, according to their real parts. We check the validity of the truncation of the expansion in \refeq{se1order} by performing an iteration using an expansion up to second order:
\BEA
\ser{jk}({\cal M}_{h_a}^2)&\simeq&\ser{jk}(M_{h_a}^2)+i\im \left[ {\cal M}_{h_a}^2 \right] \ser{jk}^{\prime}(M_{h_a}^2)\non\\
&+&\frac{1}{2}\left(i \im \left[ {\cal M}_{h_a}^2 \right]\ser{jk}^{\prime\prime}(M_{h_a}^2)\right)^2,
\label{se2order}
\EEA
and confirming that the resultant Higgs masses show no significant changes.
\subsection{Wave function normalisation factors}
\label{section:zfactors}

In order to ensure that the S-matrix is correctly normalised, the residues 
of the propagators have to be set to one.  We achieve this by including finite wave function normalisation factors which are composed of the renormalised self-energies. These `Z-factors' can be collected in to a matrix $\matr{\hat Z}$ where
\BEA
\label{eq:Zfactors1}
\lim_{p^2 \to {\cal M}_{h_a}^2}-\frac{i}{p^2-{\cal M}_{h_a}^2}\left(\matr{\hat
Z}\cdot\matr{\hat \Gamma}_2\cdot\matr{\hat Z}^{T}\right)_{hh}&=&1,
\label{eq:ZfactorsA}\\
\lim_{p^2 \to {\cal M}_{h_b}^2}-\frac{i}{p^2-{\cal M}_{h_b}^2}\left(\matr{\hat
Z}\cdot\matr{\hat \Gamma}_2\cdot\matr{\hat Z}^{T}\right)_{HH}&=&1,
\label{eq:ZfactorsB}\\
\lim_{p^2 \to {\cal M}_{h_c}^2}-\frac{i}{p^2-{\cal M}_{h_c}^2}\left(\matr{\hat
Z}\cdot\matr{\hat \Gamma}_2\cdot\matr{\hat Z}^{T}\right)_{AA}&=&1,
\label{eq:ZfactorsC}
\EEA
such that
\begin{align}
\begin{pmatrix} \hat\Ga_{h_a} \\ \hat\Ga_{h_b} \\ \hat\Ga_{h_c} \end{pmatrix} = \matr{\hat Z} \cdot
\begin{pmatrix} \hat\Ga_h \\ \hat\Ga_H \\\hat \Ga_A \end{pmatrix},
\end{align}
where $\hat\Ga_{h_a}$ is a one-particle irreducible n-point vertex-function which involves a single external Higgs $h_a$, and $h_a,h_b,h_c= $ some combination of $h_1,h_2,h_3$.

The matrix $\matr{\hat Z}$ is non-unitary. We write it as
\BEA
 \matr{\hat Z}  &=
  \begin{pmatrix}
    \sqrt{Z_h} & \sqrt{Z_h} Z_{hH} & \sqrt{ Z_h} Z_{hA} \\
    \sqrt{Z_H}Z_{Hh} & \sqrt{ Z_H} & \sqrt{ Z_H} Z_{HA} \\
    \sqrt{Z_A}Z_{Ah} & \sqrt{ Z_A} Z_{AH} & \sqrt{ Z_A}
  \end{pmatrix}. 
\EEA
We find the elements of $\matr{\hat Z}$ by solving \refeq{eq:ZfactorsA}, 
which gives
\BEA
Z_h=\frac{1}{
\left.\frac{\partial}{\partial p^2}\left(\frac{i}{\De_{hh}(p^2)}\right)\right|}
_{p^2={\cal M}_{h_a}^2},&
Z_H=\frac{1}{
\left.\frac{\partial}{\partial p^2}\left(\frac{i}{\De_{HH}(p^2)}\right)\right|}
_{p^2={\cal M}_{h_b}^2},&
Z_A=\frac{1}{
\left.\frac{\partial}{\partial p^2}\left(\frac{i}{\De_{AA}(p^2)}\right)\right|}
_{p^2={\cal M}_{h_c}^2},\\
Z_{hH}=\left.\frac{\De_{hH}}{\De_{hh}}\right|_{p^2={\cal M}_{h_a}^2},&
Z_{Hh}=\left.\frac{\De_{hH}}{\De_{HH}}\right|_{p^2={\cal M}_{h_b}^2},&
Z_{Ah}=\left.\frac{\De_{hA}}{\De_{AA}}\right|_{p^2={\cal M}_{h_c}^2},\\
Z_{hA}=\left.\frac{\De_{hA}}{\De_{hh}}\right|_{p^2={\cal M}_{h_a}^2},&
Z_{HA}=\left.\frac{\De_{HA}}{\De_{HH}}\right|_{p^2={\cal M}_{h_b}^2},&
Z_{AH}=\left.\frac{\De_{HA}}{\De_{AA}}\right|_{p^2={\cal M}_{h_c}^2}.
\EEA
We choose $h_a=h_1$, $h_b=h_2$ and $h_c=h_3$. $Z_i$ is, in general, complex. Other choices for the Z-factors are possible, such as that in \citere{Hahn:2006np}, where we use the limit $p^2=M_{h_{1,2,3}}^2$. However, this does not allow the same freedom for choosing $a,b,c$.

Since the elements of $\matr{\hat Z}$ involve evaluating self-energies at 
complex momenta, we use again
the expansion given in \refeq{se1order}. In order to make sure that the 
neglected higher order terms in  \refeq{se1order} are small, we also 
calculate $\matr{\hat Z}$ using \refeq{se2order}, and check that this does 
not significantly change the result. 

The wave function normalisation factors are included in the calculation by multiplying the irreducible vertex factor $\hat\Gamma$ by $\matr{\hat Z}$ once for each external Higgs boson involved in the process.


\subsection{Goldstone or gauge bosons mixing contributions to the Higgs propagators}
\label{section:GZ}

A complete 1-loop prediction for a process involving a 
neutral Higgs propagator in the 
MSSM with complex parameters will, in general, contain terms involving the self-energies  
$\hat{\Sigma}_{hG}$, $\hat{\Sigma}_{HG}$, 
$\hat{\Sigma}_{AG}$ and $\hat{\Sigma}_{hZ}$, $\hat{\Sigma}_{HZ}$, 
$\hat{\Sigma}_{AZ}$, such as those shown in \reffi{fig:hiGhiZ}. 
These terms are required to ensure that the 1-loop result is 
gauge-parameter independent and free of unphysical poles. As we will
illustrate for an example, it is essential to treat these
mixing contributions strictly at one-loop level in order to ensure the
cancellation of the unphysical contributions (some care is necessary to
achieve this, since the loop corrected masses used for the external
particles and the Z-factor prescription outlined above
automatically incorporate leading higher-order
contributions).

\unitlength=1.0bp%
\begin{figure}
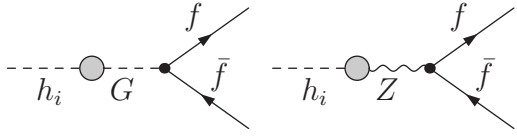

\BEA
\begin{feynartspicture}(432,100)(2,1)

\FADiagram{}
\FAProp(0.,10.)(13.,10.)(0.,){/ScalarDash}{0}
\FALabel(3.5,9.18)[t]{$h_i$}
\FALabel(9.5,9.18)[t]{$G$}
\FAProp(20.,15.)(13.,10.)(0.,){/Straight}{-1}
\FALabel(16.2244,13.0779)[br]{$f$}
\FAProp(20.,5.)(13.,10.)(0.,){/Straight}{1}
\FALabel(16.7756,8.07785)[bl]{$\bar{f}$}
\FAVert(7.,10.){-1}
\FAVert(13.,10.){0}

\FADiagram{}
\FAProp(0.,10.)(7.,10.)(0.,){/ScalarDash}{0}
\FAProp(7.,10.)(13.,10.)(0.,){/Sine}{0}
\FALabel(3.5,9.18)[t]{$h_i$}
\FALabel(9.5,9.18)[t]{$Z$}
\FAProp(20.,15.)(13.,10.)(0.,){/Straight}{-1}
\FALabel(16.2244,13.0779)[br]{$f$}
\FAProp(20.,5.)(13.,10.)(0.,){/Straight}{1}
\FALabel(16.7756,8.07785)[bl]{$\bar{f}$}
\FAVert(7.,10.){-1}
\FAVert(13.,10.){0}

\end{feynartspicture}
\non\EEA
\caption{The Goldstone and Z boson propagator corrections to the $h_i\rightarrow f\bar{f}$ decay, where $h_i=h,H$ or $A$}
\label{fig:hiGhiZ}
\end{figure}

As an example, we consider diagrams involving mixing contributions for 
a neutral Higgs decaying to two fermions, as in \reffi{fig:hiGhiZ}. 
We use here the lowest order Z-boson propagator with explicit gauge parameter 
dependence in the $R_\xi$ gauge,
%
\begin{equation}
\left(-g_{\mu\nu} + \frac{p_{\mu}p_{\nu}}{p^2}\right) \frac{i}{p^2-M_Z^2}
-\frac{p_{\mu}p_{\nu}}{p^2}\frac{i\xi_Z}{(p^2-\xi_ZM_Z^2)},
\end{equation}
%
and the G-boson propagator
\BEA
\frac{i}{p^2-\xi_ZM_Z^2}.
\EEA
The vertex $\Gamma^{\mu,{\rm tree}}_{\rm Zff}$ involving on-shell fermions is related to $\Gamma^{\rm tree}_{\rm Gff}$ by
\BEA
p_{\mu}\Gamma^{\mu,{\rm tree}}_{\rm Zff}&=&-iM_Z \Gamma^{\rm tree}_{\rm Gff}.
\EEA

The relation between the $hG$ and $hZ$
self-energies is given in \refeq{eq:hGhZ} below. Using this, we can express 
the $h \rightarrow f \bar{f}$
decay 
(\reffi{fig:hiGhiZ} with $h_i=h$) as 
\BEA
i\hat{\Sigma}_{hG}(p^2)\frac{i}{p^2-M_Z^2\xi_Z}\Gamma^{\rm tree}_{\rm Gf\bar{f}}+ip^{\nu}\hat{\Sigma}_{hZ}(p^2)\frac{-i\xi_Zp_{\mu}p_{\nu}}{p^2(p^2-M_Z^2\xi_Z)}\Gamma^{\mu,{\rm tree}}_{\rm Zf\bar{f}}
&=&-\frac{\Gamma^{\rm tree}_{\rm Gf\bar{f}}}{p^2}\hat{\Sigma}_{hG}(p^2),
\label{eq:goldstone}
\EEA
where $p^2$ denotes the momentum of the propagator involving the neutral
Higgs boson. 
Note that the expression in \refeq{eq:goldstone}
does not contain a pole at $p^2=M_Z^2\xi_Z$. 
Similarly, \refeq{eq:AGAZ} below gives the relation between the $AG$ and
$AZ$ self-energies. When substituted into the expression for the decay
$A\rightarrow f \bar{f}$ via a self-energy (\reffi{fig:hiGhiZ} with $h_i=A$), this gives
\BEA
&&i\hat{\Sigma}_{AG}(p^2)\frac{i}{p^2-M_Z^2\xi_Z}\Gamma^{\rm tree}_{\rm Gf\bar{f}}+ip^{\nu}\hat{\Sigma}_{AZ}(p^2)\frac{-i\xi_Zp_{\mu}p_{\nu}}{p^2(p^2-M_Z^2\xi_Z)}\Gamma^{\mu,{\rm tree}}_{\rm Zf\bar{f}}\non\\
&&=-\frac{\Gamma^{\rm tree}_{\rm Gf\bar{f}}}{p^2}\left(\hat{\Sigma}_{AG}(p^2)-(p^2-m_A^2)f_0(p^2)\frac{M_Z^2\xi_Z}{p^2-M_Z^2\xi_Z}\right),
\EEA
and the quantity $f_0$ is defined in \refeq{eq:fzero} below. The
expression above shows that it is essential to use the tree-level mass
for the incoming momentum, i.e.\ $p^2 = m_A^2$, in order to ensure the 
cancellation of the unphysical pole at $p^2=M_Z^2\xi_Z$. Therefore, in
the following, we treat the contributions involving mixing between $h_i$
and $G,Z$ bosons strictly at one-loop order, which implies, in particular,  evaluating those contributions at an incoming momentum corresponding to 
the tree level mass, rather than the loop corrected mass.

\subsection{Resummation of SUSY QCD contributions}
\subsubsection{The $\Delta m_b$ correction}
\label{section:deltamb}

\begin{figure}
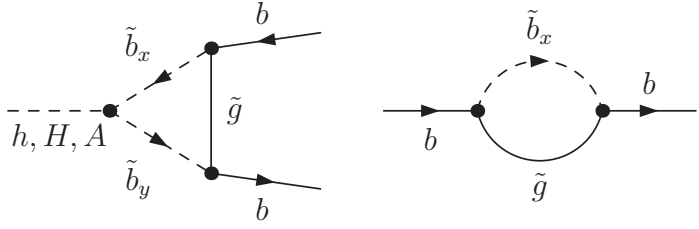

\begin{center}
\begin{tabular}{cc}
\unitlength=0.3bp%
\begin{feynartspicture}(432,504)(1,1)
\FADiagram{}
\FAProp(0.,10.)(6.5,10.)(0.,){/ScalarDash}{0}
\FALabel(3.25,9.18)[t]{$h,H,A$}
\FAProp(20.,15.)(13.,14.)(0.,){/Straight}{1}
\FALabel(16.2808,15.5544)[b]{$b$}
\FAProp(20.,5.)(13.,6.)(0.,){/Straight}{-1}
\FALabel(16.2808,4.44558)[t]{$b$}
\FAProp(6.5,10.)(13.,14.)(0.,){/ScalarDash}{-1}
\FALabel(9.20801,13.1807)[br]{$\tilde b_x$}
\FAProp(6.5,10.)(13.,6.)(0.,){/ScalarDash}{1}
\FALabel(9.20801,6.81927)[tr]{$\tilde b_y$}
\FAProp(13.,14.)(13.,6.)(0.,){/Straight}{0}
\FALabel(14.024,10.)[l]{$\tilde g$}
\FAVert(6.5,10.){0}
\FAVert(13.,14.){0}
\FAVert(13.,6.){0}
\end{feynartspicture}
&
\unitlength=0.3bp%
\begin{feynartspicture}(432,504)(1,1)
\FADiagram{}
\FAProp(0.,10.)(6.,10.)(0.,){/Straight}{1}
\FALabel(3.,8.93)[t]{$b$}
\FAProp(20.,10.)(14.,10.)(0.,){/Straight}{-1}
\FALabel(17.,11.07)[b]{$b$}
\FAProp(6.,10.)(14.,10.)(0.8,){/Straight}{0}
\FALabel(10.,5.98)[t]{$\tilde g$}
\FAProp(6.,10.)(14.,10.)(-0.8,){/ScalarDash}{1}
\FALabel(10.,14.27)[b]{$\tilde b_x$}
\FAVert(6.,10.){0}
\FAVert(14.,10.){0}
\end{feynartspicture}
\\
\end{tabular}
\caption{
\label{fig:deltamb}SUSY QCD corrections to the relation between the bottom
quark mass and the bottom Yukawa coupling
induced by gluino and sbottom quark loops, which can be 
enhanced at large $\tan \beta$, ($x,y=1,2$)}
\end{center}
\end{figure}

The tree level relation ($\mb = \lambda_b v_1$) between the bottom quark mass and the
bottom Yukawa coupling $\lambda_b$ receives large $\tb$-enhanced radiative corrections, 
which need to be properly taken into account~\cite{NUPHA.B303.172,hep-ph/9306309,PHRVA.D49.6168,hep-ph/9402253,Carena:1999py,hep-ph/9912463,hep-ph/0305101,arXiv:0808.0087,arXiv:1001.1935}.
In SUSY QCD, such contributions arise from loops containing gluinos and
sbottoms, as shown in \reffi{fig:deltamb}. For heavy SUSY mass scales 
the interaction of the neutral Higgs bosons with bottom quarks can be
expressed in terms of an effective Lagrangian~\cite{Carena:1999py}
\BEA
{\cal L}_{\rm eff}&=&-\lambda_b \bar{b}_R 
\left[ H_{11}+\frac{\Delta m_b}{t_{\beta}}H_{22}^*\right]b_L
+\mathrm{h.c.} ,
\label{eq:DelbLag}
\EEA
where the shorthand $t_{\beta} \equiv \tb$ has been used. Accordingly,
the relation between the bottom quark mass and the bottom Yukawa
coupling receives the loop-induced contribution $\Delta m_b$ such that 
\begin{equation}
m_b=\lambda_b v_1 \left(1+\Delta m_b\right) . 
\label{eq:Delta_b}
\end{equation}

We consider here the general case, in which $\Delta m_b$ is
allowed to be complex. Inserting the relation (\refeq{eq:Delta_b}) and
neglecting the terms involving Goldstone boson contributions leads to
\BEA
{\cal L}_{\rm eff}= \bar{b}\frac{1}{1+y}&&\left(\left[1-\frac{1}{t_{\alpha}t_{\beta}}y+i\gamma_5x\left(1+\frac{1}{t_{\alpha}t_{\beta}}\right)\right]v_{\rm h\bar{b}b}^{\rm tree}h\right.\non\\
&&+\left[1+\frac{t_{\alpha}}{t_{\beta}}y+i\gamma_5x\left(1-\frac{t_{\alpha}}{t_{\beta}}\right)\right]v_{\rm H\bar{b}b}^{\rm tree}H\non\\
&&+\left.\left[1-\frac{1}{t_{\beta}^2}y+i\gamma_5x\left(1+\frac{1}{t_{\beta}^2}\right)\right]v_{\rm A\bar{b}b}^{\rm tree}A\right)b+...,
\label{eq:Leffdeltamb}
\EEA
where $t_\al \equiv \tan\al$. The quantities $x,y$ are real and given by
\BEA
x&=&\frac{{\rm Im} \Delta m_b}{1+{\rm Re}\Delta m_b},\non\\
y&=&{\rm Re}\Delta m_b+x{\rm Im} \Delta m_b , \label{eq:xy}
\EEA
and $v_{\rm h\bar{b}b}^{\rm tree}$, $v_{\rm H\bar{b}b}^{\rm tree}$,
$v_{\rm A\bar{b}b}^{\rm tree}$ are defined by 
\BEA
{\cal L}^{\rm tree}&=&\bar{b}\left[v_{\rm h\bar{b}b}^{\rm tree}h+v_{\rm H\bar{b}b}^{\rm tree}H+v_{\rm A\bar{b}b}^{\rm tree}A\right]b+...\\
&=&\bar{b}\left[-\frac{\lambda_b^{(0)}}{\sqrt{2}}(-s_{\alpha})
h-\frac{\lambda_b^{(0)}}{\sqrt{2}}(c_{\alpha})H-\frac{\lambda_b^{(0)}}{\sqrt{2}}(i\gamma_5)(-s_{\beta})A\right]b+
\ldots ,
\label{Ltree}
\EEA
with 
\BEA
\lambda_b^{(0)}&=m_b/v_1&=m_be/\left(\sqrt{2}c_{\beta}s_WM_W\right).
\label{eq:lambdab0}
\EEA 
Note that, in this convention, $v_{\rm A\bar{b}b}^{\rm tree}$ contains a $\gamma_5$ dependence.

In order to find $\Delta m_b$, we perform a Feynman-diagrammatic calculation of the leading 1-loop gluino contributions to the 
$h_i\to b\bar{b}$ decays, using the $p^2=0$ approximation and $i=h,H,A$. 
Comparing this calculation for the renormalised decay width to the 
1-loop expansion of \refeq{eq:Leffdeltamb} yields for the contribution
of gluino and sbottom loops to $\Delta m_b$
\BEA
\Delta m_b^{\tilde{g}} &=& \frac{2}{3}\frac{\alpha_s}{\pi}\mu^*M_3^*t_{\beta}I\left(m^2_{\tilde{b}_1},m^2_{\tilde{b}_2},m^2_{\tilde{g}}\right),\\
I\left(a,b,c\right)&=&-\frac{ab {\rm Log}\left(\frac{b}{a}\right)+ac{\rm Log}\left(\frac{a}{c}\right)+bc {\rm Log}\left(\frac{c}{b}\right)}{\left(a-c\right)\left(c-b\right)\left(b-a\right)}
\label{dmbgluino}.
\EEA

In the $h_i\to b \bar{b}$ decay, diagrams involving charged higgsinos also 
contain $\tan {\beta}$ enhanced 
contributions~\cite{Carena:1999py,Dittmaier:2006cz}. We treat these 
analogously to the $\Delta m_b^{\tilde{g}}$ corrections above. 
Comparison with the 1-loop Feynman-diagrammatic calculation in the 
complex MSSM leads to
\BEA
\Delta m_b^{\tilde{h}} &=& \frac{\alpha_t}{4 \pi} A_t^* \mu^* t_{\beta} I\left(m^2_{\tilde{t}_1},m^2_{\tilde{t}_2},\left|\mu\right|^2\right)
\label{dmbhiggsino},
\EEA
where
\BEA
\alpha_t&=&\frac{\lambda^{(0),2}_t}{4 \pi},\\
\lambda^{(0)}_t&=& \frac{m_t}{v_2} =\frac{m_t e}{\sqrt{2}s_{\beta}s_W M_W}.
\EEA

The effective Lagrangian of \refeq{eq:DelbLag} properly resums the 
leading $\tb$-enhanced gluino and higgsino contributions given 
above~\cite{Carena:1999py}. In our calculation we therefore use 
a $\Delta m_b$ correction of
\BEA
\Delta m_b&=&\Delta m_b^{\tilde{g}}+\Delta m_b^{\tilde{h}}\\
&=&\frac{2}{3}\frac{\alpha_s}{\pi}\mu^*M_3^*t_{\beta}I\left(m^2_{\tilde{b}_1},m^2_{\tilde{b}_2},m^2_{\tilde{g}}\right)+
\frac{\alpha_t}{4 \pi}A_t^* \mu^* t_{\beta} I\left(m^2_{\tilde{t}_1},m^2_{\tilde{t}_2},\left|\mu\right|^2\right).\label{eq:dmb}
\EEA

It is also possible to incorporate effects from loops involving winos into $\Delta m_b$ as in \citere{Carena:1999py} (or even winos and binos as in \citere{Dittmaier:2006cz}). We do not include these, since they are numerically small~\cite{Carena:1999py} and, in the $\CPX$ scenario, are less important than the higgsino contributions. Since we will explicitly calculate the 1-loop diagrams involving winos and binos when calculating the full 1-loop $h_a\to b \bar{b}$ 
decay width (the $\Delta m_b$ contributions are subtracted at one-loop
order such that a double-counting from the $\Delta m_b$ resummation is
avoided), the effect of leaving them out of the $\Delta m_b$ contribution is of 
sub-leading 2-loop order.
For the scale of $\als$ in \refeq{eq:dmb} we choose the top-quark mass,
i.e.\ we use $\alpha_s(m_t^2)$ in $\Delta m_b$.

\subsubsection{The use of an effective $b$-quark mass in the 
calculation of the neutral Higgs self-energies}
\label{section:deltambinRSE}

\begin{figure*}
\begin{center}
\resizebox{\textwidth}{!}{%
  \includegraphics{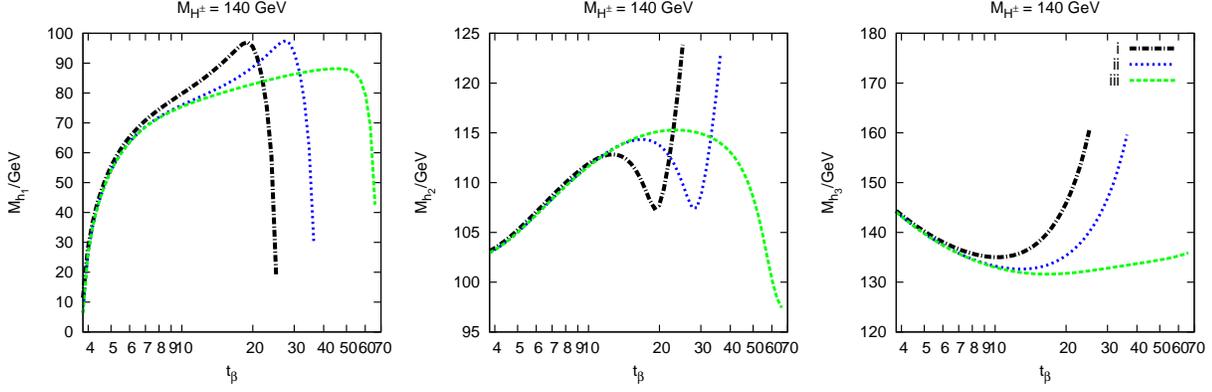}
}
\end{center}
\caption{Predictions for the neutral Higgs masses,
$M_{h_1}$, $M_{h_2}$, $M_{h_3}$,
obtained using the effective bottom quark mass
employed for the numerical analyses in this paper, (iii)
$m_{b,{\rm eff}}=m_b^{\drbarm, SM}(m_t^{OS})/|1+\Delta m_b^{\rm FH}|$ in
comparison with two other choices for the bottom mass, (i) 
$m_b^{\msbarm, SM}(m_b)$ and (ii) $m_b^{\drbarm, SM}(m_t^{OS})$. The results
are shown for the $\CPX$ scenario with $\MHpm=140\gev$.}
\label{fig:FHDeltaMB}       
\end{figure*}

It is desirable to incorporate leading $\tb$-enhanced contributions also
into the calculation of the Higgs self-energies. A direct application of
the effective Lagrangian of \refeq{eq:DelbLag} within loop calculations
is in general not possible, as it would spoil (among other things) 
the UV-finiteness of the theory. It is therefore convenient to absorb
the leading $\tb$-enhanced contributions into an effective bottom quark
mass that is used everywhere in the calculation, although this procedure
does not reproduce the decoupling properties of the effective Lagrangian
of \refeq{eq:DelbLag} in the limit where $\MHp \gg \MZ$. It has been
shown in \citeres{hep-ph/0206101,hep-ph/0411114} that the one-loop
result with an appropriately chosen effective bottom quark mass in
general approximates very well the result containing the diagrammatic
two-loop contributions.

Because of the large value of $\mu$ in the CPX scenario, contributions
from the bottom / sbottom sector can be important already for moderate
values of $\tb$. 
We use in the following an effective bottom quark mass that is defined 
as
\begin{equation}
m_{b,{\rm eff}}= \frac{m_b^{\drbarm, SM}(m_t^{OS})}{|1+\Delta m_b^{\rm FH}|},
\label{eq:mb_eff}
\end{equation}
where `$\drbarm$' or `$OS$' indicates the renormalisation scheme in which the mass is defined i.e.\ the $\drbarm$ or $OS$ renormalisation scheme respectively, and `SM' indicates that only the Standard Model contribution is included. $\Delta m_b^{\rm FH}$ is the $\Delta
m_b$ correction calculated internally by {\fh} for use in its Higgs
decays and Higgs production cross sections.

Since in the CPX scenario $|1+\Delta m_b| > 1$, the incorporation of the 
$\tb$-enhanced $\Delta m_b$ corrections leads to a reduction of the numerical 
value of $m_{b,{\rm eff}}$ in this scenario
and thus to numerically more stable results. This
is illustrated in \reffi{fig:FHDeltaMB}, where the results for the
neutral Higgs masses obtained using the effective bottom quark mass
defined in \refeq{eq:mb_eff} are compared with the predictions arising
from choosing $m_b^{\msbarm, SM}(m_b)$ or $m_b^{\drbarm, SM}(m_t^{OS})$
for the bottom quark mass. For large values of $\tb$, depending on the
choice of the bottom mass, an onset of very large corrections from the 
bottom / sbottom sector is visible in the predictions for the Higgs
masses. Since perturbative predictions are not reliable in the parameter
region where these large corrections from the bottom / sbottom sector 
occur, we limit the numerical analyses in this paper to the region 
$\tb < 30$. (Without resummation of the $\tb$-enhanced contributions as in \refeq{eq:mb_eff}, the onset of very large corrections could already occur at lower values of $\tb$.)



\section{Conversion of parameters between on-shell and 
$\drbarm$ renormalisation schemes}
\label{sec:convren}

As mentioned above, higher-order contributions in the Higgs sector of
the MSSM have been obtained using different approaches. The results
implemented in the public code {\feynhiggs} are based on the
Feynman-diagrammatic approach employing the on-shell renormalisation
scheme, while the results implemented in the public code {\cpsuperh} are
based on a renormalisation group improved effective potential calculation 
employing $\drbarm$ renormalisation. As the parameters in the two
renormalisation schemes are defined differently, a parameter conversion
is necessary for a meaningful comparison of results obtained in the two
schemes. 

\subsection{Parameter shifts}

\begin{center}
\begin{figure}
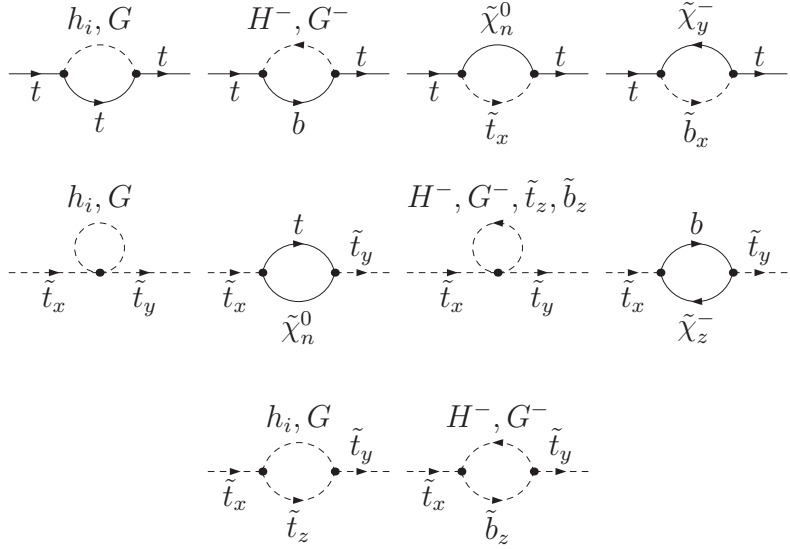

\begin{center}
\input alphatdiag
\caption{The diagrams used to calculate the shifts shown in 
\refeqs{eq:convren1} -- (\ref{eq:convren2}) 
(which convert between $\drbarm$ and on-shell parameters) at $\mathcal{O}(\alpha_t)$. ($x=1,2,\, y=1,2,\ z=1,2,\ i=1,2,3,\ n=1,2,3,4$)}
\label{fig:convren2}
\end{center}
\end{figure}
\end{center}

Since both schemes incorporate partial 2-loop contributions, a parameter conversion of
the top/stop sector parameters (that enter at the 1-loop level), is
required. In \citeres{hep-ph/0001002,Brignole:2001jy} this issue
has been discussed for the case where all the MSSM parameters are real. 
In the following we consider the general case of arbitrary complex
parameters. We can obtain the leading terms at 
$\mathcal{O}(\alpha_s)$ from considering loops involving gluons, gluinos, stops and tops as shown in 
\reffi{fig:convren}. In order to obtain the leading terms at 
$\mathcal{O}(\alpha_t)$, we must also consider loops involving neutralinos, charginos, Higgs bosons, Goldstone bosons, sbottoms and b-quarks, as shown in \reffi{fig:convren2}.

\begin{center}
\begin{figure}
\begin{center}
\input alphasdiag
\caption{The diagrams used to calculate the shifts shown in 
\refeqs{eq:convren1} -- (\ref{eq:convren2}), 
(which convert between $\drbarm$ and on-shell parameters) at $\mathcal{O}(\alpha_s)$. ($x=1,2,\, y=1,2,\ z=1,2$)}
\label{fig:convren}
\end{center}
\end{figure}
\end{center}

We label the difference between the parameters $p$ in the different renormalisation schemes by $\Delta p$, where 
\BEA
p^{\rm \drbarm}&=&p^{\rm on-shell}+\Delta p(\mu_{\rm ren}) .
\label{eq:ideaconvren}
\EEA
Since the $\drbarm$ parameters depend on the
renormalisation scale $\mu_{\rm ren}$, the shift $\Delta p$ is also a
function of $\mu_{\rm ren}$.
The parameter shift $\Delta p$ is related to the counterterms by
\BEA
\Delta p(\mu_{\rm ren}) &=\delta p^{\rm on-shell}-\delta p^{\rm \drbarm}&=
\delta p^{\rm on-shell}-\left[\delta p^{\rm on-shell}\right]^{\rm div},
\EEA
where the superscript `div' denotes that only terms proportional to 
$\frac{2}{4-D}-\gamma_E+\log(4\pi)$ are kept. Therefore this means that 
$\Delta p(\mu_{\rm ren})=\left[\delta p^{\rm on-shell}\right]^{\rm fin}$, 
where the superscript `fin' denotes the finite pieces remaining once terms 
proportional to $\frac{2}{4-D}-\gamma_E+\log(4\pi)$ have been subtracted out.

For the stop sector, we can directly adapt the counterterms used in {\fh} 
in \citere{Heinemeyer:2007aq} 
(see \refapp{sec:apprencon} for the explicit form of the
counterterms) to get the parameter shifts
\BEA
\Delta M_L^2&=&-2m_t\Delta m_t
+U_{11}^*U_{11}\Delta m^2_{\tilde{t}_1}+U_{21}^*U_{21}\Delta m^2_{\tilde{t}_2}+U_{11}^*U_{21}\Delta Y_{\tilde{t}}+U_{21}^*U_{11}\Delta Y^*_{\tilde{t}}\label{eq:convren1},\\
\Delta M_{\tilde{t}_R}^2 &=&-2m_t\Delta m_t
+U_{12}^*U_{12}\Delta m^2_{\tilde{t}_1}+U_{22}^*U_{22}\Delta m^2_{\tilde{t}_2}+U_{12}^*U_{22}\Delta Y_{\tilde{t}}+U_{22}^*U_{12}\Delta Y^*_{\tilde{t}},\\
\Delta A^*_{t}&=&e^{-i\phi_{A_t}}\left(\Delta \left|A_t\right|-i \left|A_t\right|\Delta\phi_{A_t} \right),\\
\Delta K_t&=&-\left(A_t^*-\mu \cot \beta\right)\Delta m_t+m_t \cot \beta \Delta\mu \nonumber \\
&&+U_{11}^*U_{12}\Delta m^2_{\tilde{t}_1}+U_{21}^*U_{22}\Delta m^2_{\tilde{t}_2}+
U_{11}^*U_{22}\Delta Y_{\tilde{t}}+U_{21}^*U_{12}\Delta Y^*_{\tilde{t}},
\\
\Delta \left|A_t\right|&=&\frac{1}{m_t}{\rm Re}\left[e^{i\phi_{A_t}}\Delta K_t\right],\\
\Delta \phi_{A_t}&=&-\frac{1}{m_t\left|A_t\right|}{\rm Im}\left[e^{i\phi_{A_t}}\Delta K_t\right],
\label{eq:convren2}
\EEA
where
\BEA
\Delta m_t&=&\frac{1}{2}{\rm Re}\left[ m\left(\Sigma_t^L(m^2)+\Sigma_t^R(m^2)\right)+\Sigma_t^l(m^2)+\Sigma_t^r(m^2)\right]^{\rm fin},\\
\Delta m_{\tilde{t}_1}^2&=&{\rm Re}\left[\Sigma_{\tilde{t}_{11}}(m^2_{\tilde{t}_1})\right]^{\rm fin},\\
\Delta m_{\tilde{t}_2}^2&=&{\rm Re}\left[\Sigma_{\tilde{t}_{22}}(m^2_{\tilde{t}_2})\right]^{\rm fin},\\
\Delta Y_{\tilde{t}}&=&\frac{1}{2}\left[\widetilde{\rm Re}\Sigma_{\tilde{t}_{12}}(m^2_{\tilde{t}_1})+\widetilde{\rm Re}\Sigma_{\tilde{t}_{12}}(m^2_{\tilde{t}_2})\right]^{\rm fin},
\EEA
and the components of the top self-energy are defined by 
$\Sigma(p)=\pslash\omega_-\Sigma^L(p^2)+
\pslash\omega_+\Sigma^R(p^2)+\omega_-\Sigma^l(p^2)+\omega_+\Sigma^r(p^2)$,
and ${\rm \widetilde{Re}}$ indicates that the imaginary parts of the loop integrals are discarded.

In the following we will define the $\drbarm$ parameters at the scale
$\mu_{\rm ren}=\sqrt{M_{\rm SUSY}^2+m_t^2}=:M_S$, see \citere{Lee:2003nta}
and the discussion in \citere{hep-ph/0001002}. 
We evaluate the strong coupling constant at the scale of the top mass, 
$\alpha_s(m_t^2)$. Since the two-loop
corrections of $\mathcal{O}(\alpha_t\alpha_s)$ and
$\mathcal{O}(\alpha_t^2)$
implemented in {\fh} have been obtained using a Yukawa approximation 
(the corresponding contributions are implemented in 
{\cpsh} only to leading logarithmic accuracy),
we employ the same kind of approximation to derive the
parameter conversions at $\mathcal{O}(\alpha_s)$ and
$\mathcal{O}(\alpha_t)$.
In particular, we neglect the D-terms in the stop mass matrix (i.e.,
the terms proportional to $M_Z^2$), the D-terms and the b-quark mass in the sbottom
mass matrix, and we neglect terms proportional to 
$M_1$, $M_2$, $M_Z$, $M_W$ in the neutralino and chargino mass matrices. In addition, we
make the approximation $M_{H\pm}=M_A$ when deriving the parameter shifts.
 
For the evaluation of the parameter shifts at $\mathcal{O}(\alpha_t)$ we
also need to consider a shift to the Higgsino mass parameter $\mu$,
since {\fh} takes $\mu^{\drbarm}(m_t)$ as input while {\cpsh} takes
$\mu^{\drbarm}(M_S)$ as input. Therefore we use the relation
\BEA
\mu(m_t)&=&\mu(M_S)+
\frac{3\alpha_{t}}{8\pi}\mu\log\left(\frac{m_t^2}{M_S^2}\right).
\label{eq:mueshift}
\EEA

The difference between using $\drbarm$ or
on-shell quantities as input to the shifts $\De p$ is of higher order.

\subsection{Simple approximation of parameter shifts}
\label{sec:basic}

It is useful to find a simple approximation for the $\mathcal{O}(\alpha_s)$ contribution to the parameter shifts 
$\Delta p$. It turns out that the shifts in $M_L$ and $M_{\tilde{t}_R}$ 
in general are less numerically significant than the shifts in $A_t$.
Therefore we investigate an approximate treatment in which the 
shifts in $M_L$ and $M_{\tilde{t}_R}$ are neglected. Furthermore, 
since we have neglected the D-terms in the stop mass matrix and use 
$M_{\rm SUSY}^{\drbarm}(M_S) =M^{\drbarm}_{L} (M_S)=
M^{\drbarm}_{\tilde{t}_R}(M_S)$, the stop mixing matrix defined in \refeq{eq:squarkmixmatrix} has the simple form
\BEA
\matr{U}_{\tilde{q}}=\frac{1}{\sqrt{2}}\ML 1 & -e^{-i\phi_{X_t}} \\ e^{i\phi_{X_t}} & 1 \MR.
\EEA
Accordingly, the relation between $|X_t^{\rm on-shell}|$ and 
$|X_t^{\drbarm}(M_S)|$ simplifies to
\BEA
|X_t^{\rm on-shell}|&=&|X_t^{\drbarm}|
\left(1+\frac{\Delta m_t}{m_t}\right)
-\frac{1}{2m_t}\left(\Delta m^2_{\tilde{t}_2}-\Delta m^2_{\tilde{t}_1}\right),
\label{deltaXtshift}
\EEA
where
\BEA
\Delta m_t
&=&
\frac{\alpha_s}{6 \pi m_t}
\left[g_1^t+g_2^t+m_t^2
\left(6\log\frac{m_t^2}{\mu^2_{\rm ren}}-10\right)
+2m_{\tilde{g}}^2\left(\log\frac{m_{\tilde{g}}^2}{\mu^2_{\rm ren}}-1\right)
\right]
\label{eq:appoxconvenbeg},\\
\Delta m^2_{\tilde{t}_1}-\Delta m^2_{\tilde{t}_2}
&=&
\frac{2\alpha_s}{3 \pi}\left(g_1^{\tilde{t}}-g_2^{\tilde{t}}\right),\\
g_i^t
&=&
-m_{\tilde{t}_i}^2\left(\log\frac{m_{\tilde{t}_i}^2}{\mu^2_{\rm ren}}-1\right)
+f_i{\rm Re}\left[\mathcal{B}_0[m_t^2,m_{\tilde{g}}^2,m_{\tilde{t}_i}^2]\right],\\
g_i^{\tilde{t}}
&=&
2m_{\tilde{t}_i}^2\left(\log\frac{m_{\tilde{t}_i}^2}{\mu^2_{\rm ren}}-2\right)
-f_i{\rm Re}\left[\mathcal{B}_0[m_{\tilde{t}_i}^2,m_{\tilde{g}}^2,m_t^2]\right],\\
f_i
&=&
m_{\tilde{g}}^2+m_t^2-m_{\tilde{t}_i}^2-(-1)^i2m_{\tilde{g}}m_t \cos (\phi_{M_3}-\phi_{X_t}),
\label{eq:appoxconvenend}
\EEA
and $\mathcal{B}_0=\left[B_0\right]^{\rm fin}$, with the scalar integral $B_0$ defined as
\BEA
B_0(p_1^2,m_0^2,m_1^2)&=&\frac{(2\pi\mu)^{4-D}}{i\pi^2}\int d^Dq \frac{1}{(q^2-m_0^2+i\epsilon)((q+p_1)^2-m_1^2+i\epsilon)}. 
\EEA

\subsection{Numerical examples in the $\CPXdrbar$ scheme}

\begin{figure*}
\begin{center}
\resizebox{\textwidth}{!}{%
  \includegraphics{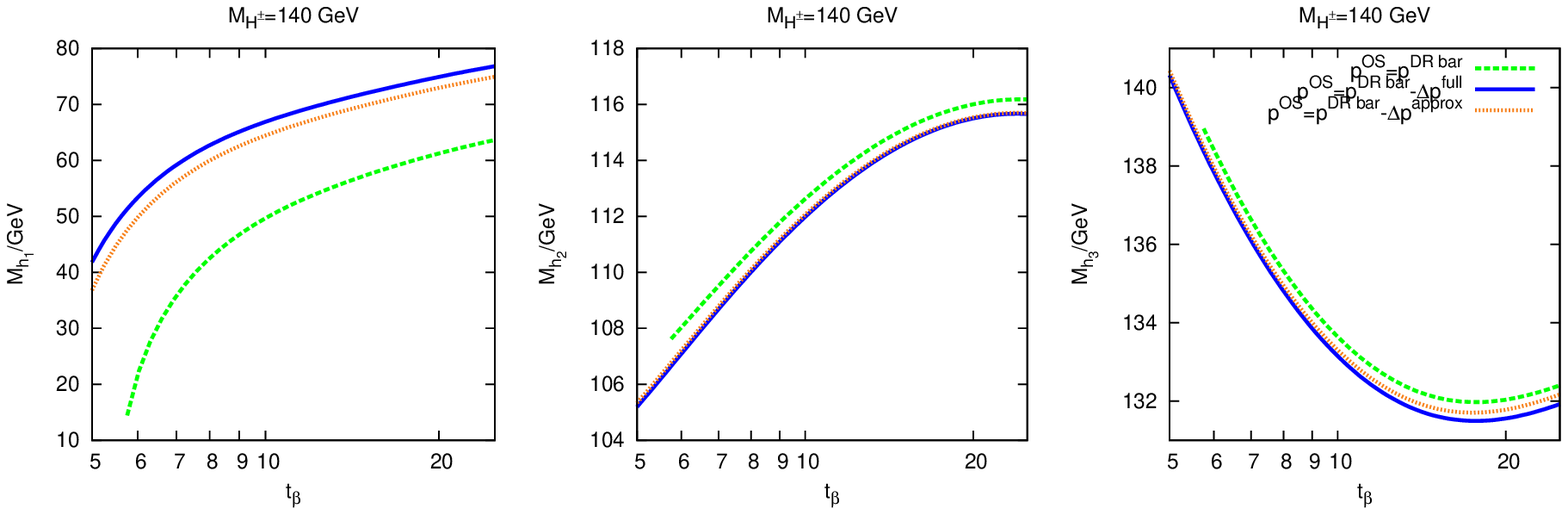}
}
\end{center}
\caption{Impact of parameter conversions on the neutral Higgs masses 
$M_{h_1}$, $M_{h_2}$, $M_{h_3}$. The predictions for the Higgs masses,
evaluated with the program {\fh}, are shown as a function of $\tb$ in
the $\CPXdrbar$ scenario. Green (dashed): Result where the numerical
values of the input parameters $A_t^{\drbarm}(M_S)$, 
$M_L^{\drbarm}(M_S)$, $M_{\tilde{t}_R}^{\drbarm}(M_S)$ are directly
inserted into {\fh} without a parameter conversion (i.e., $\Delta p=0$).
Blue (solid): Output of {\fh} if on-shell values of the parameters
are inserted that are obtained from the ${\drbarm}$ input values using
the full expressions for $\Delta p$ given in 
\refeqs{eq:convren1}--(\ref{eq:convren2}) 
at $\mathcal{O}(\alpha_s)$. Orange (dotted): Output of {\fh} if
the parameter conversion is calculated using the approximation described in
\refse{sec:basic}.}
\label{fig:convren3}       
\end{figure*}

In \reffi{fig:convren3} we investigate the impact of the $\mathcal{O}(\alpha_s)$ parameter
conversions on the predictions for the neutral Higgs masses
$M_{h_1}$, $M_{h_2}$, $M_{h_3}$ (see \refse{sec:comp_cpsh} for a discussion of the $\mathcal{O}(\alpha_t)$ terms). We compare the case where the numerical
values of the input parameters $A_t^{\drbarm}(M_S)$,
$M_L^{\drbarm}(M_S)$, $M_{\tilde{t}_R}^{\drbarm}(M_S)$ in the 
$\CPXdrbar$ scheme (for $\MHpm=140\gev$) are directly inserted as 
input into the program {\fh} (in which the parameters are interpreted as
on-shell quantities) with the case where a proper conversion of the 
${\drbarm}$ input values to on-shell parameters has been carried out
at $\mathcal{O}(\alpha_s)$
according to \refeqs{eq:convren1}--(\ref{eq:convren2}). One can see that
the parameter shifts have a very large numerical impact on the
prediction of the lightest Higgs mass of more than $20 \gev$ in the
region of small $\tb$, while the corresponding effects on the
predictions for $M_{h_2}$ and $M_{h_3}$ are typically below the GeV level. 
The result for $M_{h_1}$ indicates the well-known fact that corrections
of \order{\alt\als} in the MSSM Higgs sector can be numerically very
important. The numerical effects found here in the $\CPXdrbar$ scenario
are even larger than the corresponding shifts in the case of real
parameters as discussed in \citere{hep-ph/0001002}. 
As a consequence, it is obvious that a proper conversion of parameters 
at least for the
prediction of the lightest Higgs mass is indispensable for a meaningful
comparison of results obtained in different renormalisation schemes. 

Also shown in \reffi{fig:convren3} is the result obtained from employing
the approximate parameter conversion as given in
\refse{sec:basic}. One can see that the result for the approximate
treatment is close to the one obtained with the full parameter
conversion. This indicates that the main impact of the parameter
conversion is indeed caused by the shift in
the absolute value of the trilinear coupling $\Delta |A_t|$, as expected
from the discussion above. From the expressions given in
\refse{sec:basic} one can furthermore see 
that there is a significant dependence on
the phase $(\phi_{M_3}-\phi_{X_t})$ and that the gluino mass $|M_3|$
plays an important role. 
We will use the full $\mathcal{O}(\alpha_s)$ parameter
conversions throughout the rest of the paper. However, we note that
this approximate parameter conversion can be useful in situations 
where the inclusion of the full $\mathcal{O}(\alpha_s)$ parameter
conversions is impractical.

\subsection{Reparametrisation of $m_t$ in the neutral Higgs self-energies}
\label{sec:choiceofmtinFH}

The difference between parametrising the neutral Higgs self-energies in terms of the on-shell top mass and parametrising the neutral Higgs self-energies in terms of the $\msbarm$ top mass is formally a three-loop effect. Previously, we have chosen to use an on-shell top mass. We shall now investigate the numerical effect of parametrising in terms of the $\msbarm$ top mass $\overline{m}_t:= m_t^{\msbarm,SM}(m_t)$. In order to simplify the following discussion, we shall assume no resummation of $\tan \beta$-enhanced terms.

So far, we have been using neutral Higgs self-energies expressed in terms of the on-shell top mass: 
\BEA
\hat{\Sigma}(m_t^{OS})&=& \hat{\Sigma}^{(1)}_{\rm Yuk}(m_t^{OS})+ \hat{\Sigma}^{(1)}_{\rm non-Yuk}(m_t^{OS})+ \hat{\Sigma}^{(2)}_{\rm Yuk}(m_t^{OS}) +h.o.t.
\EEA
where `$h.o.t.$' stands for `higher order terms', and the on-shell top mass and the {\msbar} top mass are related through
\BEA
\overline{m}_t&=&m_t^{OS}+\alpha_s x\non\\
x&=&-\frac{4}{3\pi}m_t
\label{eq:mtosanddrbar}
\EEA
The difference between using an on-shell top mass and the {\msbar} top mass in $x$ is not important, since it will affect the calculation only at the 4-loop level. (We shall use $m_t^{OS}$).

We substitute for $m_t^{OS}$ in $\hat{\Sigma}(m_t^{OS})$
\BEA
\hat{\Sigma}(m_t^{OS})&=&\hat{\Sigma}^{(1)}_{\rm Yuk}(\overline{m}_t-\alpha_s x)+ \hat{\Sigma}^{(1)}_{\rm non-Yuk}(\overline{m}_t-\alpha_s x)+ \hat{\Sigma}^{(2)}_{\rm Yuk}(\overline{m}_t-\alpha_s x) +...
\EEA
and perform an expansion around $x=0$, keeping terms up to $\mathcal{O}(\alpha_t\alpha_s)$
\BEA
\hat{\Sigma}(m_t^{OS})&=& \hat{\Sigma}^{(1)}_{\rm Yuk}(\overline{m}_t)- \alpha_s x \hat{\Sigma}^{(1)'}_{\rm Yuk}(\overline{m}_t)+ \hat{\Sigma}^{(1)}_{\rm non-Yuk}(\overline{m}_t)+ \hat{\Sigma}^{(2)}_{\rm Yuk}(\overline{m}_t) +...
\EEA
The reparametrisation according to \refeq{eq:mtosanddrbar} has generated an additional term $-\alpha_s x \hat{\Sigma}^{(1)'}_{\rm Yuk}(m)$, which is of two-loop order. In order to calculate this additional term, we require the explicit expressions for the leading 1-loop Yukawa corrections to the neutral Higgs self-energies, which we give explicitly in \refeq{eq:beginyukse}--\refeq{eq:endyukse} below.

We first substitute for the stop masses their expressions in terms of
the top mass and the soft SUSY-breaking parameters, in order to ensure 
that the top mass dependence is explicit everywhere. We employ here the
Yukawa approximation, i.e.\ \refeq{eq:stopmasses} simplifies to
\BEA
m_{\tilde q_{1,2}}^2 &=& \mq^2
  + \edz \Bigl[
M_L^2 + M_{\tilde{q}_R}^2 \mp \sqrt{[M_L^2 - M_{\tilde{q}_R}^2]^2 + 4 \mq^2 |\Xq|^2}~\Bigr].
\label{eq:Yukstopmasses}
\EEA
Then we make the substitution $m_t \to m_t- \alpha_s x$ and expand
around $x=0$ to obtain terms at $\mathcal{O}({\alpha_t\alpha_s})$. We then edit the {\fh} code to include these additional terms and make use of the option in {\fh} where the tree level stop sector parameters are calculated using $\overline{m}_t$ rather than $m_t^{OS}$.\footnote{Note that the method described here differs from the {\fh} option {\tt runningMT = 1} at the 2-loop order.}

\begin{figure}
\begin{center}
\resizebox{\textwidth}{!}{%
\includegraphics{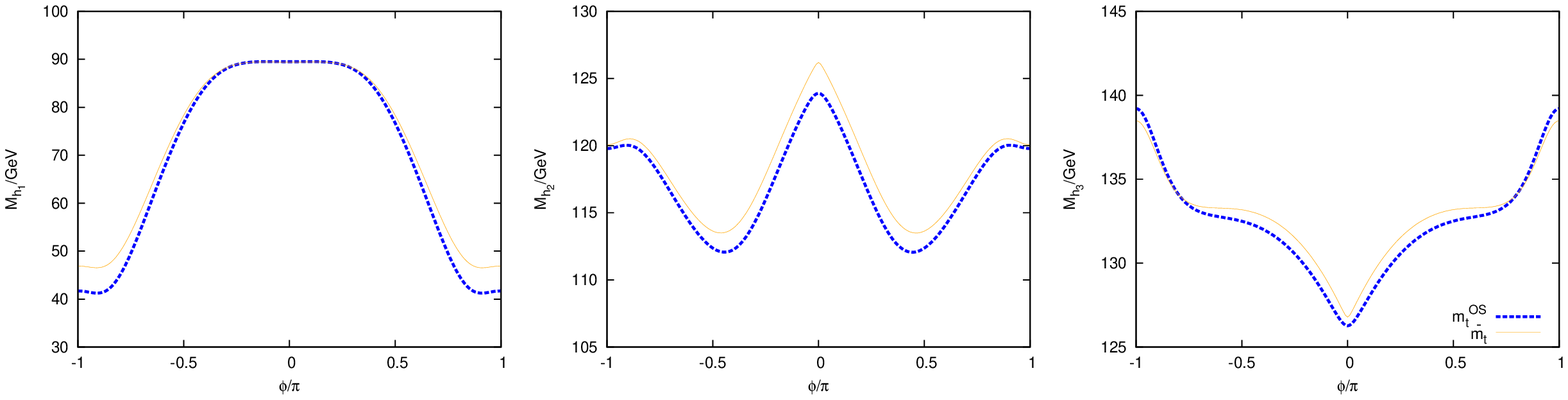}
}
\end{center}
\caption{Neutral Higgs masses in the $CPX$ scenario at $\tb=11$ and $M_{H^{\pm}}=140\gev$ as function of phase $\phi=\phi_{A_t}=\phi_{M_3}$. Blue (dashed): the Higgs self-energy calculation is parameterised in terms of $m_t^{OS}$, orange (solid): the calculation is parametrised in terms of $\overline{m}_t$.
\label{fig:changemtintwoloop}}
\end{figure}

\subsubsection{Numerical results}
\reffi{fig:changemtintwoloop} shows the neutral Higgs masses in the $CPX$ scenario at $\tb=11$ and $M_{H^{\pm}}=140\gev$ as function of phase $\phi=\phi_{A_t}=\phi_{M_3}$, with the Higgs self-energy calculation parameterised in terms of $m_t^{OS}$ (blue, dashed) and in terms of $\overline{m}_t$ (orange, solid). These results include the resummation of $\tan \beta$-enhanced terms. We can see that, even in the CP-conserving limit, parameterising the calculation in terms of $\overline{m}_t$ rather than $m_t^{OS}$ can increase the lightest Higgs mass by $5.3\gev$. However, note that we will be interested in the maximally CP-violating case $\phi=\pi/2$, where we can see that the effect on $M_{h_1}$ is more modest: a $1.6\gev$ absolute increase or a $2 \%$ relative increase.

\section{Higgs cascade decay}
\label{sec:hihjhk}

 We use the general expression for 2-body neutral Higgs boson decays\footnote{In this convention, the capital letter $\Gamma$ denotes a decay width when it has an argument which  explicitly contains the symbol `$\to$' (e.g.\ $\Gamma (h_a\rightarrow h_b h_c)$) and a vertex function when it does not (e.g.\ $\Gamma^{\rm tree}_{Gf\bar{f}}$).},
\BEA
\Gamma (h_a\rightarrow h_b h_c)&=&\frac{\mathcal{S}\rho|\mathcal{M}|^2}{8\pi M_{h_a}^2},
\label{eq:triplehdecaywidth}
\EEA
where $\mathcal{M}$ is the matrix element. 

For identical final state particles, $h_b=h_c$,  
\BEA
\rho&=&\frac{M_{h_a}}{2}\sqrt{1-\frac{4M_{h_b}^2}{M_{h_a}^2}},
\EEA
and the symmetry factor $\mathcal{S}$ is $\frac{1}{2}$. For the case $h_b \ne h_c$, 
\BEA
\rho&=&\frac{1}{2 M_{h_a}}\sqrt{M_{h_a}^4+M_{h_b}^4+M_{h_c}^4-2
\left(M_{h_a}^2M_{h_b}^2+M_{h_b}^2M_{h_c}^2+M_{h_c}^2M_{h_a}^2\right)},
\EEA
and the symmetry factor is 1. 

Since the lowest order contribution involves only scalar particles, 
the tree level decay width has a very simple form. For example, the $h\to AA$ tree level decay width is given by
\BEA
\Gamma^{\rm tree} (h\rightarrow AA)&=&\frac{\lambda_{hAA}^{\rm tree,2}}{32\pi m_{h_a}}\sqrt{1-\frac{4m_{A}^2}{m_{h}^2}},{\rm \, with \,} \lambda_{hAA}^{\rm tree}=-c_{2\beta}s_{\alpha+\beta}\frac{eM_W}{2c_W^2s_W}.
\EEA

\subsection{Calculation of the genuine $h_i\to h_jh_k$ vertex contributions}

We calculate the full 1PI (one-particle irreducible) 1-loop vertex
corrections to the $h_i\to h_jh_k$ decay width within the
Feynman-diagrammatic approach, taking into account the phases of all
supersymmetric parameters. $h_i,h_j,h_k$ are some combination of the
tree level Higgs fields $h,H,A$. The programs {\feynarts} and
{\formcalc} are used to draw and evaluate the Feynman diagrams using
dimensional reduction, and {\looptools} is used to evaluate the majority of the integrals. We use $m_b=m_b^{\msbarm, SM}(m_t^{OS})$ and a top mass of $m_t=m_t^{\msbarm, SM}(m_t^{OS})=m_t^{OS}/\left(1+\frac{4}{3\pi}\alpha_s(m_t^{OS})\right)$ in the $t,\tilde{t},b,\tilde{b}$ masses which enter the genuine vertex corrections in order to absorb some of the higher order SM QCD corrections. We use a unit CKM matrix and assume no squark generation mixing.

\subsubsection{Leading corrections (Yukawa terms)}
\label{section:massesYukCPV}

At low to moderate values of $\tan \beta$, the leading corrections to the $h_i \to h_j h_k$ vertex are the Yukawa terms from the $t,\tilde{t}$ sector. These arise from the diagrams shown in \reffi{fig:h2h1h1vertex}.

\unitlength=1bp%
\begin{figure}
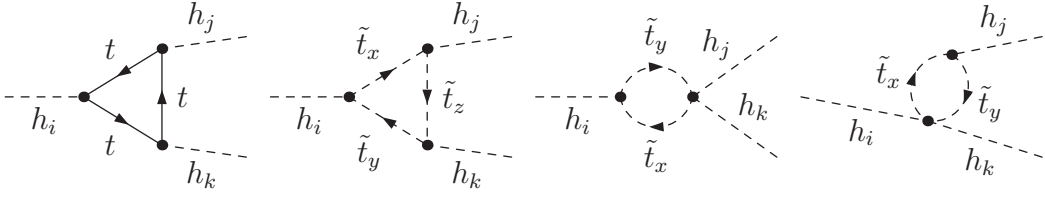

\begin{feynartspicture}(432,100)(4,1)

\FADiagram{}
\FAProp(0.,10.)(6.5,10.)(0.,){/ScalarDash}{0}
\FALabel(3.25,9.18)[t]{$h_i$}
\FAProp(20.,15.)(13.,14.)(0.,){/ScalarDash}{0}
\FALabel(16.3162,15.3069)[b]{$h_j$}
\FAProp(20.,5.)(13.,6.)(0.,){/ScalarDash}{0}
\FALabel(16.3162,4.69307)[t]{$h_k$}
\FAProp(6.5,10.)(13.,14.)(0.,){/Straight}{-1}
\FALabel(9.20801,13.1807)[br]{$t$}
\FAProp(6.5,10.)(13.,6.)(0.,){/Straight}{1}
\FALabel(9.20801,6.81927)[tr]{$t$}
\FAProp(13.,14.)(13.,6.)(0.,){/Straight}{-1}
\FALabel(14.274,10.)[l]{$t$}
\FAVert(6.5,10.){0}
\FAVert(13.,14.){0}
\FAVert(13.,6.){0}

\FADiagram{}
\FAProp(0.,10.)(6.5,10.)(0.,){/ScalarDash}{0}
\FALabel(3.25,9.18)[t]{$h_i$}
\FAProp(20.,15.)(13.,14.)(0.,){/ScalarDash}{0}
\FALabel(16.3162,15.3069)[b]{$h_j$}
\FAProp(20.,5.)(13.,6.)(0.,){/ScalarDash}{0}
\FALabel(16.3162,4.69307)[t]{$h_k$}
\FAProp(6.5,10.)(13.,14.)(0.,){/ScalarDash}{1}
\FALabel(9.20801,13.1807)[br]{$\tilde t_{x}$}
\FAProp(6.5,10.)(13.,6.)(0.,){/ScalarDash}{-1}
\FALabel(9.20801,6.81927)[tr]{$\tilde t_{y}$}
\FAProp(13.,14.)(13.,6.)(0.,){/ScalarDash}{1}
\FALabel(14.274,10.)[l]{$\tilde t_{z}$}
\FAVert(6.5,10.){0}
\FAVert(13.,14.){0}
\FAVert(13.,6.){0}

\FADiagram{}
\FAProp(0.,10.)(7.,10.)(0.,){/ScalarDash}{0}
\FALabel(3.5,9.18)[t]{$h_i$}
\FAProp(20.,15.)(13.,10.)(0.,){/ScalarDash}{0}
\FALabel(16.2244,13.0779)[br]{$h_j$}
\FAProp(20.,5.)(13.,10.)(0.,){/ScalarDash}{0}
\FALabel(16.7756,8.07785)[bl]{$h_k$}
\FAProp(7.,10.)(13.,10.)(0.833333,){/ScalarDash}{-1}
\FALabel(10.,6.43)[t]{$\tilde t_{x}$}
\FAProp(7.,10.)(13.,10.)(-0.833333,){/ScalarDash}{1}
\FALabel(10.,13.57)[b]{$\tilde t_{y}$}
\FAVert(7.,10.){0}
\FAVert(13.,10.){0}

\FADiagram{}
\FAProp(0.,10.)(10.5,8.)(0.,){/ScalarDash}{0}
\FALabel(5.00675,8.20296)[t]{$h_i$}
\FAProp(20.,15.)(12.5,13.5)(0.,){/ScalarDash}{0}
\FALabel(15.995,15.0448)[b]{$h_j$}
\FAProp(20.,5.)(10.5,8.)(0.,){/ScalarDash}{0}
\FALabel(14.8585,5.74034)[t]{$h_k$}
\FAProp(12.5,13.5)(10.5,8.)(0.8,){/ScalarDash}{-1}
\FALabel(8.32332,12.0797)[r]{$\tilde t_{x}$}
\FAProp(12.5,13.5)(10.5,8.)(-0.8,){/ScalarDash}{1}
\FALabel(14.6767,9.4203)[l]{$\tilde t_{y}$}
\FAVert(12.5,13.5){0}
\FAVert(10.5,8.){0}

\label{diag}
\end{feynartspicture}
\caption{The leading vertex corrections to the decay $h_i \to h_j h_k$, involving $t/\Stop$ loops.
($x,y,z$ = 1,2)}
\label{fig:h2h1h1vertex}
\end{figure}

As a first step, we select only terms proportional to $m_t^4/(M^3_Ws_W^3)$ (`Yukawa terms') and perform the calculation at zero incoming momentum i.e.\ $\ser{ij}(p^2=0)$. In this way, we obtain compact analytical expressions for the leading corrections to the $h_i h_j h_k$ vertex. We find that there are no counterterms contributing to the $h_i  h_j h_k$ vertex in this approximation.  

Note that, for consistency, the stop masses $m_{\tilde{t}_1}$ and $m_{\tilde{t}_2}$ must also be calculated in the Yukawa approximation
according to \refeq{eq:Yukstopmasses}.
We therefore arrive at the following expressions for the leading Yukawa corrections in the  $t/\tilde{t}$ sector, which we can express as corrections to an effective coupling $i\lambda^{\rm eff}=i\lambda^{\rm tree}+i\Delta \lambda^{\rm Yuk}$.

For vertices involving the CP-even tree-level Higgs bosons for the case $m_{\tilde{t}_1}\neq m_{\tilde{t}_2}$ (the expressions for $m_{\tilde{t}_1}= m_{\tilde{t}_2}$ are given in \refapp{app:yukequalsquarkmasses}):
\BEA
\Delta \lambda^{\rm Yuk}_{\phi_1\phi_1\phi_1}&=&-\frac{3 e^3 m_t^4}{32 \pi ^2
   M_W^3 s_W^3 s_{\beta }^3}
   \frac{\RemuXt}{m_{\tilde{t}_1}^2-m_{\tilde{t}_2}^2}\left\{4m_t^2 \ResqmuXt\Efunc +3\mu  \mu ^{*}
   \Cfunc \right\}
   \label{eq:phi1phi1phi1Yuk}\non,\\
\EEA   
\BEA  
\Delta \lambda^{\rm Yuk}_{\phi_1\phi_1\phi_2}&=&-\frac{3 e^3 m_t^4}{32 \pi ^2 M_W^3 s_W^3 s_{\beta }^3}\frac{1}{m_{\tilde{t}_1}^2-m_{\tilde{t}_2}^2}   \left\{
   \mu  \mu ^{*} \left(2 \logMStebyMStz-3
   \ReAtCXt\Cfunc\right)
   \right.
   \nonumber\\&&
   -4m_t^2 \ResqmuXt \left(
   \Dfunc+\ReAtCXt\Efunc \right)
   \nonumber\\&&   
   \left.
   +2\ImsqmuXt\Cfunc 
   \right\},
\EEA
\BEA
\Delta \lambda^{\rm Yuk}_{\phi_1\phi_2\phi_2}&=&\frac{3 e^3 m_t^4}{16 \pi ^2 M_W^3 s_W^3 s_{\beta }^5}\frac{1}{m_{\tilde{t}_1}^2-m_{\tilde{t}_2}^2}\left\{2 \mu\mu ^{*}s_{\beta } c_{\beta }   \logMStebyMStz
   \right.
  \nonumber\\
  && 
  -s_{\beta }^2 \RemuXt 
   \left(\frac{m_t^2}{m_{\tilde{t}_1}^2
   m_{\tilde{t}_2}^2}
   \left(m_{\tilde{t}_1}^2-m_{\tilde{t}_2}^2\right)-3 \logMStebyMStz+\frac{3}{2}
   \Cfunc A_t A_t^*\  
   \right. 
  \nonumber\\
  &&    
  \left. +2m_t^2\ReAtCXt \left(2
   \Dfunc+\Efunc \ReAtCXt\right) 
   \right)  
   \nonumber\\&&
   \left.- c_{\beta} s_{\beta}\ImsqmuXt\Cfunc\right\}, 
\EEA
\BEA
\Delta \lambda^{\rm Yuk}_{\phi_2\phi_2\phi_2}&=&\frac{3 e^3 m_t^4}{16 \pi ^2 M_W^3 s_W^3 s_{\beta }^3}\frac{1}{m_{\tilde{t}_1}^2-m_{\tilde{t}_2}^2}\left\{\left(m_{\tilde{t}_1}^2-m_{\tilde{t}_2}^2\right)\left(
   2-3 \logMSteMStzbyMTsq\right)\right.
  \nonumber\\
  &&
  -3 \logMStebyMStz  \left(A_t
   A_t^*+\ReAtCXt\right) -\frac{
   m_t^2}{m_{\tilde{t}_1}^2 m_{\tilde{t}_2}^2}\left(m_{\tilde{t}_1}^4-m_{\tilde{t}_2}^4\right)
  \nonumber\\
  &&   
  +3\ReAtCXt
\left(\frac{
   m_t^2}{m_{\tilde{t}_1}^2 m_{\tilde{t}_2}^2}\left(m_{\tilde{t}_1}^2-m_{\tilde{t}_2}^2\right)+\frac{1}{2}
   \Cfunc A_t A_t^*\right.
  \nonumber\\
  &&    
 \left.\left.+\frac{2}{3} m_t^2 \ReAtCXt
   \left(3\Dfunc+\Efunc \ReAtCXt\right)\right)\right\} ,
\EEA
where
\BEA
Y_t&=&A_t+t_{\beta}\mu^*\\
\Cfunc&=&C_0\left(0,0,0,m_{\tilde{t}_1}^2,m_{\tilde{t}_1}^2,m_{\tilde{t}_2}^2\right)-C_0\left(0,0,0,m_{\tilde{t}_1}^2,m_{\tilde{t}_2}^2,m_{\tilde{t}_2}^2\right)\non\\
&=&-2\left(\frac{1}{m_{\tilde{t}_1}^2-m_{\tilde{t}_2}^2}-\frac{\left(m_{\tilde{t}_1}^2+m_{\tilde{t}_2}^2\right)}{\left(m_{\tilde{t}_1}^2-m_{\tilde{t}_2}^2\right)^2}\logMStebyMStz \right),\\
\Dfunc&=&\text{D}_0\left(0,0,0,0,m_{\tilde{t}_1}^2,m_{\tilde{t}_1}^2,m_{\tilde{t}_1}^2,m_{\tilde{t}_2}^2\right)-\text{D}_0\left(0,0,0,0,m_{\tilde{t}_1}^2,m_{\tilde{t}_2}^2,m_{\tilde{t}_2}^2,m_{\tilde{t}_2}^2\right)\nonumber\\
&=&-\frac{1}{2
   \left(m_{\tilde{t}_1}^2-m_{\tilde{t}_2}^2\right)}\left(\frac{m_{\tilde{t}_1}^2+m_{\tilde{t}_2}
   ^2}{m_{\tilde{t}_1}^2 m_{\tilde{t}_2}^2}-\frac{4}{m_{\tilde{t}_1}^2-m_{\tilde{t}
   _2}^2}\logMStebyMStz\right),\\
\Efunc&=&\text{E}_0\left(0,0,0,0,0,m_{\tilde{t}_1}^2,m_{\tilde{t}_1}^2,m_{\tilde{t}_1}^2,m_{\tilde{t}_2}^2,m_{\tilde{t}_2}^2\right)\non\\
&-&\text{E}_0\left(0,0,0,0,0,m_{\tilde{t}_1}^2,m_{\tilde{t}_1}^2,m_{\tilde{t}_2}^2,m_{\tilde{t}_2}^2,m_{\tilde{t}_2}^2\right)\nonumber\\
&=&\frac{1}{2 }\left(\frac{1}{m_{\tilde{t}_1}^2
   m_{\tilde{t}_2}^2\left(m_{\tilde{t}_1}^2-m_{\tilde{t}_2}^2\right)}\right.\non\\
   &+&\left.\frac{12}{\left(m_{\tilde{t
   }_1}^2-m_{\tilde{t}_2}^2\right)^3}\left(  1-\frac{\left(m_{\tilde{t}_1}^2+m_{\tilde{t}_2}^2\right)}{\left(m_{\tilde{t}_1}^2-m_{\tilde{t}_
   2}^2\right)}\logMStebyMStz \right)\right),
\EEA
and $m_{\tilde q_{1,2}}^2$ are the stop masses in the Yukawa approximation, as given by \refeq{eq:Yukstopmasses}. $\Cfunc$, $\Dfunc$ and $\Efunc$ are functions of $C_0$, $D_0$ and $E_0$ scalar integrals, respectively. Since we are describing a process with 3 external legs, $D_0$ and $E_0$ do not appear explicitly in the Feynman diagrams. However, these functions are very useful for simplifying the vertex expressions. 

The 1-loop corrections to a $h_ih_jh_k$ vertex involving at least one CP-odd eigenstate (again, for $m_{\tilde{t}_1}\neq m_{\tilde{t}_2}$) are given by
\BEA
\Delta \lambda^{\rm Yuk}_{\phi_1\phi_1A}&=&\frac{3 e^3 m_t^4}{32 \pi ^2 M_W^3 s_W^3 s_{\beta }^4}\frac{\ImmuXt}{m_{\tilde{t}_1}^2-m_{\tilde{t}_2}^2}\left\{\right.
  \nonumber\\
  &&
  \left.
   \left(\mu 
   \mu ^{*}-2c_{\beta } s_{\beta } \RemuXt\right)\Cfunc+4m_t^2 \ResqmuXt\Efunc
  \right\},
\EEA   
\BEA  
%
%
\Delta \lambda^{\rm Yuk}_{\phi_1\phi_2A}&=&-\frac{3e^3 m_t^4}{32 \pi ^2 M_W^3 s_W^3 s_{\beta }^4}\frac{\ImmuXt}{m_{\tilde{t}_1}^2-m_{\tilde{t}_2}^2}\left\{+2 c_{\beta } s_{\beta}\logMStebyMStz \right.
    \nonumber\\
  && 
    + \left(2\RemuAt-c_{\beta } s_{\beta } \left(\mu  \mu
   ^{*}+A_t A_t^{*}\right)\right)\Cfunc
    \nonumber \\
  &&  
  \left.+4m_t^2\RemuXt \left(
   \Dfunc+\ReAtCXt\Efunc \right)
  \right\},
\EEA   
\BEA   
\Delta \lambda^{\rm Yuk}_{\phi_2\phi_2A}&=&-\frac{3 e^3 m_t^4}{32 \pi ^2 M_W^3 s_W^3 s_{\beta }^4}\frac{\ImmuXt}{m_{\tilde{t}_1}^2-m_{\tilde{t}_2}^2}\left\{ -2\frac{ m_t^2}{m_{\tilde{t}_1}^2 m_{\tilde{t}_2}^2}\left(m_{\tilde{t}_1}^2-m_{\tilde{t}_2}^2\right)\right.
 \nonumber \\
  &&   
+2 \left(2 s_{\beta
   }^2+1\right) \logMStebyMStz- \left(2s_{\beta }^2\ReAtCXt+A_t
   A_t^*\right)\Cfunc
  \nonumber\\
  &&  
  \left.
  -4m_t^2 \ReAtCXt \left(2
   \Dfunc+ \ReAtCXt\Efunc\right)
  \right\} ,
\EEA   
\BEA  
\Delta \lambda^{\rm Yuk}_{\phi_1AA}&=&\frac{3 e^3 m_t^4}{32 \pi ^2 M_W^3 s_W^3 s_{\beta }^5}\frac{1}{m_{\tilde{t}_1}^2-m_{\tilde{t}_2}^2}\left\{c^2_{\beta } s^2_{\beta }\RemuXt\left(2\logMStebyMStz- Y_{t} Y_{t}^*\Cfunc\right) \right.
  \nonumber\\
  &&   
  \left.+2\ImsqmuXt\left(c_{\beta } s_{\beta  }\Cfunc - 2m_t^2 \RemuXt\Efunc\right)\right\} ,
\EEA   
\BEA  
\Delta \lambda^{\rm Yuk}_{\phi_2AA}&=&\frac{3 e^3 m_t^4}{32 \pi ^2 M_W^3 s_W^3 s_{\beta }^5}\frac{1}{m_{\tilde{t}_1}^2-m_{\tilde{t}_2}^2}\left\{s_{\beta }^2  c_{\beta }^2\left(-2 \left(m_{\tilde{t}_1}^2-m_{\tilde{t}_2}^2\right) \logMSteMStzbyMTsq\right.
   \right.
   \nonumber\\
  && \left.
-2 \left( \ReAtCXt+ Y_{t} Y_{t}^*\right) 
  \logMStebyMStz  +  Y_{t}Y_{t}^*\ReAtCXt \Cfunc\right)
  \nonumber\\
  && 
   \left.+ 2\ImsqmuXt \left(s_{\beta }^2\Cfunc+2m_t^2 \left(
   \Dfunc+ \ReAtCXt\Efunc\right)\right)\right\}\non,\\
\EEA   
\BEA  
\Delta \lambda^{\rm Yuk}_{AAA}&=&-\frac{3  e^3 m_t^4}{32 \pi ^2 M_W^3 s_W^3 s_{\beta }^6}\frac{\ImsqmuXt}{m_{\tilde{t}_1}^2-m_{\tilde{t}_2}^2}\left\{3c_{\beta }^2 s_{\beta }^2\left(2 \logMStebyMStz- Y_t Y_t^* \Cfunc\right)\right.
   \nonumber\\
  && 
  \left.
   -4 m_t^2 \ImsqmuXt\Efunc \right\}
   \label{eq:AAAYuk}.
\EEA
 
These compact, momentum independent expressions have the advantage that they are extremely easy to implement into a computer code. In this form, we are also able to see that, despite including the effect of complex phases, these corrections are themselves entirely real.

In order to convert these corrections to the $h,H,A$ basis, we use the mixing matrix from \refeq{basischange}. For example, $\lambda_{hAA}^{\rm eff}=\lambda_{hAA}^{\rm tree} - s_{\alpha} \Delta \lambda_{\phi_1AA}^{\rm Yuk}+ c_{\alpha}\Delta \lambda_{\phi_2AA}^{\rm Yuk}$.

In the MSSM with real parameters and $m_{\tilde{t}_1}\neq
m_{\tilde{t}_2}$, these corrections reduce to the form (here we drop the
subscripts on $\Cfunc, \Dfunc, \Efunc$ for brevity)
\BEA
\Delta \lambda^{\rm Yuk,CPC}_{\phi_1\phi_1\phi_1}&=&-\frac{3 e^3 m_t^4}{32 \pi ^2
   M_W^3 s_W^3 s_{\beta }^3}
   \frac{\mu^3 X_t}{m_{\tilde{t}_1}^2-m_{\tilde{t}_2}^2}\left\{4m_t^2 X_t^2\Efuncshort +3
   \Cfuncshort \right\}
   \non,\\
\EEA   
\BEA  
\Delta \lambda^{\rm Yuk,CPC}_{\phi_1\phi_1\phi_2}&=&-\frac{3 e^3 m_t^4}{32 \pi ^2 M_W^3 s_W^3 s_{\beta }^3}\frac{\mu^2}{m_{\tilde{t}_1}^2-m_{\tilde{t}_2}^2}   \left\{
2 \logMStebyMStz-3
   \AtXt\Cfuncshort
   -4m_t^2 X_t^2 \left(
   \Dfuncshort+\AtXt\Efuncshort \right)
   \right\}\non,\\
\EEA
\BEA
\Delta \lambda^{\rm Yuk,CPC}_{\phi_1\phi_2\phi_2}&=&\frac{3 e^3 m_t^4}{32 \pi ^2 M_W^3 s_W^3 s_{\beta }^3}\frac{2\mu}{m_{\tilde{t}_1}^2-m_{\tilde{t}_2}^2}\left\{(2  A_t+X_t)  \logMStebyMStz
   \right.
  \nonumber\\
  && 
  - 
   \frac{X_tm_t^2}{m_{\tilde{t}_1}^2
   m_{\tilde{t}_2}^2}
   \left(m_{\tilde{t}_1}^2-m_{\tilde{t}_2}^2\right)-\frac{3}{2}X_t
    A_t^2\Cfuncshort \  
  \left. -2m_t^2A_tX_t^2 \left(2
   \Dfuncshort+ \AtXt\Efuncshort \right) 
    \right\}\non, \\
\EEA
\BEA
\Delta \lambda^{\rm Yuk,CPC}_{\phi_2\phi_2\phi_2}&=&\frac{3 e^3 m_t^4}{32 \pi ^2 M_W^3 s_W^3 s_{\beta }^3}\frac{2}{m_{\tilde{t}_1}^2-m_{\tilde{t}_2}^2}\left\{\left(m_{\tilde{t}_1}^2-m_{\tilde{t}_2}^2\right)\left(
   2-3 \logMSteMStzbyMTsq\right)\right.
  \nonumber\\
  &&
  -3 \logMStebyMStz  A_t\left(A_t+X_t\right) -\frac{
   m_t^2}{m_{\tilde{t}_1}^2 m_{\tilde{t}_2}^2}\left(m_{\tilde{t}_1}^4-m_{\tilde{t}_2}^4\right)
  \nonumber\\
  &&   
  +3\AtXt
\left(\frac{
   m_t^2}{m_{\tilde{t}_1}^2 m_{\tilde{t}_2}^2}\left(m_{\tilde{t}_1}^2-m_{\tilde{t}_2}^2\right)+\frac{1}{2}
    A_t^2\Cfuncshort \left.+\frac{2}{3} m_t^2 \AtXt
   \left(3\Dfuncshort+ \AtXt \Efuncshort\right)\right)\right\}\non ,\\
\EEA
\BEA  
\Delta \lambda^{\rm Yuk,CPC}_{\phi_1AA}&=&\frac{3 e^3 m_t^4}{32 \pi ^2 M_W^3 s_W^3 s_{\beta }^3}\frac{c^2_{\beta }}{m_{\tilde{t}_1}^2-m_{\tilde{t}_2}^2}\left\{ \muXt\left(2\logMStebyMStz- Y_{t}^2\Cfuncshort\right)\right\} ,
\EEA   
\BEA  
\Delta \lambda^{\rm Yuk,CPC}_{\phi_2AA}&=&\frac{3 e^3 m_t^4}{32 \pi ^2 M_W^3 s_W^3 s_{\beta }^3}\frac{c_{\beta }^2}{m_{\tilde{t}_1}^2-m_{\tilde{t}_2}^2}\left\{  -2 \left(m_{\tilde{t}_1}^2-m_{\tilde{t}_2}^2\right) \logMSteMStzbyMTsq
   \right.
   \nonumber\\
  && \left.
-2 \left( \AtXt+ Y_{t}^2\right) 
  \logMStebyMStz  +  Y_{t}^2\AtXt \Cfuncshort\right\}\non\\
\EEA  
and $\Delta \lambda^{\rm Yuk,CPC}_{\phi_1\phi_1A}=\Delta \lambda^{\rm Yuk,CPC}_{\phi_1\phi_2A}=\Delta \lambda^{\rm Yuk,CPC}_{\phi_2\phi_2A}=\Delta \lambda^{\rm Yuk,CPC}_{AAA}=0$. These expressions correspond exactly to the results for the leading Yukawa corrections to the triple Higgs vertex in the MSSM with real parameters published in \citere{Barger:1991ed}.

For completeness, we also give the leading Yukawa corrections to the neutral Higgs self-energies in the complex MSSM, which we used when investigating the effect of reparametrising the Higgs self-energies in terms of the $\msbarm$ top mass, as described in \refse{sec:choiceofmtinFH}. These expressions can be found by considering diagrams involving $t,\tilde{t},b,\tilde{b}$ loops only and selecting those terms proportional to $m^2_t/M^2_W$ (`Yukawa terms'). The resulting corrections will be finite and proportional to $m_t^4$. As before, the renormalisation constants $\delta t_{\beta}$, $\delta M_W^2$, $\delta M_Z^2$ and $\delta Z_{ij}$, are all zero in this approximation and the incoming momentum is taken to be zero. These expressions also involve the Higgs tadpoles and the charged Higgs self-energy (since $\MHpm$ is the input parameter rather than $M_A$), which is also taken at zero incoming momentum, such that
\BEA
\delta \MHpm^2&=&\Si_{H^-H^+}\left(0\right).
\EEA
The leading corrections to the renormalised neutral Higgs energies in the Yukawa approximation for $m_{\tilde{t}_1}\neq m_{\tilde{t}_2}$ are thus given by~\cite{Heinemeyer:2001qd}:
\BEA
\hat{\Sigma}^{(1)\phi_1\phi_1}_{\rm Yuk}&=&\frac{3 e^2  m_t^4 }{16 \pi^2  M_W^2 s_W^2 s_{\beta }^2}\left(\frac{\ResqmuXt}{m_{\tilde{t}_1}^2-m_{\tilde{t}_2}
   ^2}\Cfunc + \mu \mu ^{*}
   \frac{\CfuncL}{2} \right),\label{eq:beginyukse}\\
\hat{\Sigma}^{(1)\phi_1\phi_2}_{\rm Yuk}&=&-\frac{3 e^2  m_t^4 }{16 \pi^2  M_W^2 s_W^2 s_{\beta }^2}\left(\frac{ \mu \mu ^{*}}{t_{\beta }}
   \frac{\CfuncL }{2}\right.
   \nonumber\\&&
   \left.+\frac{\RemuXt}{m_{\tilde{t}_1}^2
   -m_{\tilde{t}_2}^2}
   \left( \ReAtCXt\Cfunc-2 \logMStebyMStz\right)\right),\\
\hat{\Sigma}^{(1)\phi_2\phi_2}_{\rm Yuk}&=&-\frac{3 e^2  m_t^4 }{16 \pi^2  M_W^2 s_W^2 s_{\beta }^2}\left( 2\logMSteMStzbyMTsq-\frac{\mu \mu ^{*}
    }{t_{\beta }^2}\frac{\CfuncL}{2}\right.
   \nonumber\\
&& \left.  
   +\frac{\ReAtCXt}{m_{\tilde{t}_1}^2-m_{\tilde{
   t}_2}^2}\left(4 \logMStebyMStz
   -
   \ReAtCXt\Cfunc\right)\right),\\
\hat{\Sigma}^{(1)\phi_1A}_{\rm Yuk}&=&-\frac{3 e^2  m_t^4 }{32 \pi^2 
    M_W^2 s_W^2 s_{\beta }^3} \frac{\ImmusqXtsq}{m_{\tilde{t}_1}^2-m_{\tilde{t}_2}^2}\Cfunc,\\
\hat{\Sigma}^{(1)\phi_2A}_{\rm Yuk}&=&\frac{3 e^2  m_t^4 }{16\pi^2 
    M_W^2 s_W^2 s_{\beta }^3}\frac{\ImmuXt }{\left(m_{\tilde{t}_1}^2-m_{\tilde{t}_2}^2\right)}\left(
   \ReAtCXt\Cfunc-2\logMStebyMStz \right),\\
\hat{\Sigma}^{(1)AA}_{\rm Yuk}&=&\frac{3 e^2  m_t^4 }{16 \pi^2  M_W^2 s_W^2 s_{\beta }^4}\left(\frac{ \ImsqmuXt}{m_{\tilde{t}_1}^2-m_{\tilde{t}_2}
   ^2}\Cfunc+ \mu  \mu ^{*}
  \frac{\CfuncL}{2} \right)\label{eq:endyukse},
\EEA
where
\BEA
\CfuncL&=&\text{C}_0\left(0,0,0,m_{\tilde{t}_1}^2,m_{\tilde{t}_2}^2,M_L^2\right).
\EEA
Note that the $C_0$ integrals do not appear automatically, as no 3-point functions are calculated. However, substituting $C_0$ integrals for combinations of the $A_0$ and $B_0$ integrals which appear naturally in the calculation (all at zero momentum) does make the self-energy expressions more compact. The expressions in the case where $m_{\tilde{t}_1}=m_{\tilde{t}_2}$ are given in \refapp{app:yukequalsquarkmasses}.

\subsubsection{Full 1-loop 1PI vertex corrections}
For the full 1PI 1-loop corrections to the $h_ih_jh_k$ vertex, we need the relevant counterterms.  Note that, for triple Higgs vertices with an external Higgs boson $A$, the field renormalisation constant $\delta Z_{AG}$ is required in order to ensure that the vertex is UV-finite. We have extended the {\feynarts} model files in order to include these counterterms.

\begin{figure}[htb!]
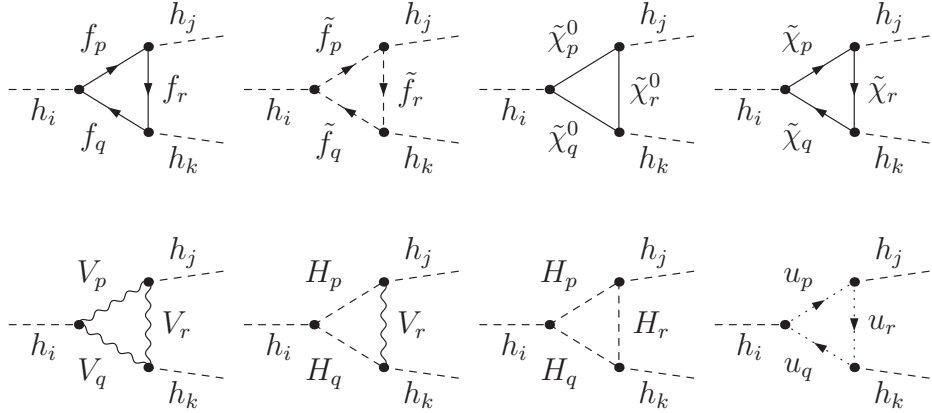

\begin{center}
\input hAhBhCdiag 
\caption[Examples of generic diagrams (showing only one of the
topologies) contributing to the processes
$h_i \to h_j h_k$.]{Examples of generic diagrams (showing only one of the
topologies) contributing to the processes
$h_i \to h_j h_k$. $h_i,h_j,h_k$ are the physical Higgs fields at tree level ($h,H,A$), $f$ are SM fermions, $\tilde{f}$ are their superparters, 
$\tilde{\chi}^0,\tilde{\chi}$ are neutralinos and charginos, 
$V$ are vector bosons, $H$ denote the neutral and charged Higgs bosons and 
the Goldstone bosons, $u$ are Faddeev--Popov ghost fields.
}
\label{fig:fd_hhh}
\end{center}
\end{figure}

Examples of Feynman diagrams contributing to these vertex corrections are shown in \reffi{fig:fd_hhh}.

We also investigated the effect of including loop-corrected masses and 
couplings of the Higgs bosons in the one-loop contributions to the 
$h_ih_jh_k$ vertex, instead of the tree level quantities. In order to ensure the UV divergences cancelled, we transformed the couplings of the internal Higgs to the other particles using a unitary approximation to the matrix $\matr{\hat Z}$ (by setting the momentum in the neutral Higgs self-energies $\Sigma_{h_ih_j}$ to $(m_{h_i}^2+m_{h_j}^2)/2$), which we implemented into a {\feynarts} model file. For consistency, the loop corrected Higgs masses of these internal Higgs bosons were also calculated using this unitary rotation matrix. (Note that we continued to use the full Higgs masses and Higgs propagator corrections for the external Higgs bosons). These corrections were numerically insignificant in the examples investigated.

\subsection{Combining the 1PI vertex corrections with propagator corrections to obtain the full $h_a\to h_bh_c$ decay width}

We can combine vertices involving the tree level Higgs bosons $h_i,h_j,h_k$ with the wave-function normalisation factors contained in the matrix $\matr{Z}$, which contain self-energies from the program {\fh}, in order to obtain processes involving the loop-corrected states $h_a,h_b,h_c$ as the external particles (as discussed in \refse{section:zfactors}).

These `Z-factors' can be used in conjunction with tree level\footnote{This method can also be used for the effective vertices $\lambda^{\rm eff}_{h_ah_bh_c}$ i.e.\ $\lambda^{\rm eff}_{h_ah_bh_c} =\matr{\hat{Z}}_{ck}\matr{\hat{Z}}_{bj}\matr{\hat{Z}}_{ai}\lambda^{\rm eff}_{h_ih_jh_k}$.} vertices $\lambda^{\rm tree}_{h_ih_jh_k}$ using (sum over $i,j,k$)
\BEA
\lambda^{\rm tree}_{h_ah_bh_c}&=&\matr{\hat{Z}}_{ck}\matr{\hat{Z}}_{bj}\matr{\hat{Z}}_{ai}\lambda^{\rm tree}_{h_ih_jh_k}.
\EEA

The decay width is then given by \refeq{eq:triplehdecaywidth} with $|\mathcal{M}|^2=|\lambda^{\rm tree}_{h_ah_bh_c}|^2$ or $|\mathcal{M}|^2=|\lambda^{\rm eff}_{h_ah_bh_c}|^2$. Note that this means that our decay width will contain pieces of type (1-loop)$\times$(1-loop), which is necessary since the tree level coupling is often smaller than the leading loop corrections to the coupling.

We obtain our full result by combining the complete genuine 1-loop vertex corrections $\Gamma_{h_ih_jh_k}^{\rm 1PI, 1-loop}$ and the corrections involving 1-loop Goldstone and Z boson self-energy contributions  $\Gamma_{h_ih_jh_k}^{\rm G,Z \,se}$ with the Z-factors, such that (sum over $i,j,k$)
\begin{equation}
\Gamma_{h_ah_bh_c}^{\rm full}=\matr{\hat{Z}}_{ck}\matr{\hat{Z}}_{bj}\matr{\hat{Z}}_{ai}\left[\Gamma_{h_ih_jh_k}^{\rm tree}+\Gamma_{h_ih_jh_k}^{\rm 1PI, 1-loop}\left(M^2_{h_a},M^2_{h_b},M^2_{h_c}\right)+\Gamma_{h_ih_jh_k}^{\rm G,Z \,se}\left(m^2_{h_i},m^2_{h_j},m^2_{h_k}\right)\right]
\label{eq:fullhihjhk}.
\end{equation}

The genuine 1-loop vertex corrections $\Gamma_{h_ih_jh_k}^{\rm
1PI,1-loop}$ contain the full momentum dependence and therefore depend
on the loop-corrected masses $M^2_{h_a},M^2_{h_b},M^2_{h_c}$ at the
external legs. However, as discussed in detail in \refse{section:GZ}, unphysical poles from diagrams involving Goldstone and Z boson self-energies can be avoided by approximating the external momenta to the tree level values in the corresponding contributions, i.e.\ $\Gamma_{h_ih_jh_k}^{\rm G,Z \,se}$ is a function of $m^2_{h_i},m^2_{h_j},m^2_{h_k}$.  
Again, the decay width is then given by \refeq{eq:triplehdecaywidth} but with $|\mathcal{M}|^2=|\Gamma_{h_ah_bh_c}^{\rm full}|^2$.

\subsection{Numerical Results}

\subsubsection{$h_2 \to h_1 h_1$ decay width}

We will now investigate the importance of the full 1-loop genuine corrections through their numerical impact on the $h_2\to h_1 h_1$ decay width. All the results plotted in this section include the wave-function normalisation factors, through the matrix $\matr{Z}$. The case where only wave-function
normalisation factors but no genuine one-loop vertex contributions are included will be denoted `tree'.

\begin{figure} 
\begin{center}
\resizebox{0.4\textwidth}{!}{%
\includegraphics{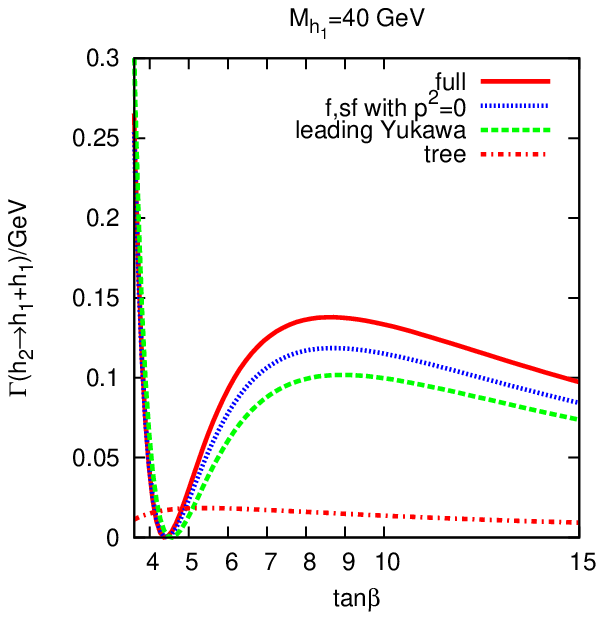}
}
\resizebox{0.4\textwidth}{!}{%
\includegraphics{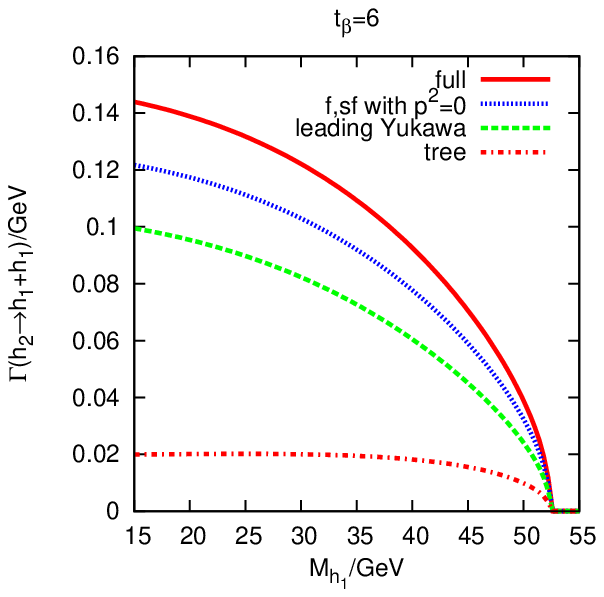}
}
\end{center}
\caption{The decay width $\Ga(h_2 \to h_1h_1)$ in the CPX scenario as a
function of $\tb$ at $M_{h_1}=40\gev$ (left) and as a function of
$M_{h_1}$ at $t_{\beta}=6$ (right). $\MHp$ is adjusted to give the
$M_{h_1}$ as required. All results include the propagator corrections.
`tree' indicates that the tree level triple Higgs vertex has been used. `leading Yukawa' includes the leading genuine vertex corrections given in \refeqs{eq:phi1phi1phi1Yuk}--(\ref{eq:AAAYuk}). `f,sf $p^2$=0' includes contributions to the genuine vertex corrections from fermions and sfermions only and approximates the external momenta to zero. The curve labelled `full' shows the result including the full genuine vertex corrections. \label{h2h1h1cpx}}
\end{figure}

\reffi{h2h1h1cpx} compares the `tree' result with the full result which
includes the genuine vertex correction and all propagator corrections,
as described by  \refeq{eq:fullhihjhk}, in the $\CPX$ scenario. In
\reffi{h2h1h1cpx} (left), we can see that `tree' and full decay widths
are very different. The full result has a peak (i.e. local maximum) at
$\tb=8.7$. There is a corresponding peak in the `tree' result at
$\tb=5.5$. However this peak is 7.5 times smaller than the peak in the
full result. The sharp increase in the full result at low $\tb$ is
because we have chosen to keep $M_{h_1}$ constant, which requires a
rapid change in $\MHp$ in this part of parameter space (the `tree' result also exhibits this behaviour, but at $\tb \sim 2.5$, which is not shown on the graph).

In \reffi{h2h1h1cpx} (right), we can see that both the `tree' and full decay widths decrease as the lightest Higgs mass $M_{h_1}$ increases, as we would expect from the kinematics. Again, the `tree' level result is heavily suppressed compared to the full result. We can conclude that calculations of triple-Higgs couplings which just combine the propagator corrections with the tree level vertex but do not take into account genuine vertex corrections are an extremely poor approximation to the full result.

\begin{figure} 
\begin{center}
\resizebox{0.5\textwidth}{!}{%
\includegraphics{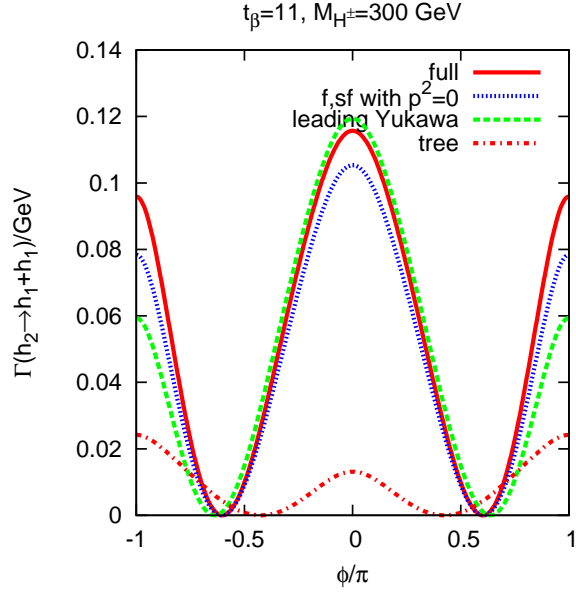}
}
\end{center}
\caption{The dependence of the $h_2\to h_1h_1$ decay width on the phase $\phi=\phi_{A_t}=\phi_{A_b}=\phi_{A_\tau}=\phi_{M_3}$ ($\CPX$ scenario at $\tb=11   $, $\MHpm= 300\gev$). 
The full result is compared with different approximations, which are the same as those specified in \reffi{h2h1h1cpx}. 
\label{h2h1h1varyargAt}}
\end{figure}

\reffi{h2h1h1varyargAt} demonstrates the pronounced dependence of the
results on the complex phase $\phi$, where
$\phi=\phi_{A_t}=\phi_{A_b}=\phi_{A_\tau}=\phi_{M_3}$, at $\tb=11$,
$\MHp=300\gev$ (all other parameters are taken from the $\CPX$
scenario). We can see once again that while the `tree' result gives
qualitatively similar behaviour, the peaks are less than a quarter of
the peaks in the full result, and the positions of the troughs are shifted. It is interesting to note that the genuine vertex corrections play a highly significant role over essentially the full range of $\phi$, including the points at $\phi=0$ and $\phi=\pi$ where the parameters are entirely real.

\reffi{h2h1h1cpx} and \reffi{h2h1h1varyargAt} also demonstrate the results of two methods for obtaining effective $h_1h_1h_1$ couplings: the leading Yukawa corrections \refeqs{eq:phi1phi1phi1Yuk}--(\ref{eq:AAAYuk}) and the fermion/sfermion corrections taken at $p^2=0$. Both approximations are a big improvement over the `tree' result. 

In \reffi{h2h1h1cpx}, at the peak at $\tb=8.7$, the result using $f,\tilde{f}$ at $p^2=0$ is within 14\% of the full result and the `leading Yukawa' approximation is within 27\% of the full result. As we can see from \reffi{h2h1h1varyargAt} at $\phi=0$, in some parts of parameter space, the `leading Yukawa' approximation accidentally performs better than the $f,\tilde{f}$ at $p^2=0$ terms, although it is a less complete approximation to the full result.

\begin{figure} 
\begin{center}
\resizebox{0.5\textwidth}{!}{%
\includegraphics{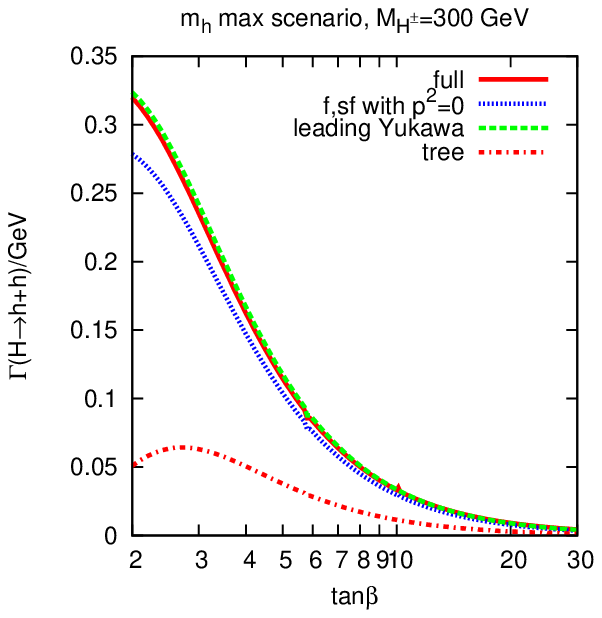}
}
\end{center}
\caption{The decay width $\Ga(H \to hh)$ in the $m_h^{\rm max}$ scenario as a function of $\tb$ at $\MHpm=300\gev$. 
The full result is compared with different approximations, which are the same as those specified in \reffi{h2h1h1cpx}. 
\label{fig:Hhhmhmax}}
\end{figure}

This calculation of the $h_2\to h_1 h_1$ decay width also applies to the MSSM with real parameters. In \reffi{fig:Hhhmhmax}, we show the results for the decay width $\Ga(H \to hh)$ in the $m_h^{\rm max}$ scenario as a function of $\tb$ at $\MHpm=300\gev$. As we saw in the $\CPX$ scenario, using the tree level triple Higgs vertices combined with the propagator corrections is a poor approximation to the full genuine triple Higgs vertex corrections combined with the propagator corrections. The `leading Yukawa' approximation and using the fermion/sfermion corrections at $p^2=0$ in the genuine vertex corrections are far better approximations to the full result.

\subsubsection{Effective coupling approximation for the lightest neutral Higgs boson}

As we have seen, the effective triple Higgs vertices obtained using the leading corrections in the Yukawa approximation (as given in \refeqs{eq:phi1phi1phi1Yuk}--(\ref{eq:AAAYuk})) and the effective triple Higgs vertices obtained using the fermion/sfermion corrections at $p^2=0$ in the genuine vertex corrections and combining with the propagator corrections have both performed rather well as approximations to the $h_2 \to h_1 h_1$ or $H \to hh$ decay width. We have also seen that it is possible to get a large enhancement in these decay widths from the genuine vertex corrections. We shall now apply these approximations to the triple coupling of the lightest Higgs boson. 

\begin{figure} 
\begin{center}
\resizebox{0.4\textwidth}{!}{%
\includegraphics{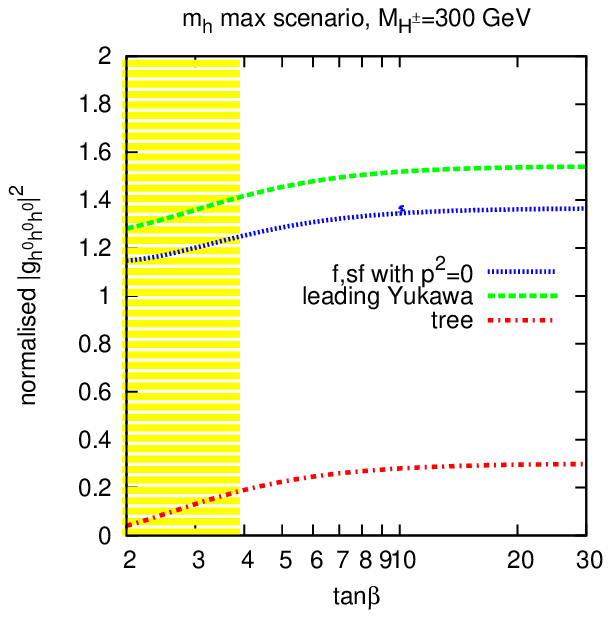}
}
\resizebox{0.4\textwidth}{!}{%
\includegraphics{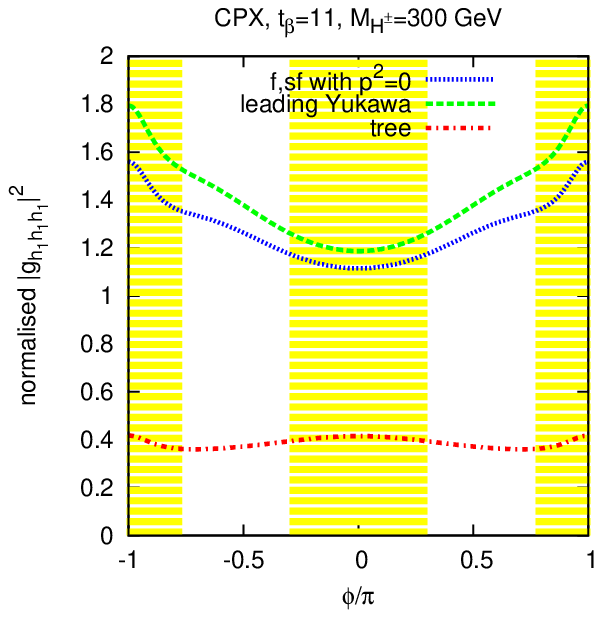}
}
\end{center}
\caption{The triple coupling of the lightest Higgs boson squared, normalised to the SM tree-level value (with equal Higgs mass). Left: $m_h^{\rm max}$ scenario at $\MHpm=300\gev$, varying $\tan {\beta}$. Right: $\CPX$ scenario, $t_{\beta}=11$, $\MHpm=300\gev$, varying the phase $\phi=\phi_{A_t}=\phi_{A_b}=\phi_{A_\tau}=\phi_{M_3}$. All results include the propagator corrections. `tree' indicates that the tree level triple Higgs vertex has been used. `leading Yukawa' includes the leading genuine vertex corrections using the Yukawa approximation, as given in \refeqs{eq:phi1phi1phi1Yuk}--(\ref{eq:AAAYuk}). `f,sf $p^2$=0' includes contributions to the genuine vertex corrections from fermions and sfermions only and approximates the external momenta to zero. Yellow (shaded): excluded at 95\% CL by the LEP Higgs searches (see \refse{sec:exclplots} for details). \label{fig:effC}}
\end{figure}

\reffi{fig:effC} shows this effective coupling squared, supplemented by
propagator corrections and normalised to the tree level SM value (with
equal Higgs mass). We can see that there is a suppression with respect
to the tree level SM value if no genuine vertex corrections are
included. This holds even in the limit of large $m_A$. This is because
the SM tree-level coupling involves the square of the physical Higgs
mass whereas the effective coupling in this limit involves the square 
of the tree-level mass of the lightest MSSM Higgs boson.

In \reffi{fig:effC}, we also show the areas which are already excluded at the 95\% CL by the LEP Higgs searches, which we have determined using the method described in \refse{sec:exclplots} below. Including the genuine vertex corrections gives an overall enhancement of the effective couplings squared of approximately 1.2 to 1.6 in the parts of this parameter space which are not yet excluded. This could have interesting implications for the sensitivity of searches at the LHC and LC to effects of the triple-Higgs coupling in the MSSM.

\section{Higgs decay to SM fermions}
\label{chapter:Hiff}

\subsection{Calculation of the $h_a\to b \bar{b} $ decay width}

We use the general expression for two-body decay widths given in \refeq{eq:triplehdecaywidth} to obtain:
\BEA
\Gamma(h_a\to b \bar{b})&=&\frac{N_c}{8\pi M_{h_a}^2}\frac{M_{h_a}}{2}\sqrt{1-\frac{4m_b^2}{M_{h_a}^2}}\left|\mathcal{M}\right|^2,
\EEA
where $N_c$ is the number of colours. The mass dependence of the squared matrix element $\left|\mathcal{M}\right|^2$ will be affected by the CP properties of the Higgs boson. For example, at lowest order, 
\BEA
\Gamma_{\rm tree}(h_a\to b \bar{b})&=&\frac{N_c}{8\pi }\frac{M_{h_a}}{2}\left( 1- \frac{4m_b^2}{M_{h_a}^2} \right)^{x} \left| \lambda^{\rm tree}_{h_ab\bar{b}} \right|^2,
\EEA
where $\left|\lambda^{\rm tree}_{hb\bar{b}}\right|^2=\lambda_b^{(0),2}s^2_{\alpha}$, $\left|\lambda^{\rm tree}_{Hb\bar{b}}\right|^2=\lambda_b^{(0),2}c^2_{\alpha}$ and $\left|\lambda^{\rm tree}_{Ab\bar{b}}\right|^2=\lambda_b^{(0),2}s^2_{\beta}$ with $x=3/2$ for the CP even states $h,H$ and $x=1/2$ for the CP odd state $A$. $\lambda_b^{(0)}$ was defined in \refeq{eq:lambdab0}.

\subsubsection{Standard Model QED corrections}
\label{section:hAbbSMQCD}

The real and virtual QED contributions to the Standard Model $H^{\rm SM}\rightarrow b \bar{b}$ decay width lead to the 1-loop correction 
\BEA
\Gamma_{\rm QED}(H^{\rm SM}\to b \bar{b})&=&\Gamma_{\rm tree}(H^{\rm SM}\to b \bar{b})\left[ 1+\frac{\alpha}{\pi}Q_b^2\left( -3\log(\frac{M_{H^{\rm SM}}}{m_b}) + \frac{9}{4} \right)\right],
\label{hbbQED}
\EEA 
for $M_H^{2,{\rm  SM}}\gg m_f$, as derived in \citere{Braaten:1980yq}. In this limit, this equation can be used for the QED corrections for both the scalar and pseudoscalar MSSM Higgs bosons\cite{hep-ph/9503443}. We will therefore use the correction term 
\BEA
\delta_{\rm QED}&:=& \frac{\alpha}{\pi}Q_b^2\left( -3\log(\frac{M_{h_a}}{m_b}) + \frac{9}{4} \right)
\label{hbbQED2}
\EEA
in our MSSM calculation.

\subsubsection{Standard Model QCD corrections}

If the factor $Q_b^2\alpha$ in \refeq{hbbQED} is replaced by the factor $C_f\alpha_s(M_H^{\rm SM})$, the expression for the 1-loop QCD correction to the $H^{\rm SM}\rightarrow b \bar{b}$ decay in the Standard Model is obtained, as shown in \citere{Braaten:1980yq}. $C_f$ is the quadratic Casimir operator ($C_f=\frac{4}{3}$) and $\alpha_s(M_H^{2,{\rm SM}})$ is the running coupling. Including the tree level result gives
\BEA
\Gamma_{\rm QCD}(H^{\rm SM}\to b \bar{b})&=&
\left[\frac{\Gamma^{\rm tree}(H^{\rm SM}\to b \bar{b})}{m_b^{2,OS}}\right]
m_b^{2,OS}\left[1+ \frac{\alpha_s(\mu_{\rm ren}^2)}{\pi}C_f\left( -3\log(\frac{M_{H^{\rm SM}}}{m_b}) + \frac{9}{4} \right)\right]\non,\\
\label{hbbQCD1}
\EEA 
where we have factored out the Yukawa coupling from the term in the square bracket.

In the mass range we are interested in, $M_H^{\rm SM}\gg m_b$, this leading 
logarithmic approximation is not sufficient. However, the large logarithmic corrections can be absorbed into a running bottom quark mass. Substituting the relation between the on-shell b-quark mass and the running SM $\msbarm$ b-quark mass into \refeq{hbbQCD1} gives
\BEA
\Gamma_{\rm QCD}(H^{\rm SM}\to b \bar{b})&=& 
\left[\frac{\Gamma^{\rm tree}(H^{\rm SM} \to b \bar{b})}{m_b^{2,OS}}\right]
m_b^{2,\msbarm,SM}(\mu_{\rm ren})\left[1+ \frac{\alpha_s(\mu_{\rm ren}^2)}{4\pi}\left( -16\log(\frac{M_{H}}{\mu_{\rm ren}}) + \frac{68}{3} \right)\right]\non,\\
\label{hbbQCD2}
\EEA 
and we can choose $\mu_{\rm ren}=M_H^{\rm SM}$ in order to cancel the logarithmic term. 

In practice, we parametrise our calculation in terms of $m_b^{\msbarm,SM}(M_{h_a})$.  Therefore, in order to encompass the full 1-loop Standard Model-type QCD corrections in our calculation, we will need to add a correction
\BEA
\delta_{\rm QCD}:=\frac{17}{3}\frac{\alpha_s(M_{h_a}^2)}{\pi}
\label{hbbQCD3}
\EEA
to the $h_a\to b \bar{b}$ decay width.

Our method differs from that of \citere{Heinemeyer:2000fa}, which includes some higher order terms in $\alpha_s(\mu_{\rm ren}^2)$ and $m_b^{\msbarm,SM}(\mu_{\rm ren})$ and an extra term proportional to $\alpha_s(m_b^2)/\alpha_s(M_{h_a}^2)$. Our method also differs from \citere{hep-ph/0305101}, which includes terms proportional to $\alpha_s^2$. However, some of these terms depend on the CP properties of the Higgs, and thus are not trivially extendable to the complex MSSM. Both \citere{Heinemeyer:2000fa} and \citere{hep-ph/0305101} restrict their analyses to the MSSM with real parameters.

\subsubsection{Full 1-loop 1PI $h_i\to b\bar{b}$ vertex corrections}

In order to calculate the 1-particle irreducible vertex corrections at 1-loop, we have once again extended the {\it FeynArts} model file to include the relevant counterterms. We have evaluated the complete contributions in the MSSM, apart from the QED- and QCD-type corrections, for which we use the results given in \refeq{hbbQED2} and \refeq{hbbQCD3}, respectively. (Note that we include the photon contribution to the W boson mass counterterm). We include all complex phases. As discussed above, we use $m_b^{\msbarm,SM}(M_{h_a})$ in these corrections in order to absorb some of the higher order terms. As before, we use a unit CKM matrix and include no squark generation mixing.

\subsubsection{Resummed $\Delta m_b$ corrections to $h_i\to b \bar{b}$}

In order to resum the leading SUSY QCD (and higgsino) corrections for the case of large $\tan \beta$ in the limit of heavy SUSY particles, we use the effective vertices $v_{\rm h_i\bar{b}b}^{\Delta m_b} $ from the effective Langrangian in \refeq{eq:Leffdeltamb}.

\BEA
v_{\rm h\bar{b}b}^{\Delta m_b} & = &
\frac{1}{1+y}\left[1-\frac{1}{t_{\alpha}t_{\beta}}y+i\gamma_5x\left(1+\frac{1}{t_{\alpha}t_{\beta}}\right)\right]v_{\rm h\bar{b}b}^{\rm tree}\non\\
v_{\rm H\bar{b}b}^{\Delta m_b} & = & \frac{1}{1+y}\left[1+\frac{t_{\alpha}}{t_{\beta}}y+i\gamma_5x\left(1-\frac{t_{\alpha}}{t_{\beta}}\right)\right]v_{\rm H\bar{b}b}^{\rm tree}\non\\
v_{\rm A\bar{b}b}^{\Delta m_b} & = & \frac{1}{1+y}\left[1-\frac{1}{t_{\beta}^2}y+i\gamma_5x\left(1+\frac{1}{t_{\beta}^2}\right)\right]v_{\rm A\bar{b}b}^{\rm tree}.
\EEA

When combining this contribution with the full genuine vertex corrections, we need to avoid double-counting of the 1-loop corrections involving gluinos or higgsinos. Therefore, we subtract the 1-loop part of these effective vertices, i.e.\ we use effective couplings of the form $\left(v_{\rm h_i\bar{b}b}^{\Delta m_b}-v_{\rm h_i\bar{b}b}^{\Delta m_b,{\rm 1-loop}}\right)$ where
\BEA
v_{\rm h\bar{b}b}^{\Delta m_b,{\rm 1-loop}} & = &\left[{\rm Re}\Delta m_b\left(-1-\frac{1}{t_{\alpha}t_{\beta}}\right)+i\gamma_5{\rm Im}\Delta m_b\left(1+\frac{1}{t_{\alpha}t_{\beta}}\right)\right]v_{\rm h\bar{b}b}^{\rm tree},\\
v_{\rm H\bar{b}b}^{\Delta m_b,{\rm 1-loop}} & = &\left[{\rm Re}\Delta m_b\left(-1+\frac{t_{\alpha}}{t_{\beta}}\right)+i\gamma_5{\rm Im}\Delta m_b\left(1-\frac{t_{\alpha}}{t_{\beta}}\right)\right]v_{\rm H\bar{b}b}^{\rm tree},\\
v_{\rm A\bar{b}b}^{\Delta m_b,{\rm 1-loop}} & = &\left[{\rm Re}\Delta m_b\left(-1+\frac{1}{t_{\beta}^2}\right)+i\gamma_5{\rm Im}\Delta m_b\left(1+\frac{1}{t_{\beta}^2}\right)\right]v_{\rm A\bar{b}b}^{\rm tree}.
\EEA

\subsubsection{Combining these contributions with propagator corrections to obtain the full $h_a\to b \bar{b}$ decay width}

The tree level, 1-loop 1PI $h_i\to b\bar{b}$ vertex function and the additional $\Delta m_b$ corrections are then combined with the propagator corrections from the $h_ih_j$ self-energies (via $\matr{\hat{Z}}$) and from the $h_iG$ and $h_iZ$ self-energies:
\BEA
 \Gamma_{h_ab\bar{b}}&=&\matr{\hat{Z}}_{ai}\left[\Gamma_{h_ib\bar{b}}^{\Delta m_b,{\rm no\, 1-loop}} 
+\Gamma_{h_ib\bar{b}}^{\rm 1PI, 1-loop}\left(M^2_{h_a}\right)+\Gamma_{h_ib\bar{b}}^{\rm G,Z \,se}\left(m^2_{h_i}\right)\right].
\EEA

The arguments to $\Gamma_{h_ib\bar{b}}^{\rm 1PI,1-loop}$ and $\Gamma_{h_ib\bar{b}}^{\rm G,Z \,se}$ denote the external momenta used. $\Gamma_{h_ib\bar{b}}^{\Delta m_b,{\rm no\, 1-loop}}$ denotes the use of $\left(v_{\rm h_i\bar{b}b}^{\Delta m_b}-v_{\rm h_i\bar{b}b}^{\Delta m_b,{\rm 1-loop}}\right)$.
$\Gamma_{h_ab\bar{b}}$ is combined with the external fermion wavefunctions, then we take the squared modulus and sum over external spins to get $|\mathcal{M}_{h_ab\bar{b}}|^2$.

The full $h_a\to b \bar{b}$ decay width is thus found using
\BEA
\Gamma^{\rm full}(h_i\to b \bar{b})&=&\left[1+\delta_{\rm QCD}+\delta_{\rm QED}\right]\frac{N_c}{8\pi M_{h_a}^2}\frac{M_{h_a}}{2}\sqrt{1-\frac{4m_b^2}{M_{h_a}^2}}\left|\mathcal{M}_{h_ab\bar{b}}\right|^2
\label{eq:fullhAbb},
\EEA
which is an extension of the method used to combine QED, QCD and Z-factor corrections in \citere{Heinemeyer:2000fa}.

\subsubsection{Numerical Results}

\begin{figure} 
\begin{center}
\resizebox{1\textwidth}{!}{%
\includegraphics{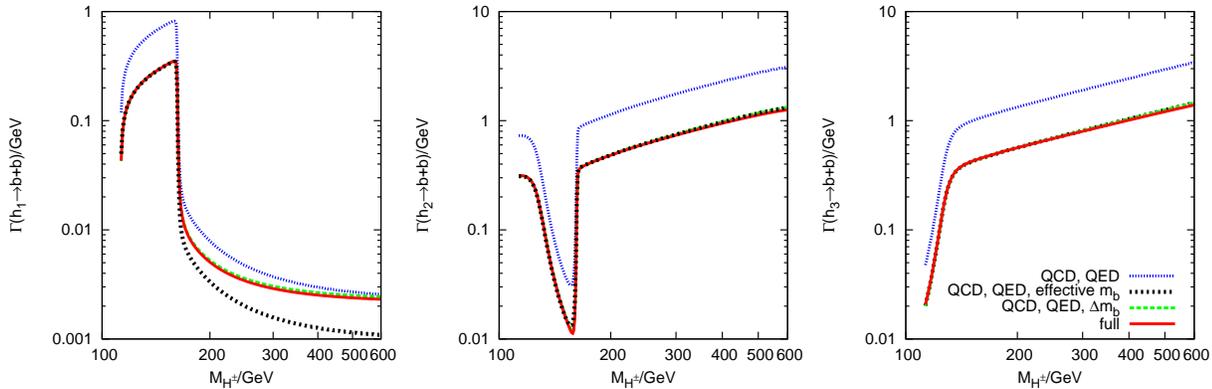}
}
\end{center}
\caption{The decay widths of the neutral Higgs bosons to two b-quarks in the $\CPX$ scenario with $\tan \beta=20$. See text for explanation of the various approximations used. \label{fig:habb1}}
\end{figure}

\reffi{fig:habb1} illustrates the decay widths $\Gamma(h_1\to b
\bar{b})$ (left), $\Gamma(h_2\to b \bar{b})$ (centre) and $\Gamma(h_3\to
b \bar{b})$ (right) as a function of the charged Higgs mass for the CPX
scenario with $\tan \beta=20$. All results include the propagator
corrections, incorporated via the matrix $\matr{Z}$, and the Goldstone and gauge boson mixing contribution. We note that the $h_1$ and $h_2$ decay widths have steep gradients at $\MHpm\sim 157\gev$ due to a `cross-over' effect in the masses (i.e.\ $M_{h_1}$ and $M_{h_2}$ approach each other). At $\MHpm\sim 150\gev$, $h_1$ is mostly $A$, $h_2$ is mostly $h$ and $h_3$ is mostly $H$, whereas at $\MHpm\sim 200\gev$, $h_1$ is mostly $h$, $h_2$ is mostly $A$ and $h_3$ is mostly $H$.

The impact of including the resummed $\Delta m_b$ contributions is very significant, as we can see from comparison of the result using the propagator, SM QCD and QED corrections (`QED, QCD') and the result when the resummed $\Delta m_b$ contributions are included in addition (`QED, QCD, $\Delta m_b$'). \reffi{fig:habb1} also includes the full decay widths (`full'), which combine the propagator, QED, SM QCD, SUSY QCD corrections with the full electroweak genuine vertex corrections, as described by \refeq{eq:fullhAbb}, and the effect of including the 
full corrections turns out to be relatively small but not negligible. At $\MHpm=600\gev$, these extra electroweak corrections are 5.6\%, 5.7\% and 5.0\% for the decay of $h_1,h_2,h_3$, respectively.

It is possible to approximate the `QED, QCD, $\Delta m_b$' result using
an effective $b$-quark mass of $m_b^{\msbarm,SM}(m_t)/|1+\Delta m_b|$
(with no additional $\Delta m_b$ contributions), which we call `QED,
QCD, effective $m_b$'. The resulting decay widths are also shown in
\reffi{fig:habb1}. In general, this approximation performs well.
However, note that in the decoupling limit, where the $h_1$ is SM-like,
the effect of the SUSY QCD corrections should cancel out in the $h_1$
decay width. In the approximate `QED, QCD, effective $m_b$' result, this
does not occur, giving a offset which is significant in relative terms,
although small in absolute terms. In the region $\MHpm =200\gev$ to
$\MHpm =600\gev$ shown here, this offset between the `QED, QCD,
effective $m_b$' result and the approximate result is more than 44\% of
the `QED, QCD, $\Delta m_b$' result, while the absolute difference is 
less than $0.002\gev$.

\begin{figure} 
\begin{center}
\resizebox{0.5\textwidth}{!}{%
\includegraphics{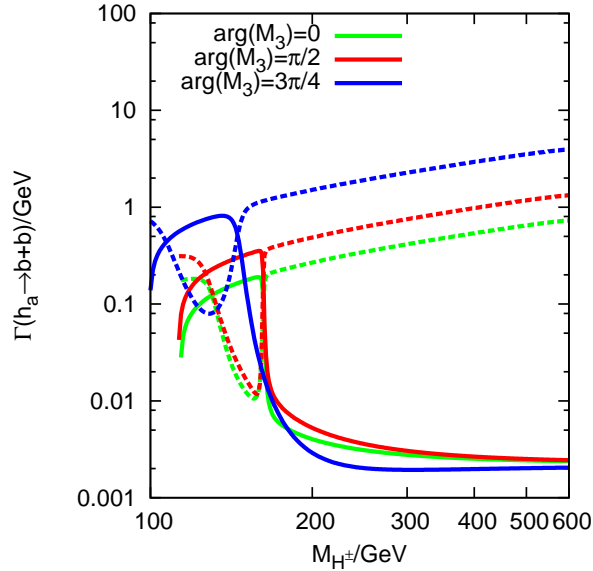}
}
\end{center}
\caption{The decay widths of the two lightest neutral Higgs bosons to two b-quarks in the $\CPX$ scenario with $\tan \beta=20$ at various values of $\phi_{M_3}$. Solid: $h_a=h_1$, dashed: $h_a=h_2$.\label{habb3}}
\end{figure}

The $h_a\to b\bar{b}$ decay width in the $\CPX$ scenario is highly
dependent on $\phi_{M_3}$. \reffi{habb3} shows the decay widths of the
two lightest Higgs bosons into b-quarks at $\phi_{M_3}=0$,
$\phi_{M_3}=\pi/2$ and $\phi_{M_3}=3\pi/4$. Whereas the inclusion of the
resummed $\Delta m_b$-type corrections suppresses the decay width at
$\phi_{M_3}=0$, at $\phi_{M_3}=\pi/2$ and $\phi_{M_3}=3\pi/4$ these
corrections have caused an enhancement. The value of $\Delta m_b$ in
these plots is $1.05-0.12i$, $-1.16i$,$-0.75-0.86i$ for $\phi_{M_3}=0$,
$\phi_{M_3}=\pi/2$ and $\phi_{M_3}=3\pi/4$, respectively.






\subsection{Calculation of the $h_a\to\tau^+ \tau^- $ decay width}

The calculation of the $h_a\to\tau^+ \tau^- $ decay width is similar to that of the $h_a \to b \bar{b}$  decay width, with the simplification that no QCD corrections are required. We have calculated the full 1-loop genuine vertex corrections and supplemented these with propagator corrections (including 1-loop mixing with Goldstone and Z bosons) and QED corrections. As before, we have included all complex phases.


\section{Higgs cascade decay branching ratios}
\label{section:higgsBR}

Accurate predictions for Higgs branching ratios are vital for Higgs phenomenology. In particular, they are required as part of calculations of cross sections of collider processes involving the production and decay of an on-shell Higgs boson, which are often performed using the narrow width approximation. In \refse{sec:exclplots}, we will use Higgs branching ratios for the CPX scenario in conjunction with the LEP topological cross section limits. In order to understand the resulting exclusions, it will be necessary to refer to the behaviour of the contributing branching ratios.  

We combine the $h_a \to h_b h_c$ decay widths calculated in \refse{sec:hihjhk} with the $h_a \to b \bar{b}$ and $h_a \to \tau^- \tau^+$ decay widths calculated in \refse{chapter:Hiff}. As we have discussed, these decay widths include the full 1-loop genuine vertex corrections and are combined with propagator corrections\footnote{Notice also that there are some points within the CPX parameter space that are shown here without a branching ratio value and that the edge of the allowed parameter region is uneven. These are points where either the mass calculation or the Z-factor calculation did not produce a stable result because the terms involving double derivatives of self-energies were non-negligible, as described in \refeq{se2order}.} obtained using neutral Higgs self-energies from the program {\feynhiggs}, which include the leading 2-loop contributions. The 1-loop propagator mixing with Goldstone and Z bosons is also consistently incorporated. These results take into account the full phase dependence of the supersymmetric parameters. For the $h_a \to b \bar{b}$ decay width, the $\Delta  m_b$ corrections are resummed in a way that preserves the phase dependence. We take all other decay widths from the program {\feynhiggs} (these decay modes are subdominant in most of the regions of MSSM parameter space).

\begin{figure}
\begin{center}
\resizebox{\textwidth}{!}{%
\includegraphics{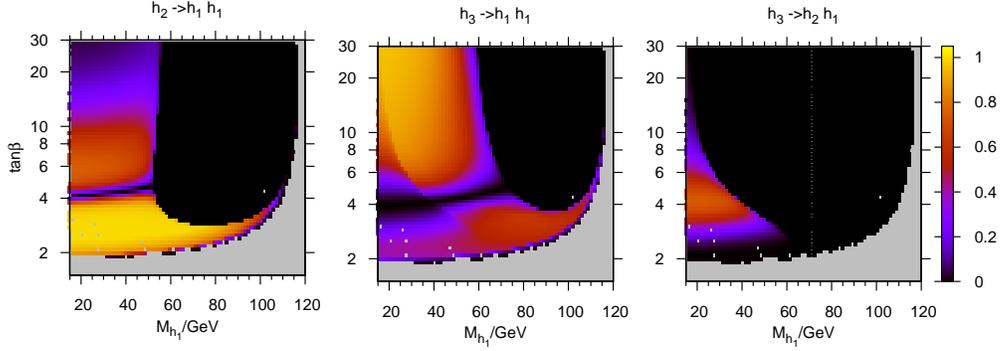}
}
\end{center}
\caption{The branching ratios of neutral Higgs bosons into other neutral Higgs bosons in the $\CPX$ scenario.
}
\label{Br2}
\end{figure}

\reffi{Br2} (left plot) illustrates the pronounced dependence 
of the $h_2\to h_1 h_1$ branching ratio on $\tb$ and $M_{h_1}$. We see that this decay mode is significant and often dominant in almost all of the regions where it is kinematically allowed. We can see that the characteristics of the $h_2\to h_1 h_1$ branching ratio are largely determined by the behaviour of the $h_2\to h_1 h_1$ decay widths (in \reffi{h2h1h1cpx} we showed this decay width for two slices of $\CPX$ parameter space). Note, in particular, the narrow `knife-edge' region of very low $h_2\to h_1 h_1$ branching ratio, which occurs at $\tb\sim 4.5$, where the $h_2\to h_1 h_1$ decay width tends to zero. The behaviour of the $h_2\to b \bar{b}$ branching ratio is also heavily dependent on the $h_2\to h_1 h_1$ decay width where the $h_2\to h_1 h_1$ decay is allowed kinematically, since in this region, the $h_2\to h_1 h_1$ decay usually makes up the majority of the total decay width. Over the majority of $\CPX$ parameter space, BR($h_2\to h_1h_1$)+BR($h_2\to b\bar{b})$+BR($h_2\to \tau^+\tau^-$)$\sim 1$, and BR($h_2\to \tau^+\tau^-$) is comparatively small. 

The Higgs cascade decays for the heaviest neutral Higgs $h_3$ also dominate 
in the majority of the region where they are kinematically allowed. The 
$h_3\to h_1 h_1$ branching ratio (middle plot of \reffi{Br2}) also has a 
narrow region at $\tb\sim 4-5$ in which, while the $h_3\to h_1 h_1$ decay 
is kinematically allowed, the decay width is nevertheless suppressed, characteristically similar to the suppressed region we observed in the $h_2\to h_1 h_1$ branching ratio. In particular, topologies involving $h_3$ can be relevant to the LEP exclusions in the region $10 \lsim \tb \lsim 30$, $M_{h_1}\lsim 60 \gev$ for variations of the CPX scenario. In this region of parameter space, 
the $h_3\to h_1 h_1$ decay width is a crucial contribution also to the
$h_3\to b\bar{b}$ branching ratio. The $h_3\to h_2 h_1$ decay width (the 
$h_3\to h_2 h_1$ branching ratio is shown in the right plot of
\reffi{Br2}), on the other hand, is of less phenomenological interest in this scenario, since this decay width dominates the total $h_3$ width in a region which, as we will see, can be excluded at the 95\% CL using limits on the cross sections of topologies involving $h_1$ and $h_2$.

\newpage
\section{Normalised $e^+e^-$ Higgsstrahlung and pair production cross sections}
\label{sec:Higgsprod} 

In order to examine the effect of our predictions for the Higgs
branching ratios on the size of the regions of $\CPX$ parameter space
which can be excluded by the LEP Higgs searches, we need to consider the
Higgsstrahlung and Higgs pair production cross sections, normalised to a
reference cross section $\sigma_{\rm ref}$. In the MSSM with real
parameters, the corrections to the LEP Higgsstrahlung and LEP pair production processes have been studied in detail \cite{hep-ph/9303309,Driesen:1995ew,Driesen:1995ib,Akeroyd:2001aka,Heinemeyer:2001iy,Beccaria:2005un}. In the MSSM with CP violation, there has also been considerable interest in accurate predictions of these cross sections, although the full 1-loop corrections are not yet available~\cite{Demir:1998dp,Carena:2000yi,Akeroyd:2001kt,Arhrib:2002ti,Ham:2007gw}. 

For the Higgsstrahlung topologies $e^+e^-\rightarrow h_a Z$, the reference cross section $\sigma_{\rm ref}$ is the tree level SM cross section for the process $e^+e^-\rightarrow Z\rightarrow H Z$, for a SM-like Higgs of mass $M_H=M_{h_a}$. For the pair production topologies $e^+e^-\rightarrow h_a h_b$, the reference cross section $\sigma_{\rm ref}$ is the tree level MSSM cross section for the process $e^+e^-\rightarrow Z\rightarrow h^0 A^0$, where the MSSM coupling factor $\cos^2{(\beta-\alpha)}$ has been divided out and the masses of $h^0$ and $A^0$ taken as $M_{h_a}$ and $M_{h_b}$ respectively. This reference cross section can also be expressed in terms of the Standard Model Higgsstrahlung production cross section,
\BEA
\sigma_{\rm ref}=\bar{\lambda}\sigma^{\rm SM}(e^+e^-\rightarrow H Z),
\EEA
where $\bar{\lambda}$ is a kinematic factor which takes into account the different kinematic dependences of the SM Higgsstrahlung and the pair production process, i.e.\
\BEA
\bar{\lambda}&=&\lambda^{3/2}_{h_ah_b}/\left(12 M_Z^2/s+\lambda_{ZH}\right)/\lambda^{1/2}_{ZH},\\
\lambda_{xy}&=&\left[1-\left(M_x+M_y\right)^2/s\right]\left[1-\left(M_x-M_y\right)^2/s\right],
\EEA
and $H$ is a SM-like Higgs with mass $M_H$.

For the majority of our scans, we will calculate these normalised cross sections using an effective coupling for the $h_a$-$Z$-$Z$ or $h_a$-$h_b$-$Z$ vertex, which incorporates external Higgs propagator corrections. However, we will also examine the effect of including the complete $t,\tilde{t},b,\tilde{b}$ 1-loop corrections (involving also genuine vertex corrections) in these cross sections.

\subsection{Normalised effective Higgs couplings to gauge bosons}

The matrix $\matr{Z}$ can be used to create a normalised effective coupling between neutral Higgs bosons and Z bosons which takes the corrections to the external Higgs propagators into account, though the relations
\BEA
g^{\rm eff}_{h_aZZ}&=&\matr{Z}_{ai}g^{\rm tree}_{h_iZZ},\label{eq:g2hjZZ}\\
g^{\rm eff}_{h_ah_bZ}&=&\matr{Z}_{bj}\matr{Z}_{ai}g^{\rm tree}_{h_ih_jZ},
\label{eq:g2hjhiZ}
\EEA
where $g^{\rm tree}_{h_iZZ}$ are normalised to the SM coupling, such that $(g^{{\rm tree}}_{hZZ})^2=\sin^2{(\beta-\alpha)}$, $(g^{{\rm tree}}_{HZZ})^2=\cos^2{(\beta-\alpha)}$ and $(g^{{\rm tree}}_{AZZ})^2=0$. In addition, we normalise the $g^{\rm tree}_{h_ih_jZ}$ such that $(g^{{\rm tree}}_{hAZ})^2=\cos^2{(\beta-\alpha)}$ and $(g^{{\rm tree}}_{HAZ})^2=\sin^2{(\beta-\alpha)}$.  All other $g^{\rm tree}_{h_ih_jZ}$ are zero.

\begin{figure}
\begin{center}
\resizebox{\textwidth}{!}{%
\includegraphics{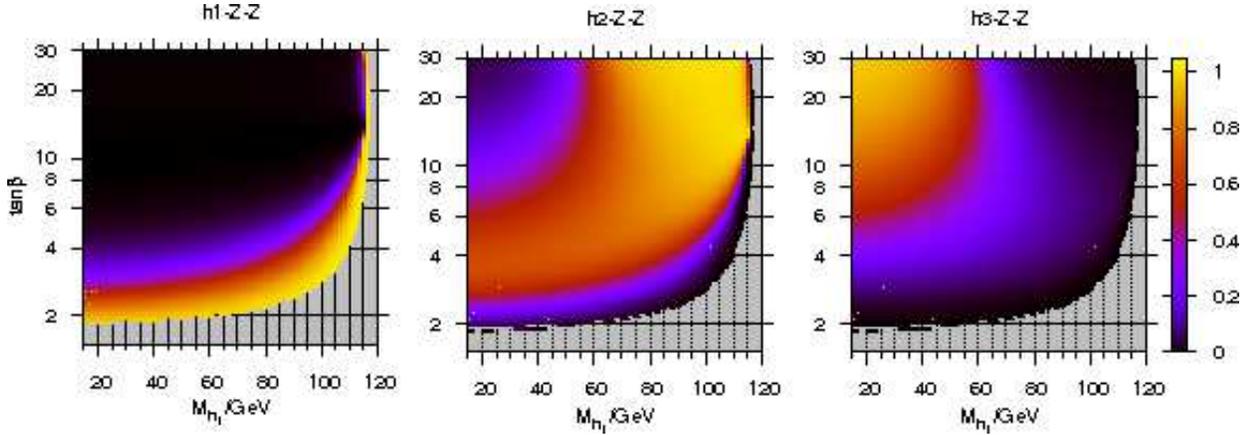}
}
\end{center}
\caption{The effective couplings of the neutral Higgs bosons to two $Z$ bosons, $|g^{{\rm eff}}_{h_1ZZ}|^2$,$|g^{{\rm eff}}_{h_2ZZ}|^2$ and $|g^{{\rm eff}}_{h_3ZZ}|^2$, which include the Higgs propagator corrections calculated using the matrix $\matr{\hat Z}$.
\label{ghiZZsq}
}
\end{figure}

\begin{figure}
\begin{center}
\resizebox{\textwidth}{!}{%
\includegraphics{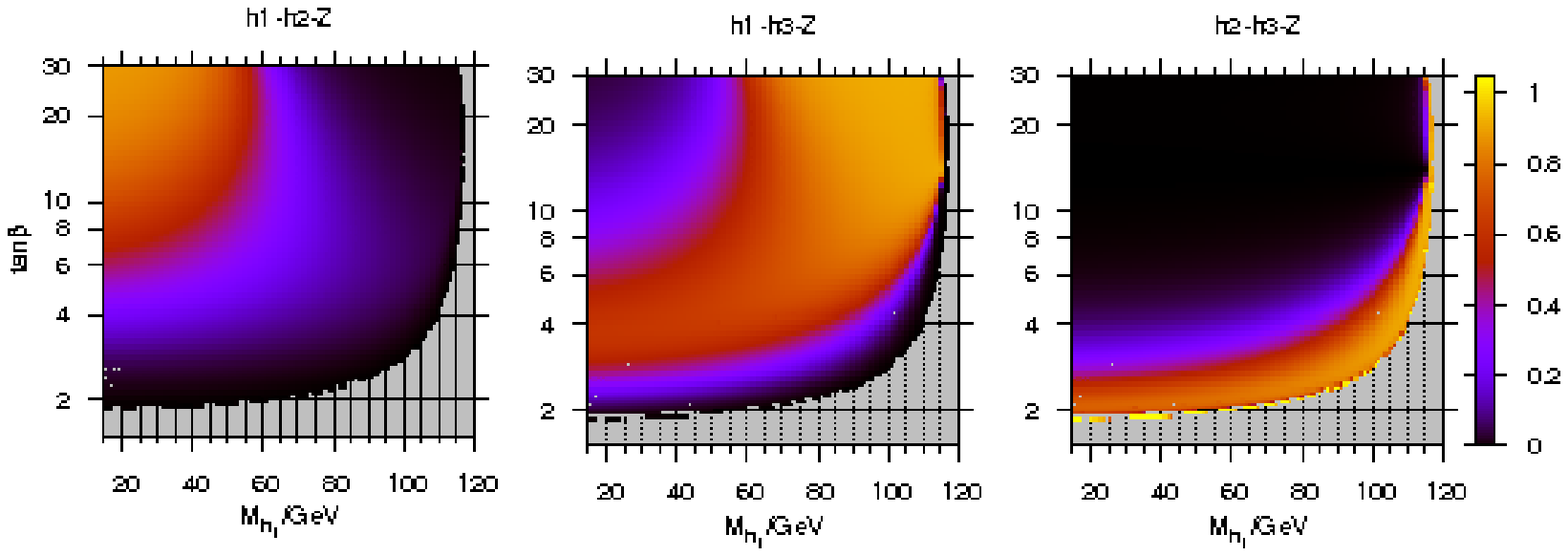}
}
\end{center}
\caption{The effective couplings of two neutral Higgs bosons to a $Z$ boson, $|g^{{\rm eff}}_{h_1h_2Z}|^2$,$|g^{{\rm eff}}_{h_1h_3Z}|^2$ and $|g^{{\rm eff}}_{h_2h_3Z}|^2$, which include the Higgs propagator corrections calculated using the matrix $\matr{\hat Z}$.
\label{ghihjZsq}
}
\end{figure}

\reffi{ghiZZsq} illustrates the normalised squared effective Higgs
couplings to gauge bosons $|g^{{\rm eff}}_{h_aZZ}|^2$ in the CPX
scenario. We can see that the $h_1$-$Z$-$Z$ coupling dominates around
the edge of the available parameter space, the $h_3$-$Z$-$Z$ coupling
dominates in a region $M_{h_1}<60 \gev$ and $7 \lsim \tan\beta$, and the $h_2$-$Z$-$Z$ coupling dominates the region in between, such that $|g^{{\rm eff}}_{h_1ZZ}|^2+|g^{{\rm eff}}_{h_2ZZ}|^2+|g^{{\rm eff}}_{h_3ZZ}|^2\sim 1$. \reffi{ghihjZsq} illustrates the behaviour of $|g^{{\rm eff}}_{h_ah_bZ}|^2$, which can be described as $|g_{h^{{\rm eff}}_aZZ}|^2\sim |g^{{\rm eff}}_{h_bh_cZ}|^2$, where $h_a,h_b,h_c$ are all different. (If a unitary approximation to the $\matr{Z}$ matrix is used, as in the LEP Higgs Working Group analysis of the $\CPX$ scenario\cite{Schael:2006cr}, these relations become equalities.)

The normalised Higgsstrahlung and pair production cross sections can then be approximated by these effective couplings, i.e
\BEA
\sigma(e^+e^-\rightarrow h_a Z)/\sigma_{\rm ref}&=&|g^{{\rm eff}}_{h_aZZ}|^2,\\
\sigma(e^+e^-\rightarrow h_a h_b)/\sigma_{\rm ref}&=&|g^{{\rm eff}}_{h_ah_bZ}|^2.
\EEA

\subsection{Loop corrections to the Higgsstrahlung and pair production cross sections}
\label{section:prodxs}

\unitlength=1bp%

\begin{figure}
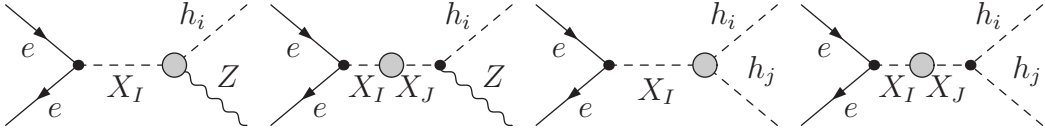

\begin{center}
\begin{feynartspicture}(432,100)(4,1)

\FADiagram{}
\FAProp(0.,15.)(6.,10.)(0.,){/Straight}{1}
\FALabel(2.48771,11.7893)[tr]{$e$}
\FAProp(0.,5.)(6.,10.)(0.,){/Straight}{-1}
\FALabel(3.51229,6.78926)[tl]{$e$}
\FAProp(20.,15.)(14.,10.)(0.,){/ScalarDash}{0}
\FALabel(16.6478,13.0187)[br]{$h_i$}
\FAProp(20.,5.)(14.,10.)(0.,){/Sine}{0}
\FALabel(17.5123,8.21074)[bl]{$Z$}
\FAProp(6.,10.)(14.,10.)(0.,){/ScalarDash}{0}
\FALabel(10.,9.18)[t]{$X_I$}
\FAVert(6.,10.){0}
\FAVert(14.,10.){-1}

\FADiagram{}
\FAProp(0.,15.)(6.,10.)(0.,){/Straight}{1}
\FALabel(2.48771,11.7893)[tr]{$e$}
\FAProp(0.,5.)(6.,10.)(0.,){/Straight}{-1}
\FALabel(3.51229,6.78926)[tl]{$e$}
\FAProp(20.,15.)(14.,10.)(0.,){/ScalarDash}{0}
\FALabel(16.6478,13.0187)[br]{$h_i$}
\FAProp(20.,5.)(14.,10.)(0.,){/Sine}{0}
\FALabel(17.5123,8.21074)[bl]{$Z$}
\FAProp(6.,10.)(14.,10.)(0.,){/ScalarDash}{0}
\FAVert(6.,10.){0}
\FALabel(8.,9.18)[t]{$X_I$}
\FAVert(10.,10.){-1}
\FALabel(12.,9.18)[t]{$X_J$}
\FAVert(14.,10.){0}

\FADiagram{}
\FAProp(0.,15.)(6.,10.)(0.,){/Straight}{1}
\FALabel(2.48771,11.7893)[tr]{$e$}
\FAProp(0.,5.)(6.,10.)(0.,){/Straight}{-1}
\FALabel(3.51229,6.78926)[tl]{$e$}
\FAProp(20.,15.)(14.,10.)(0.,){/ScalarDash}{0}
\FALabel(16.6478,13.0187)[br]{$h_i$}
\FAProp(20.,5.)(14.,10.)(0.,){/ScalarDash}{0}
\FALabel(17.5123,8.21074)[bl]{$h_j$}
\FAProp(6.,10.)(14.,10.)(0.,){/ScalarDash}{0}
\FALabel(10.,8.93)[t]{$X_I$}
\FAVert(6.,10.){0}
\FAVert(14.,10.){-1}

\FADiagram{}
\FAProp(0.,15.)(6.,10.)(0.,){/Straight}{1}
\FALabel(2.48771,11.7893)[tr]{$e$}
\FAProp(0.,5.)(6.,10.)(0.,){/Straight}{-1}
\FALabel(3.51229,6.78926)[tl]{$e$}
\FAProp(20.,15.)(14.,10.)(0.,){/ScalarDash}{0}
\FALabel(16.6478,13.0187)[br]{$h_i$}
\FAProp(20.,5.)(14.,10.)(0.,){/ScalarDash}{0}
\FALabel(17.5123,8.21074)[bl]{$h_j$}
\FAProp(6.,10.)(14.,10.)(0.,){/ScalarDash}{0}
\FAVert(6.,10.){0}
\FALabel(8.,9.18)[t]{$X_I$}
\FAVert(10.,10.){-1}
\FALabel(12.,9.18)[t]{$X_J$}
\FAVert(14.,10.){0}

\end{feynartspicture}
\caption{Additional corrections to the Higgsstrahlung and pair production processes. The grey circles indicate loops involving $t,\tilde{t},b,\tilde{b}$ and $X_I,X_J = (h,H,A,G,Z)$. Note that not all of the combinations are physical, since the tree level vertices must be CP-conserving. In addition, we neglect the electron mass, therefore no contributions where a Higgs boson couples to the electrons are included.
\label{fig:prodloopcorrdiagrams}}
\end{center}
\end{figure}

We now consider 1-loop corrections to the LEP Higgsstrahlung and pair
production processes involving the particles $t,\tilde{t},b,\tilde{b}$.
The structure of these diagrams is shown in
\reffi{fig:prodloopcorrdiagrams}, where loops involving
$t,\tilde{t},b,\tilde{b}$ are indicated by grey circles. In general,
lines labelled with $X$ can be $h,H,A,G,Z$, where the resulting diagram
conserves CP at the tree level vertices involving the gauge and Higgs
bosons. We neglect the electron mass and therefore will not include any diagrams in which a Higgs boson couples directly to the electrons. We then include the Higgs propagator factors for the Higgs on external legs through the $\matr{Z}$ matrix, as previously.

\begin{figure*}
\begin{center}
\resizebox{1\textwidth}{!}{%
  \includegraphics{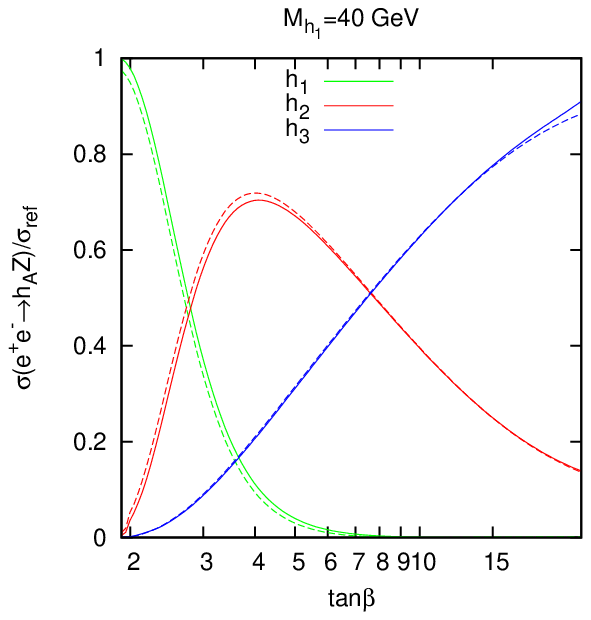}
  \includegraphics{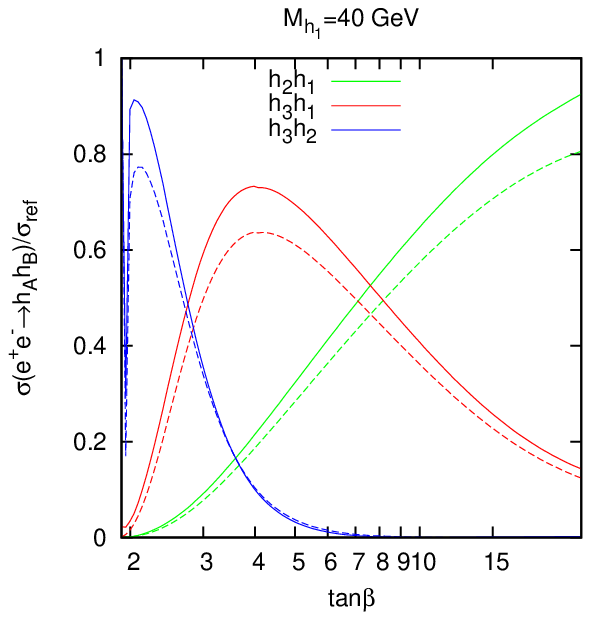}
}
\end{center}
\caption{The normalised Higgsstrahlung cross sections (left) and normalised pair production cross sections (right) in the $\CPX$ scenario at $M_{h_1}=40\gev$. Dashed lines: using \refeq{eq:g2hjZZ} or \refeq{eq:g2hjhiZ}. Solid lines: as for the dashed lines except that the $t,\tilde{t},b,\tilde{b}$ 1-loop corrections are also included (neglecting the electron mass).}
\label{fig:prodloopcorrgraph}       
\end{figure*}

The result of including these corrections on the normalised
Higgsstrahlung cross section in the $\CPX$ scenario at $M_{h_1}=40\gev$
is given in \reffi{fig:prodloopcorrgraph} (left). The dashed lines show
the result for the tree level $h_i$-$Z$-$Z$ vertex combined with Higgs
propagator factors, using \refeq{eq:g2hjZZ}. The solid lines show the
result when the full $t,\tilde{t},b,\tilde{b}$ 1-loop corrections are
also included. Similarly, the dashed lines in
\reffi{fig:prodloopcorrgraph} (right) show the normalised pair
production cross sections using \refeq{eq:g2hjhiZ}, and the solid lines
show the result if the $t,\tilde{t},b,\tilde{b}$ 1-loop corrections are
also used. The effect of including the $t,\tilde{t},b,\tilde{b}$
corrections turns out to be negligible for the Higgsstrahlung process.
However, for the pair production process, a more sizable effect is
visible, leading to an increase of the normalised cross section. This 
is due to the additional Yukawa coupling in the genuine vertex
corrections of the Higgs pair production process as compared to the
Higgsstrahlung process, so that a larger
enhancement factor is possible in this case. As will be discussed in the
follwing section, the impact of this kind of corrections on the coverage
of the LEP Higgs searches in the CPX scenario is nevertheless rather
small.

\section{Confronting the Higgs sector predictions with limits from the LEP Higgs searches}
\label{sec:LHWGresults}

\subsection{LHWG limits on the parameter space of the $\CPX$ scenario}

After the LEP programme finished in 2000, the final results from the
four LEP collaborations (ALEPH
\cite{Barate:2003sz,Barate:2000ts,Heister:2001kr}, DELPHI
\cite{Abreu:2000fw,Abdallah:2003ip}, L3 \cite{Acciarri:2000ke} and OPAL
\cite{Abbiendi:2000ac,Abbiendi:2002yk}) were combined and examined for
consistency with a background hypothesis and a signal plus background
hypothesis in a coordinated effort between the LEP Higgs Working Group
for Higgs Searches and the LEP collaborations (LHWG). The results showed
no significant excess of events which would indicate the production of a 
Higgs boson. In the Standard Model, a lower bound on the Higgs mass of
$114.4 \gev$ at the 95\% confidence level was established
\cite{Barate:2003sz}, while restrictions were placed on the available parameter space of a variety of MSSM benchmark scenarios \cite{Schael:2006cr}, including the $\CPX$ scenario\cite{Carena:2000ks}\footnote{Note that the definition of the CPX scenario used in \citeres{Carena:2000ks,Schael:2006cr} differs slightly from the definition used in the present paper, as discussed in \refse{section:cpx}.}.

For the purposes of the LHWG analysis, two different programs were used to calculate Higgs masses and branching ratios in the complex MSSM: {\fh} version 2.0 \cite{Heinemeyer:1998yj} and {\cph} \cite{Carena:2000ks}, which was a predecessor of the program {\cpsuperh}. 
These two codes had significant differences in the incorporated higher
order corrections, and it was necessary to perform a conversion between
the two sets of input parameters, due to the different renormalisation
schemes used in the two codes. As explained below, the parameter conversion used was an approximation based on a calculation performed in the MSSM with real parameters. 

Separate analyses were performed using {\fh} and {\cph}. In order to
combine these results, a conservative method was adopted, in which a
point in parameter space was regarded as excluded only if it was
excluded by both the analysis using results from {\fh} and the analysis
using results from {\cph}. The LHWG analysis resulted in three unexcluded regions of $\CPX$ parameter space at 95 \% CL: 

\begin{itemize}
\item{(A)} $60 \gev \lsim M_{h_1}$ and $3 \lsim \tan \beta$ 
\item{(B)} $30 \gev \lsim M_{h_1}\lsim 50 \gev$ and $3 \lsim \tan \beta \lsim 10$ 
\item{(C)} $0 \gev \lsim M_{h_1}\lsim 10 \gev$ and $3\lsim \tan \beta \lsim 20$
\end{itemize}

The results from the separate {\fh} and {\cph} analyses showed substantial differences. In particular, the {\fh} analysis had a larger unexcluded region of type B, and the {\cph} analysis had a larger unexcluded region of type A, while both results showed similar unexcluded regions of type C. We shall concentrate on unexcluded regions of type A and B in this paper, since constraints other than those from Higgs searches play a role in the unexcluded region C (see, e.g., \citere{Lee:2007ai} for a discussion of region C).

There was an additional complication, since {\fh} as yet does not have a reliable calculation for the loop corrections to the triple Higgs couplings in the CP-violating MSSM. For the purposes of the `{\fh}' analysis, the triple Higgs coupling was therefore obtained from {\cph}, and then combined with Higgs masses and other Higgs sector quantities as calculated by {\fh}\cite{Schael:2006cr}. As we will see, higher order corrections to the triple Higgs coupling have a great influence on the size, shape and position of the unexcluded region B.

\subsection{The effect of the new Higgs sector corrections on the limits on the parameter space of the $\CPX$ scenario}
\label{sec:exclplots}

The LEP Higgs Working Group for Higgs Searches and the LEP collaborations also published their combined results in the form of topological cross section limits at 95\% CL, which can be applied to a wide range of theoretical models. In each of these topologies, the Higgs is produced either through Higgsstrahlung or pair production and decays either to b-quarks, tau-leptons or via the Higgs cascade decay. To a very good approximation, the kinematic distributions of these processes are independent of the CP properties of the Higgs bosons involved, as discussed in \citere{Schael:2006cr}. Therefore, the same topological bounds can be used for CP-even, CP-odd or mixed CP Higgs bosons. 

In this section, we will use the topological cross section limits from LEP in conjunction with updated predictions for the Higgs masses, couplings and branching ratios. In particular, we shall be using our full 1-loop diagrammatic calculation for the $h_i\to h_j h_k$ decay processes with full phase dependence as described in  \refse{sec:hihjhk}, combined with renormalised neutral Higgs self-energies obtained from the current version of {\fh} (which includes corrections at $\mathcal{O}(\alpha_t \alpha_s)$ with full phase dependence). 

In order to utilise the cross section limits, we use the program
{\higgsbounds}. As input, it requires the Higgs masses, the normalised
production cross sections and the branching ratios BR($h_a\to h_bh_b$),
BR($h_a\to b \bar{b}$) and BR($h_a\to \tau^+ \tau^-$) for each parameter
point. We obtain the Higgs masses as described in \refse{section:Higgsmasses} 
and we calculate the Higgs branching ratios as described in \refse{sec:hihjhk}. 
We will begin by using LEP Higgsstrahlung production cross sections which include both the full propagator corrections and additional $t,\tilde{t},b,\tilde{b}$ corrections, as discussed in \refse{section:prodxs}. 
Unless otherwise stated, we will investigate the CPX scenario, as defined in \refse{section:cpx} and, unless explicitly stated, we do not consider the possible impact of theoretical uncertainties from unknown higher order corrections on the exclusion bounds in the parameter space.

{\hb} uses the provided Higgs sector predictions to determine which process has the highest statistical sensitivity for setting an exclusion limit for each parameter point, using the median expected limits based on Monte Carlo simulations with no signal. It then compares the theoretical cross section for this particular process with the experimentally observed limit for this process. In this way, only one topological limit is used for each parameter point, thus ensuring that any resulting exclusion is valid at the 95\% CL.

However, it should be noted that, in general, the dedicated analyses
carried out in \citere{Schael:2006cr} for specific MSSM benchmark
scenarios have a higher exclusion power than analyses using {\hb},
since, in a dedicated analysis, the information from different search
channels can be combined. This can, in particular, lead to an improved result in regions of parameter space where several channels have similar statistical sensitivities.

\begin{figure}
\begin{center}
\resizebox{1\textwidth}{!}{%
  \includegraphics{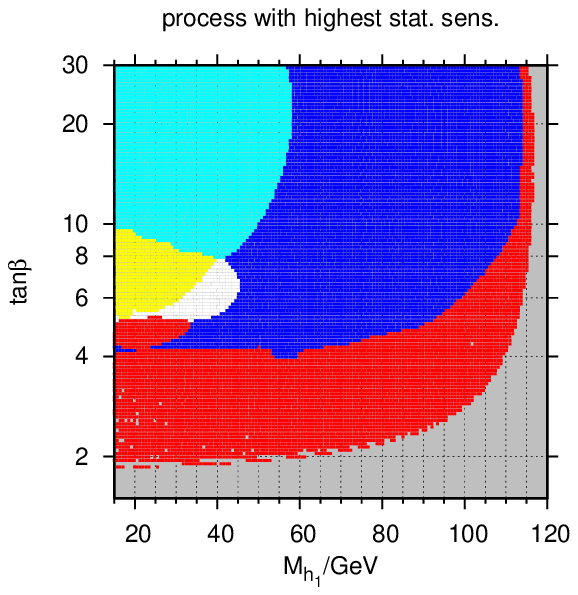}
  \includegraphics{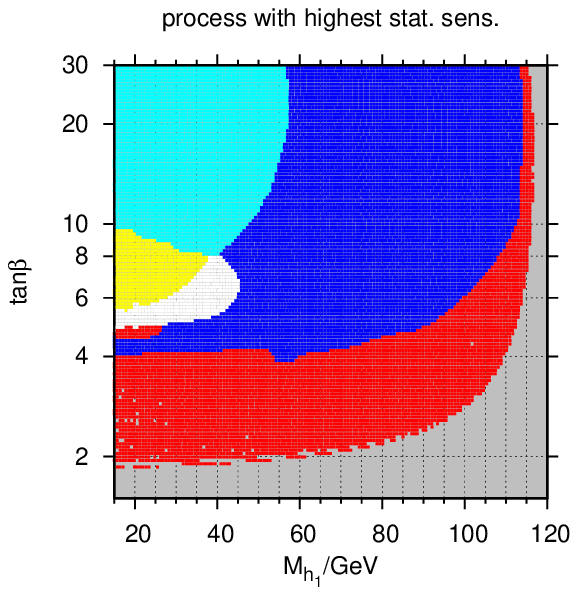}
}
\end{center}
\caption{The coverage of the LEP Higgs searches in the $M_{h_1}$--$\tb$ plane
of the CPX scenario, showing the channels that are 
predicted to have the highest statistical sensitivity for setting an
exclusion limit.
The colour codings
are: red = $\hZtobbZ$,
blue = $\HZtobbZ$,
white = $\HZtohhZtobbbbZ$,
cyan = $\Hhtobbbb$,
yellow = $\Hhtohhhtobbbbbb$,
green = $\Ahtobbbb$,
purple = other channels (${\color{Purple}\blacksquare}$). Left: full result (i.e.\ including the extra corrections to the Higgstrahlung and pair production cross sections described in \refse{section:prodxs}). Right: using effective coupling approximation for production cross sections.
\label{CPXreg}}
\end{figure}

\reffi{CPXreg} (left) indicates which channel has the highest
sensitivity and therefore which channel will be used for each point in
CPX parameter space, to determine whether or not it is excluded at the
95 \% CL. The channel $\hZtobbZ$ has the highest statistical sensitivity
at the edge of the $\CPX$ parameter space where $\tb$ is low or
$M_{h_1}$ is high, due to the fact that the coupling of the lightest
Higgs boson to two Z bosons is unsuppressed in this region, as we saw in
\reffi{ghiZZsq}. Similarly, in a band adjacent to this, where the
coupling of the second heaviest Higgs boson to two Z bosons is
unsuppressed, the Higgstrahlung processes $\HZtobbZ$ and
$\HZtohhZtobbbbZ$ have the highest statistical sensitivity. Finally, in
the upper left region of the plot, where $|g^{{\rm eff}}_{h_1h_2Z}|^2$
is high, the pair production processes $\Hhtobbbb$ and
$\Hhtohhhtobbbbbb$ have the highest statistical sensitivity. The part of
the parameter space in which the processes directly involving the
$h_2\to h_1 h_1$ decay ($\HZtohhZtobbbbZ$ and $\Hhtohhhtobbbbbb$)
dominate occurs in a region with an increased $h_2\to h_1h_1$ branching ratio at $\tb\sim 5-10$, as we saw in \reffi{Br2} (left), and which is influenced by the peak in the $h_2\to h_1h_1$ decay width as shown in \reffi{h2h1h1cpx} (left) at $\tb\sim 8$. \reffi{CPXreg} (right) differs from \reffi{CPXreg} (left) in that only the propagator corrections have been used when calculating the predictions for the LEP Higgs production cross sections (i.e.\ we use the normalised squared effective couplings $|g^{{\rm eff}}_{h_aZZ}|^2$ and $|g^{{\rm eff}}_{h_ah_bZ}|^2$ described in \refse{section:prodxs}). We see that the graphs are very similar, with the main difference being the reduced size of the $\hZtobbZ$ region at $\tb\sim 5$, $M_{h_1}\sim 25\gev$.

\begin{figure}
\begin{center}
\resizebox{1\textwidth}{!}{%
  \includegraphics{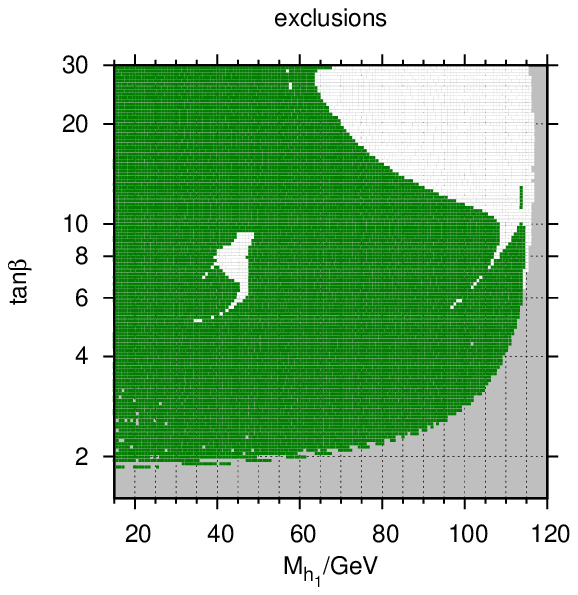}
  \includegraphics{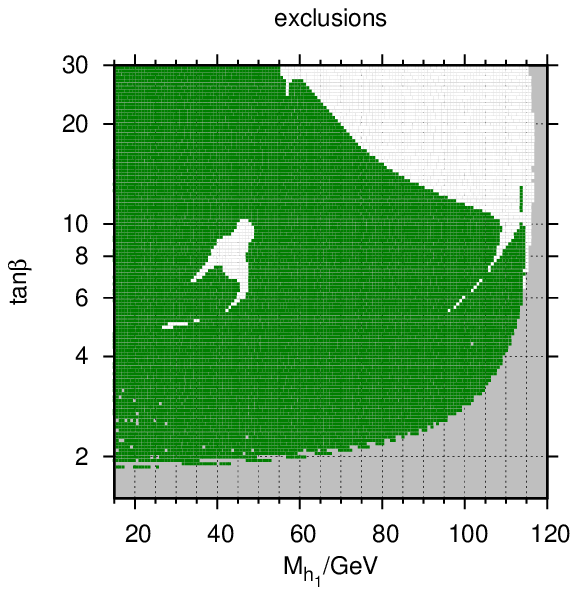}
}
\end{center}
\caption{ The coverage of the LEP Higgs searches in the $M_{h_1}$--$\tb$ plane
of the CPX scenario, showing the parameter regions excluded 
at the 95\% C.L. by the topological cross section limits obtained at LEP.
The colour codings are: green (darker grey) = LEP excluded, white = LEP allowed. Left: full result (i.e.\ including the extra corrections to the Higgstrahlung and pair production cross sections described in \refse{section:prodxs}). Right: using effective coupling approximation for production cross sections.
\label{CPXexcl}}
\end{figure}

In \reffi{CPXexcl} (left), we have compared our theoretical cross section predictions for each parameter point in the CPX scenario with the observed topological cross section limits obtained at LEP for the channel with the highest statistical sensitivity at that point, in order to obtain exclusions at 95\% CL. The Higgsstrahlung topologies where the Higgs decays to b-quarks are unable to exclude Higgs masses above $M_{h_1} \gsim 114.4 \gev$, as we would expect, since this is the limit on the mass of a SM-like Higgs boson~\cite{Barate:2003sz}. The upper edge of the excluded area in the $\HZtobbZ$ region has a similar shape to the $M_{h_2} \gsim 114.4 \gev$ contour and, at $\tb=20$, occurs at a position 
$\Delta M_{h_2}\sim 4\gev$ relative to it. As before, we call the unexcluded area in the top right region of the plot, `unexcluded region A'. It has a narrow `tail', which extends to lower $\tb$, one side of which is bounded by the limit for a SM-like $M_{h_2}$ and one side of which is bounded by the edge of the region where the channel $\hZtobbZ$ has the highest statistical significance, as shown in \reffi{CPXreg} (left). \reffi{CPXexcl} (right) differs from \reffi{CPXexcl} (left) in that only the propagator corrections have been used when calculating the predictions for the LEP Higgs production cross sections. Once again, we see that the extra $t,\tilde{t},b,\tilde{b}$ corrections only have a very small numerical effect. Therefore, we will neglect these corrections in the results that we show below.

\clearpage

\begin{figure}[h!]
\begin{center}
\begin{tabular}{cc}
\resizebox{0.27\textwidth}{!}{%
\includegraphics{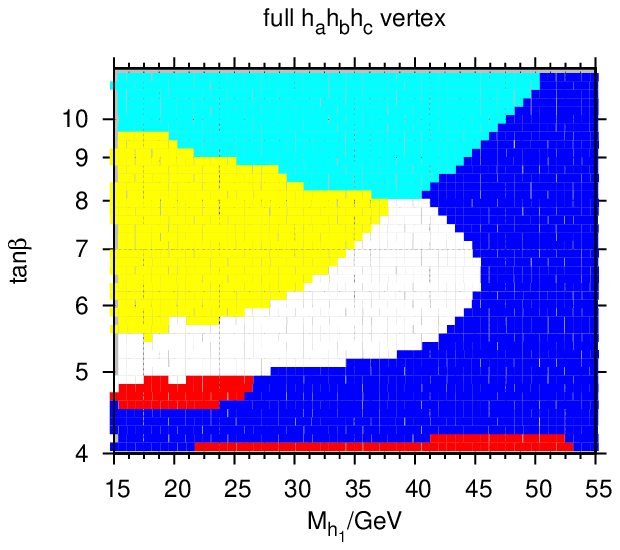}
}
&
\resizebox{0.27\textwidth}{!}{%
\includegraphics{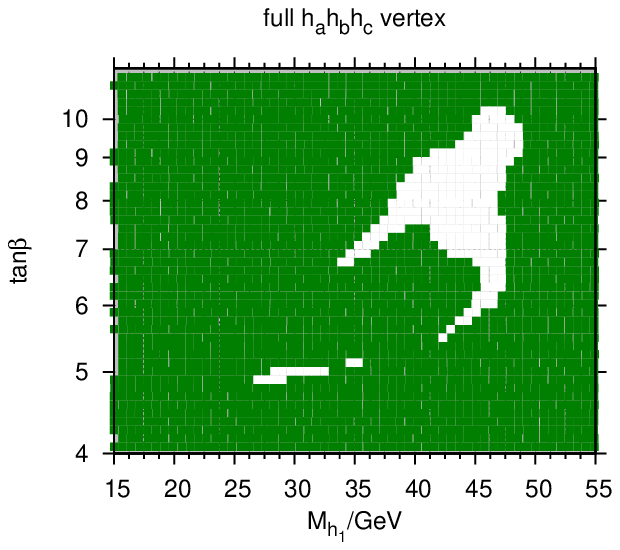}
}
\\
\resizebox{0.27\textwidth}{!}{%
\includegraphics{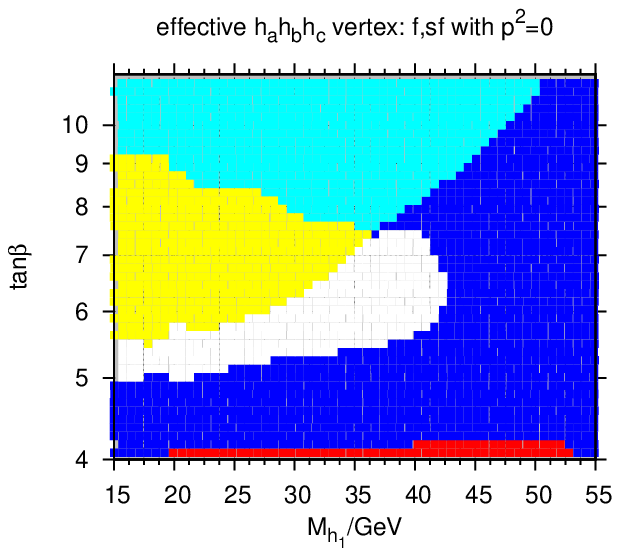}
}
&
\resizebox{0.27\textwidth}{!}{%
\includegraphics{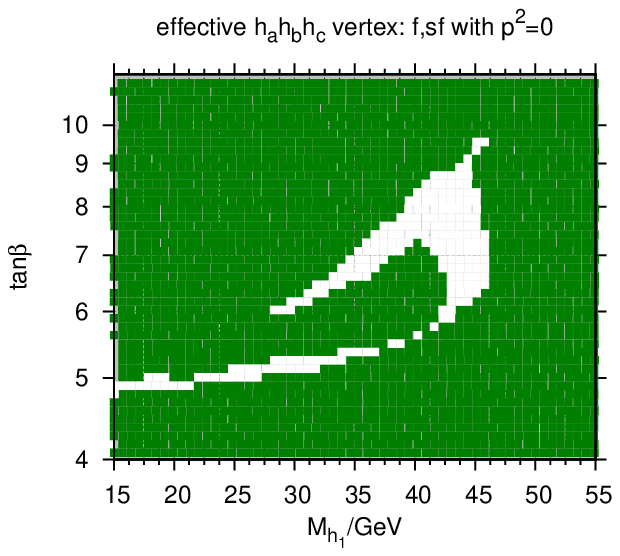}
}
\\
\resizebox{0.27\textwidth}{!}{%
\includegraphics{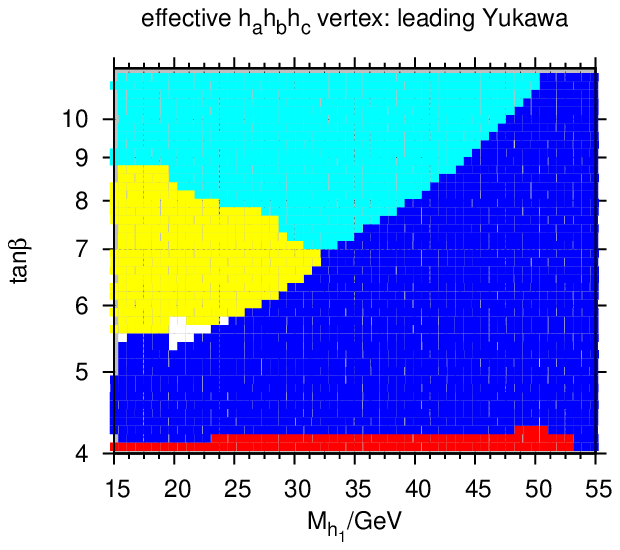}
}
&
\resizebox{0.27\textwidth}{!}{%
\includegraphics{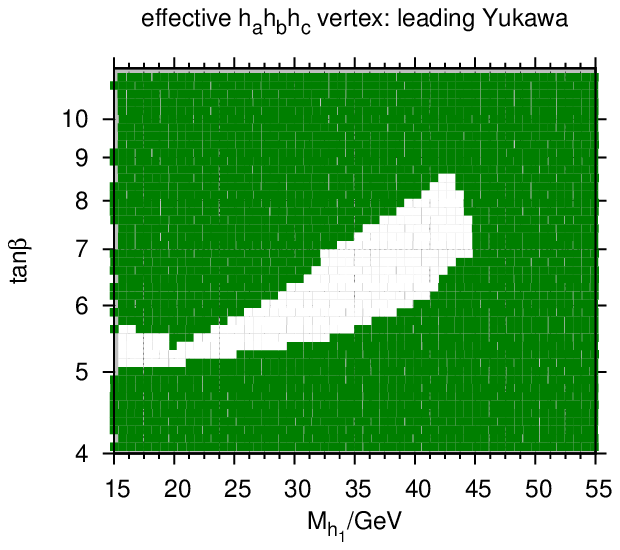}
}
\end{tabular}
\caption{ The channels with the highest statistical sensitivity (left) and LEP exclusion regions (right) for the CPX scenario at low values of $M_{h_1}$ and moderate values of $\tb$. 
The top row shows a subset of the results shown in \reffi{CPXreg} and \reffi{CPXexcl}.
In the second row, the genuine vertex corrections to the $h_2\to h_1h_1$ branching ratio include only the $f,\tilde{f}$ corrections at zero external momenta and, in the third row, the vertex corrections to the $h_2\to h_1h_1$ branching ratio have been calculated using the Yukawa approximation. In all three rows, the genuine vertex corrections to the $h_2\to h_1h_1$ branching ratio are combined with the full propagator corrections.
Left: red = $\hZtobbZ$,
blue = $\HZtobbZ$,
white = $\HZtohhZtobbbbZ$,
cyan = $\Hhtobbbb$,
yellow = $\Hhtohhhtobbbbbb$,
green = $\Ahtobbbb$,
purple = other channels (${\color{Purple}\blacksquare}$). Right: green (darker grey) = LEP excluded, white = LEP allowed.
\label{CPXzoomA} 
}
\end{center}
\end{figure}

The plots in \reffi{CPXexcl} also show an unexcluded region of type B at $M_{h_1}\sim 45\gev$ and $\tb \sim 8$. From comparison with \reffi{Br2} (left), we can see that the entire unexcluded region B in \reffi{CPXexcl} lies in an area where the $h_2\to h_1 h_1$ branching ratio is sizable.  
We examine this unexcluded region B in more detail in \reffi{CPXzoomA}
(top row), where we show an enlarged version of the relevant part of
parameter space from \reffi{CPXreg} (right) and \reffi{CPXexcl} (right).
As we can see, the two thin extensions of the unexcluded region both lie
along boundaries between areas where different processes have the
highest statistical significance. As we have discussed, we would expect
our method of combining channels (which only makes use of one observed
limit for each parameter point) to be less effective at such boundaries.
Thus, a dedicated analysis including the combination of different
channels might well be able to exclude such areas. 

The results presented here update our previous result\cite{Williams:2007dc} in two main ways. Most importantly, we now use a newer version of {\feynhiggs} to obtain the neutral Higgs self-energies, with an improved treatment of the $\tan {\beta}$ enhanced contributions. In addition, \citere{Williams:2007dc} uses $m_t=170.9\gev$. The resulting unexcluded region B has a similar size, shape and position in both analyses. However, in \citere{Williams:2007dc}, the unexcluded region A is almost non-existent, whereas the updated result shows a sizable unexcluded region of parameter space here, as we have discussed. 
 
\reffi{CPXzoomA} (top row) also shows that the bulk of the unexcluded region B lies in an area in which the channel $\HZtobbZ$ has the highest statistical sensitivity. The extent of the unexcluded region B on the higher $\tb$ side is very sensitive to the $h_2\to b\bar{b}$ branching ratio, which, as we discussed in \refse{section:higgsBR}, is critically dependent on the $h_2\to h_1h_1$ decay in this region of CPX parameter space, since the $h_2\to h_1h_1$ decay width yields the dominant contribution to the total decay width. The extent of the unexcluded region B towards lower values of $M_{h_1}$ is roughly determined by the edge of the region in which the channel $\HZtobbZ$ has the highest statistical sensitivity. This boundary is also very sensitive to the $h_2\to h_1h_1$ decay width, which has a large influence on the theoretical predictions of the other relevant channels: $\HZtobbZ$, $\Hhtobbbb$, $\Hhtohhhtobbbbbb$ and $\HZtohhZtobbbbZ$.

\begin{figure}
\begin{center}
\resizebox{0.5\textwidth}{!}{%
\includegraphics{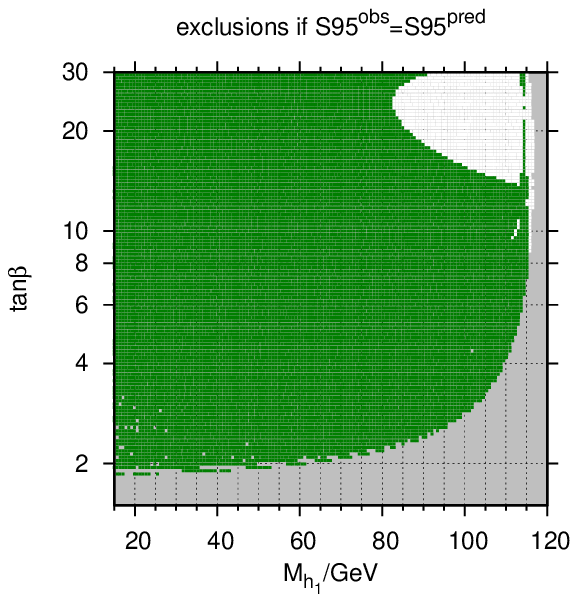}
}
\end{center}
\caption{The LEP exclusion regions for the hypothetical case in which all of the observed cross section values were measured to be the same as the ones expected in a background-only simulation. The colour codings are: green (darker grey) = LEP excluded, white = LEP allowed.
\label{ifnoexcess}}
\end{figure}

The unexcluded region B occurs within a region where the observed limit for the $e^+e^-\rightarrow (h_a)Z\rightarrow (b \bar{b})Z$ topology was more than one standard deviation above the expected limit (based on a background-only hypothesis). It is interesting to investigate the effect of this `slight excess' on the extent of the unexcluded region B. \reffi{ifnoexcess} shows what the exclusion would have been in the hypothetical situation in which the observed limit was exactly the same as the expected limit for all topologies. We see that the unexcluded region B disappears entirely and the size of the unexcluded region A has been reduced. We conclude that the presence of the `slight excess' in the LEP results for the $(h_a)Z\rightarrow (b \bar{b})Z$ topology is crucial to the existence of substantial unexcluded regions in the CPX scenario.

In order to further investigate the effects of our new genuine vertex corrections to the $h_2\to h_1h_1$ decay, we now compare the LEP exclusion regions based on the full result with cases where we have used various approximations to calculate the genuine vertex corrections to the $h_2\to h_1h_1$ decay. In \reffi{CPXzoomA} (centre), the genuine vertex corrections are approximated by the $f,\tilde{f}$ corrections at zero external momenta.  \reffi{CPXzoomA} (lower) displays the result when the leading Yukawa corrections to the triple Higgs vertex, as given by \refeqs{eq:phi1phi1phi1Yuk}--(\ref{eq:AAAYuk}), are used.

The boundary between channels related to $|g^{{\rm eff}}_{h_2ZZ}|^2$ (i.e.\ $\HZtohhZtobbbbZ$ and $\HZtobbZ$) and those related to $|g^{{\rm eff}}_{h_2h_1Z}|^2$ (i.e.\ $\Hhtobbbb$ and $\Hhtohhhtobbbbbb$) for both approximations are in about the same position as in \reffi{CPXzoomA} (upper). However, the boundaries between channels directly involving the $h_2 \to h_1 h_1$ decay and those that do not involve this decay have shifted. In particular, the region where the channel $\HZtohhZtobbbbZ$ has the highest statistical sensitivity disappears entirely if the `leading Yukawa' approximation is used. This has a considerable impact on the shape of the unexcluded region B. The $f,\tilde{f}$ at $p^2=0$ approximation, on the other hand, performs very well, yielding an unexcluded region B which, in position and shape, agrees very well with the full result.

\begin{figure}
\begin{center}
\resizebox{0.5\textwidth}{!}{%
\includegraphics{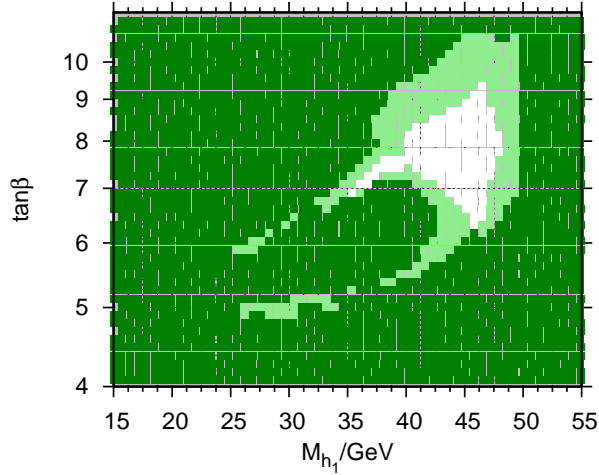}
}
\end{center}
\caption{The LEP exclusions for the CPX scenario plotted in the $M_{h_1}-\tan \beta$ plane. Results for $m_t=173.1 +1.3\gev$ and $m_t=173.1 -1.3\gev$. White: point is unexcluded in both, light green (lighter grey): point is excluded in only one, dark green (darker grey): point is excluded in both. \label{MHpTBLEPvaryMTexpt}}
\end{figure}

\begin{figure}
\begin{center}
\resizebox{0.5\textwidth}{!}{%
\includegraphics{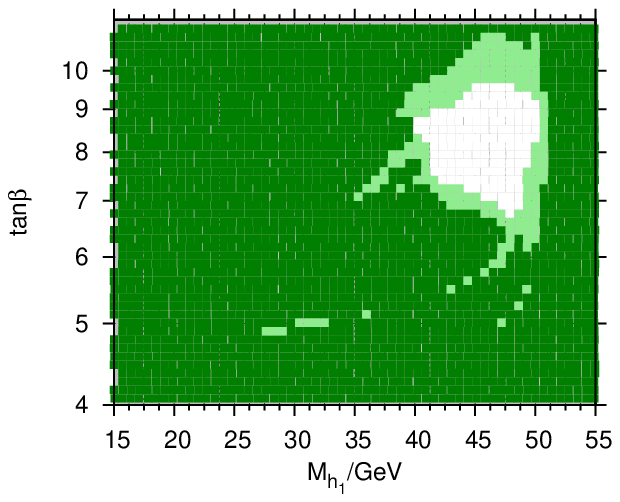}
}
\end{center}
\caption{The LEP exclusions for the CPX scenario plotted in the $M_{h_1}-\tan \beta$ plane. 
Results for (a) $m_t^{OS}$ in the 2-loop corrections in Higgs self
energies and 
$m_t^{OS}$ in the 1-loop genuine $h_ah_bh_c$ vertex corrections, and 
(b) $m_t^{\msbarm,SM}$ in the 2-loop corrections in 
Higgs self energies and $m_t^{\msbarm,SM}$ in the 1-loop genuine $h_ah_bh_c$ vertex corrections. 
White: point is unexcluded in both, light green (lighter grey): point is excluded in only one, dark green (darker grey): point is excluded in both. \label{MHpTBLEPvaryMTtheo}}
\end{figure}

It is interesting to consider the effect of the experimental uncertainty on the mass of the top quark. Since the leading corrections to the $h_2\to h_1h_1$ vertex are Yukawa corrections proportional to $m_t^4$, one would expect the unexcluded region B to exhibit a strong dependence on $m_t$. The neutral Higgs masses are also very sensitive to $m_t$, since these Yukawa corrections are also the leading corrections to the Higgs self-energies. We use the experimental value $m_t = 173.1\pm 1.3\gev$\cite{:2009ec}. \reffi{MHpTBLEPvaryMTexpt} shows the result of running scans for $m_t = 173.1+ 1.3\gev$ and $m_t = 173.1- 1.3\gev$ over the area of $\CPX$ parameter space containing the unexcluded region B. Parameter points which are excluded at both top masses are shown as dark green (darker grey), those excluded at one of the masses only are in light green (lighter grey) and those which remain unexcluded at both masses are in white. Therefore, the light green area (lighter grey area) demonstrates the uncertainty on the size and shape of the unexcluded region B due to the uncertainty from the measurement of the top mass. The unexcluded area for $m_t = 173.1+ 1.3\gev$ is approximately double the size of the unexcluded area for $m_t = 173.1- 1.3\gev$.

It is also possible to use a similar strategy to investigate the
dependence of the size of the unexcluded region B on the choice of $m_t$
at the higher loop orders in the calculation. In
\reffi{MHpTBLEPvaryMTtheo}, we use (a) $m_t^{OS}$ in the 2-loop corrections in Higgs self energies and 
$m_t^{OS}$ in the 1-loop genuine $h_ah_bh_c$ vertex corrections, 
and (b) $m_t^{\msbarm,SM}$ in the 2-loop corrections in 
Higgs self energies (as discussed in \refse{sec:choiceofmtinFH}) and
$m_t^{\msbarm,SM}$ in the 1-loop genuine $h_ah_bh_c$ vertex corrections (as previously). Points in parameter space which are excluded for the results of both calculation (a) and (b) are shown in dark green (darker grey), points excluded for the results of one calculation only are shown in light green (lighter grey), and points in white can be excluded for the results of neither calculation. In this way, we can get a rough estimate of the dependence of unexcluded region B on higher order corrections in the $t,\tilde{t}$ sector -- whilst these corrections are significant, the existence of an unexcluded region in this part of CPX parameter space is confirmed in both. Note that the region in which the results of one of the two calculations yield an unexcluded region forks into two narrow regions at its base, rather than one, reflecting the different positioning of the boundaries between regions where different processes have the highest statistical sensitivity in the two analyses contributing to this exclusion plot.

\begin{figure} 
\begin{center}
\resizebox{\textwidth}{!}{%
\includegraphics{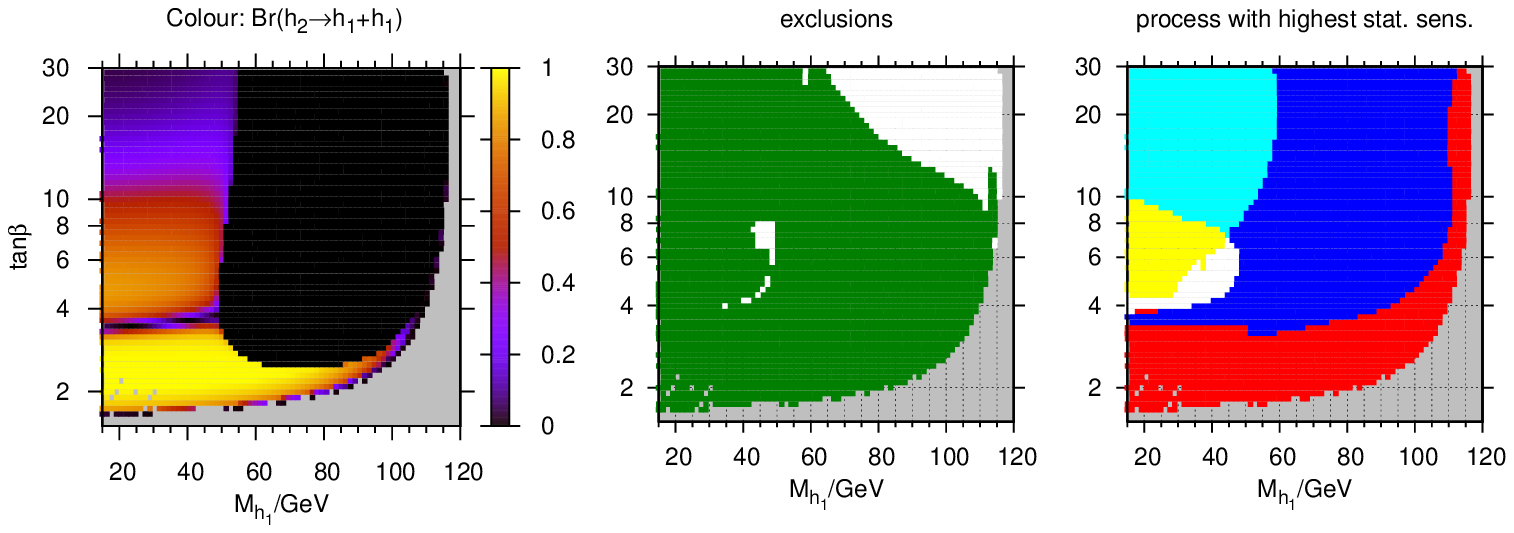}
}
\resizebox{\textwidth}{!}{%
\includegraphics{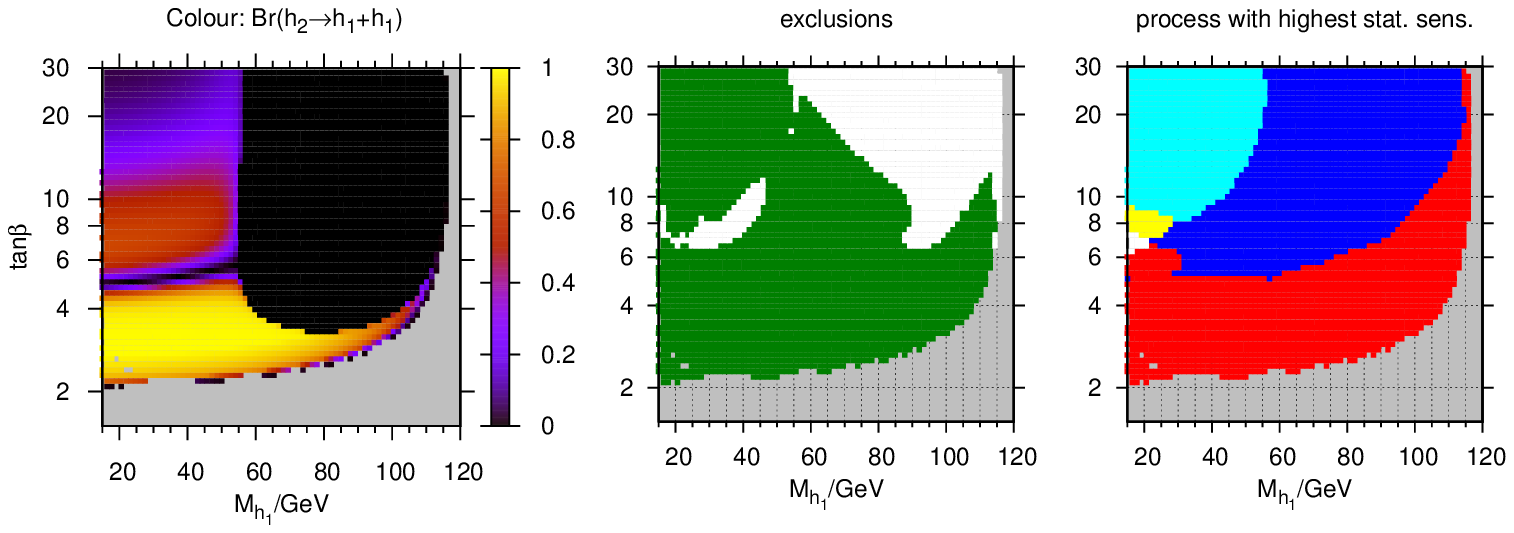}
}
\end{center}
\caption{The $h_2\to h_1 h_1$ branching ratio, LEP exclusions and channels with the highest statistical sensitivity plotted in the $M_{h_1}-\tan \beta$ plane. Upper graphs show $\phi_{A_t}=0.9 \times \pi/2$, lower graphs show $\phi_{A_t}=1.1 \times \pi/2$, other parameters taken from CPX scenario. See the captions of \reffi{CPXreg} and \reffi{CPXexcl} for the colour codes of the plots in the second and third columns.\label{MHpTBLEPbc}}
\end{figure}

The variation of $\phi_{A_t}$ has an interesting impact on the unexcluded regions. Recall that, in \reffi{h2h1h1cpx} (left), we saw that there was a peak in the $h_2\to h_1h_1$ decay width at moderate $\tb$, and a minimum at lower $\tb$. Varying $\phi_{A_t}$ by 10\% has a dramatic effect on the decay width, through changing the magnitude and position of this peak and changing its position with respect to $\tb$. In \reffi{MHpTBLEPbc}, which uses $\phi_{A_t}=0.9 \times \pi/2$ and $\phi_{A_t}=1.1 \times \pi/2$, we can see these effects reflected in the $h_2\to h_1 h_1$ branching ratio. In particular, we see that the thin horizontal minimum in BR($h_2\to h_1 h_1$) shifts to higher $\tb$ as $\phi_{A_t}$ increases. We can also see a change in the shape of the region in which the $h_2\to h_1 h_1$ decay is kinematically allowed and a slight change in the lower edge of CPX parameter space as plotted in the $M_{h_1}-\tb$ plane.

As one would expect, this strong dependence of the $h_2 \to h_1 h_1$
branching ratio on $\phi_{A_t}$ is reflected in a change in the balance
of
the processes with the highest statistical sensitivity at each point in
parameter space as $\phi_{A_t}$ increases. In particular, the size of
the
region in which either $\HZtohhZtobbbbZ$ or $\Hhtohhhtobbbbbb$ has the
highest sensitivity decreases.
The boundary between processes involving $|g^{{\rm eff}}_{h_2ZZ}|^2$ and those involving $|g^{{\rm eff}}_{h_2h_1Z}|^2$ also shifts to higher $\tb$. As a result, the unexcluded region B occurs at higher $\tb$ as $\phi_{A_t}$ increases and its shape changes significantly. The unexcluded region A increases in size as $\phi_{A_t}$ increases. 

\begin{figure} 
\begin{center}
\resizebox{\textwidth}{!}{%
\includegraphics{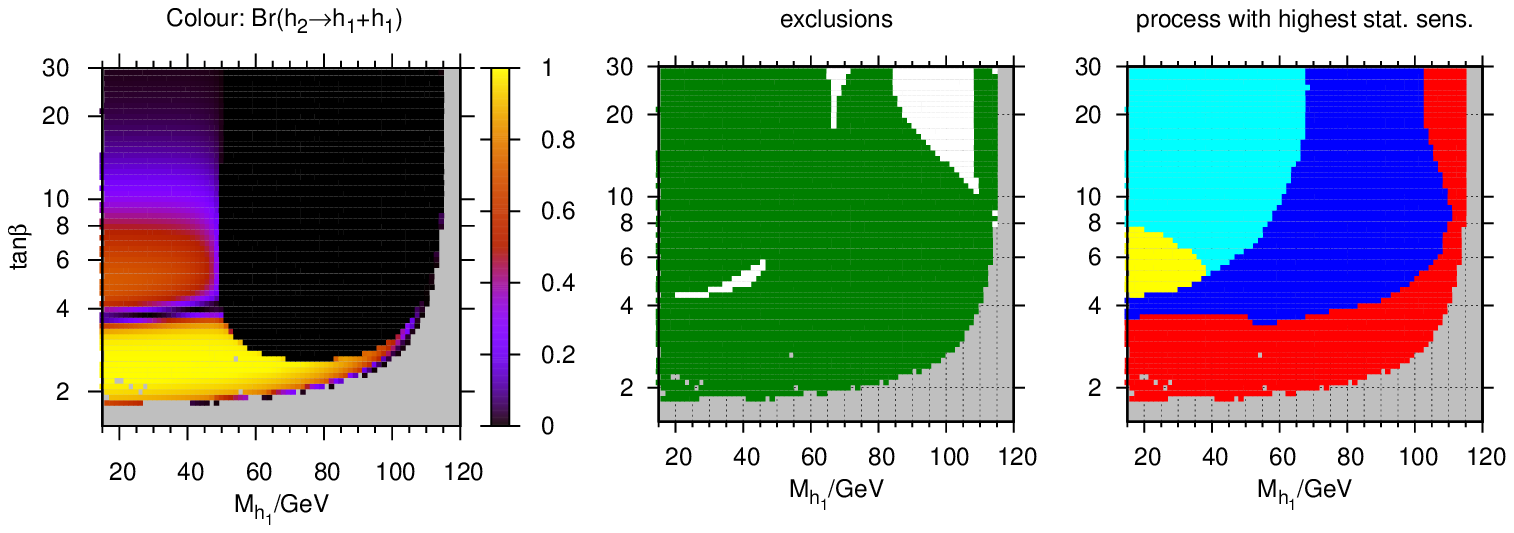}
}
\resizebox{\textwidth}{!}{%
\includegraphics{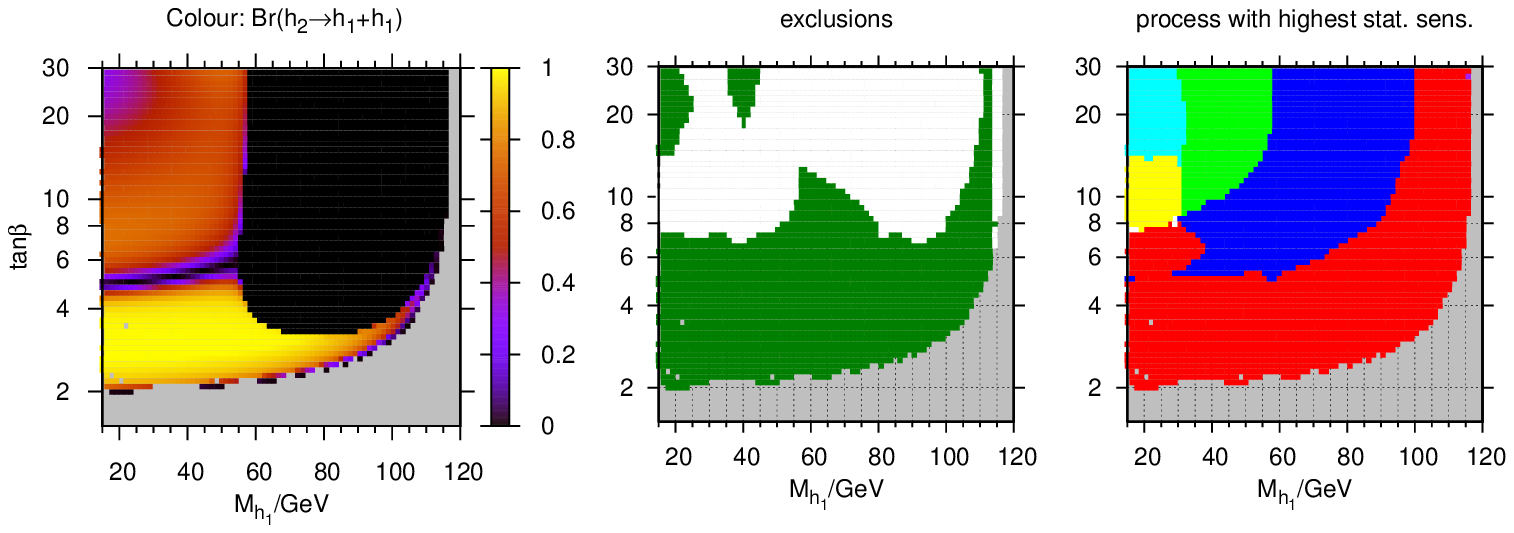}
}
\end{center}
\caption{The $h_2\to h_1 h_1$ branching ratio, LEP exclusions and channels with the highest statistical sensitivity plotted in the $M_{h_1}-\tan \beta$ plane. Upper graphs show $|A_t|=0.9 \times 900 \gev$, lower graphs show $|A_t|=1.1 \times 900 \gev$, other parameters taken from CPX scenario. See the captions of \reffi{CPXreg} and \reffi{CPXexcl} for the colour codes of the plots in the second and third columns.\label{MHpTBLEPcapBcapC}}
\end{figure}

\reffi{MHpTBLEPcapBcapC} illustrates that varying $|A_t|$ by 10\% has a
very substantial effect on the LEP exclusions in the CPX parameter
space. Increasing $|A_t|$ has a large effect on the $h_2\to h_1 h_1$
decay width. It increases the size of the peak at moderate $\tb$ in
\reffi{h2h1h1cpx} (left) and shifts the position of the minimum and the
peak shown in \reffi{h2h1h1cpx} (left) to higher values of $\tb$.
Increasing $|A_t|$ also make the slope of this graph less steep above
$\tb\sim 7$. We see these effects reflected in the $h_2\to h_1 h_1$
branching ratio in \reffi{MHpTBLEPcapBcapC}. We also see that a higher
value of $|A_t|$ yields a significant increase in the area of parameter space in which the $h_2\to h_1 h_1$ decay is kinematically allowed. 

For the case in which $|A_t|=1.1 \times 900 \gev$, the plot illustrating
the channels with the highest statistical sensitivity in
\reffi{MHpTBLEPcapBcapC} is very different from those discussed so far.
This is partly because $|g^{{\rm eff}}_{h_1h_2Z}|^2$ is reduced, which
drastically reduces the area where $\Hhtobbbb$ has the highest
statistical sensitivity. The area where the channel $\Hhtohhhtobbbbbb$
has the highest statistical sensitivity occurs at higher $\tb$ than
previously and is now unexcluded. Also, the suppression of $|g^{{\rm
eff}}_{h_1h_2Z}|^2$ means that the channel $\Ahtobbbb$ has the highest
statistical sensitivity over a large region, which can only be partially
excluded by this LEP limit. Therefore, the excluded LEP regions are
dramatically different for the CPX scenario with $|A_t|=1.1 \times 900
\gev$. It is worth noting, however, that this value of $|A_t|$ is close
to an unstable region of parameter space, in which loop corrections in
Higgs sector get extremely large. On the other hand, at $|A_t|=0.9
\times 900 \gev$, the unexcluded region B has almost disappeared and the
unexcluded region A has reduced in size as it is partially covered by
two vertical excluded regions, at $M_{h_1}\sim80\gev$ and 
$M_{h_1}\sim 110 \gev$.

The gluino mass parameter $M_3$ does not feature in the 1-loop corrections to the $h_2\to h_1 h_1$ decay or the 1-loop corrections to the Higgs masses. However, as we have seen, it can heavily influence the $h_a\to b\bar{b}$ decay width. In addition, the Higgs self-energies from {\fh} depend on $M_3$ through the $\mathcal{O}(\alpha_t\alpha_s)$ corrections and the $\Delta m_b$ corrections. Therefore, it is interesting to see if varying this parameter has a significant effect on the LEP exclusions. 

\begin{figure}
\begin{center}
\resizebox{\textwidth}{!}{%
\includegraphics{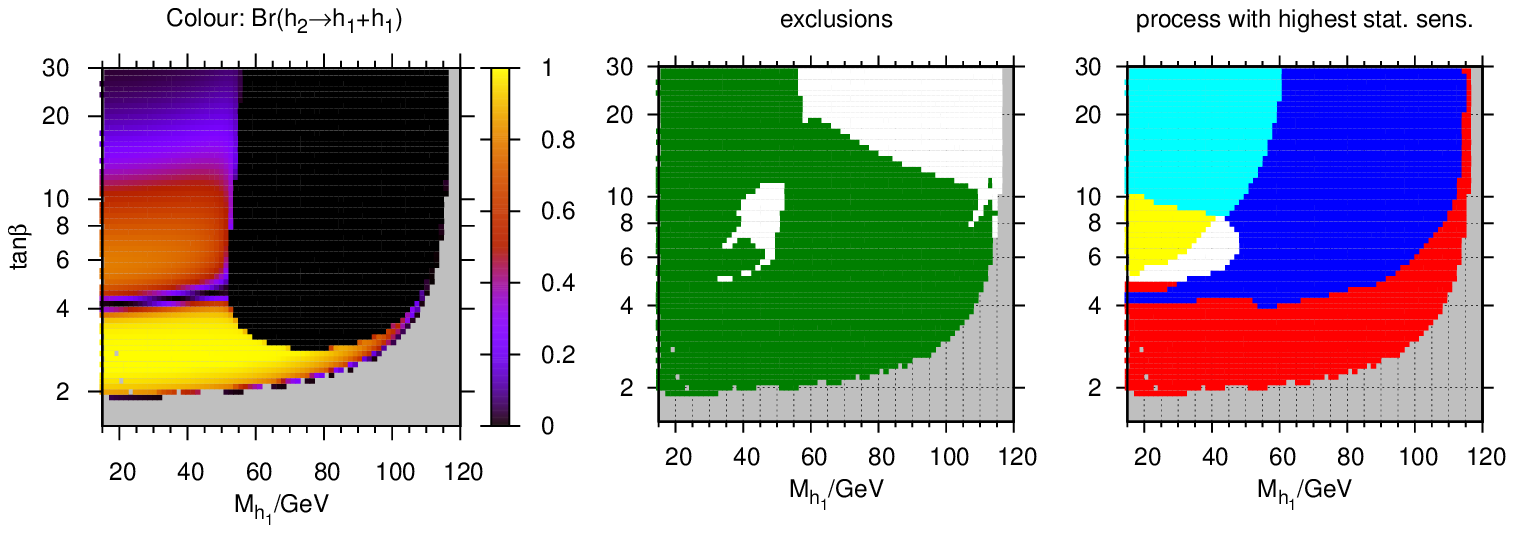}
}
\resizebox{\textwidth}{!}{%
\includegraphics{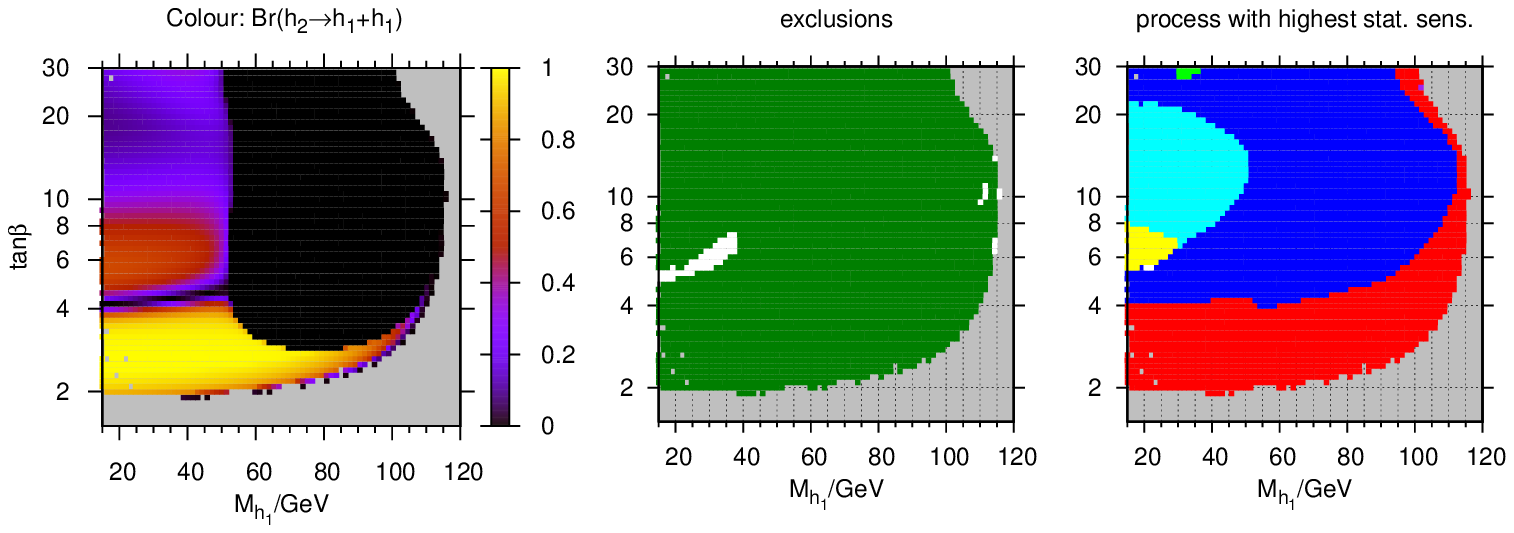}
}
\end{center}
\caption{The $h_2\to h_1 h_1$ branching ratio, LEP exclusions and channels with the highest statistical sensitivity plotted in the $M_{h_1}-\tan \beta$ plane. Upper graphs show $\phi_{M_3}= \pi/4$, lower graphs show $\phi_{M_3}=3 \pi/4$, other parameters taken from CPX scenario. See the captions of \reffi{CPXreg} and \reffi{CPXexcl} for the colour codes of the plots in the second and third columns.\label{MHpTBLEPargM3aargM3d}}
\end{figure}

In \reffi{MHpTBLEPargM3aargM3d}, we show results for $\phi_{M_3}=\pi/4$
and $\phi_{M_3}=3\pi/4$.  At $\phi_{M_3}=\pi/4$, the area of the
unexcluded regions A and B increases slightly compared to the {\CPX} setting of $\phi_{M_3}=\pi/2$. In contrast, at $\phi_{M_3}=3\pi/4$, the size of the unexcluded area B decreases and the unexcluded area A almost disappears. Note that the parameter space is not populated above $M_{h_1}\sim 100 \gev$ at $\tb=30$.

\begin{figure} 
\begin{center}
\resizebox{\textwidth}{!}{%
\includegraphics{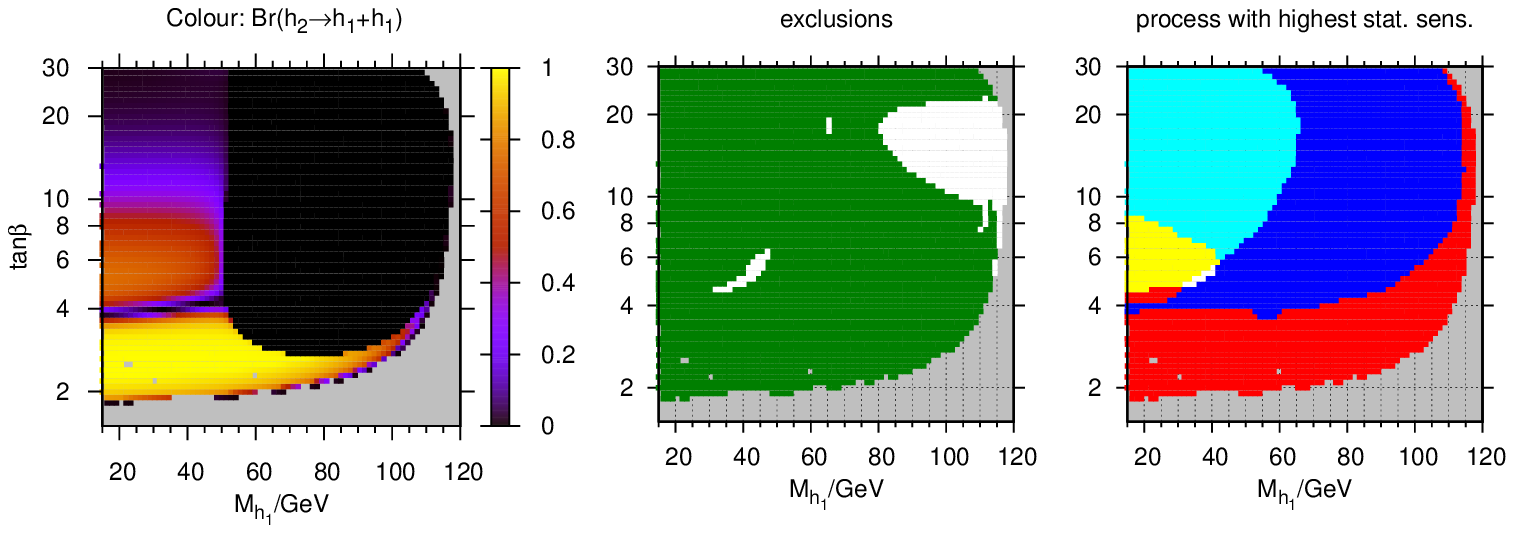}
}
\resizebox{\textwidth}{!}{%
\includegraphics{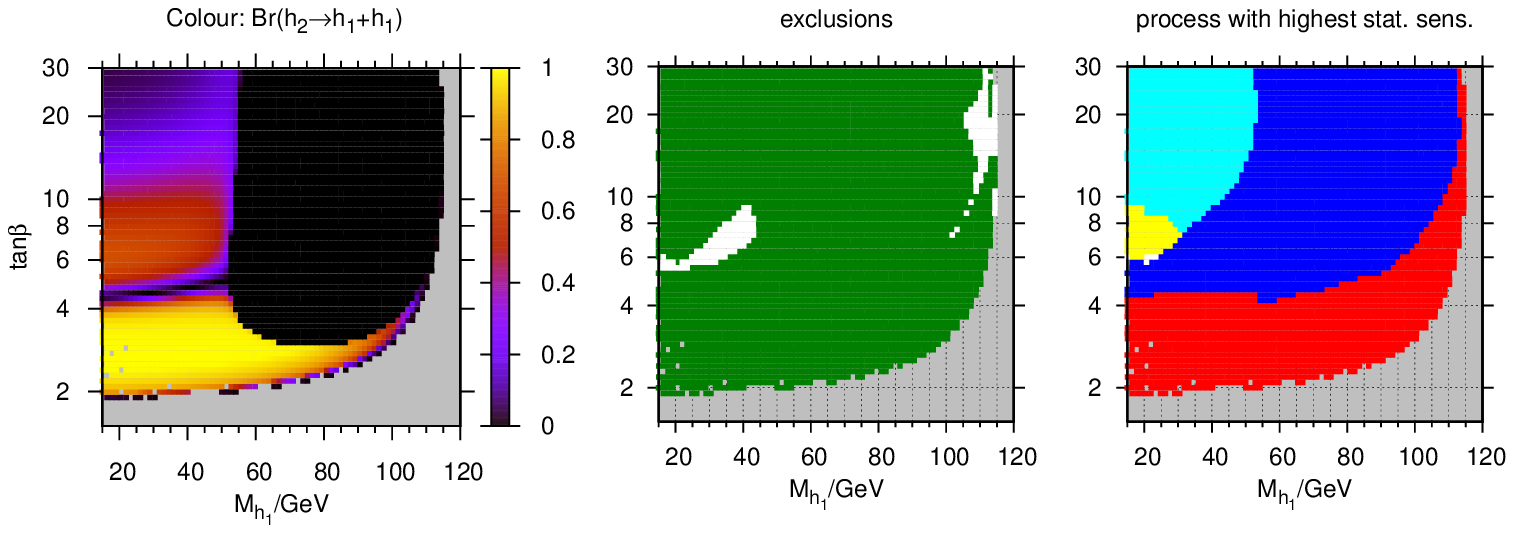}
}
\end{center}
\caption{The $h_2\to h_1 h_1$ branching ratio, LEP exclusions and channels with the highest statistical sensitivity plotted in the $M_{h_1}-\tan \beta$ plane. Upper graphs show $|M_3|=100 \gev$, lower graphs show $|M_3|=2000 \gev$, other parameters taken from CPX scenario. See the captions of \reffi{CPXreg} and \reffi{CPXexcl} for the colour codes of the plots in the second and third columns.\label{MHpTBLEPmodM3amodM3d}}
\end{figure}

\reffi{MHpTBLEPmodM3amodM3d} illustrates the results for $|M_3|=100\gev$ and $|M_3|=2000\gev$. Both unexcluded regions are reduced in both cases. Note that, for $|M_3|=100\gev$, the excluded region A is bounded from above.

\begin{figure}
\begin{center}
\resizebox{\textwidth}{!}{%
\includegraphics{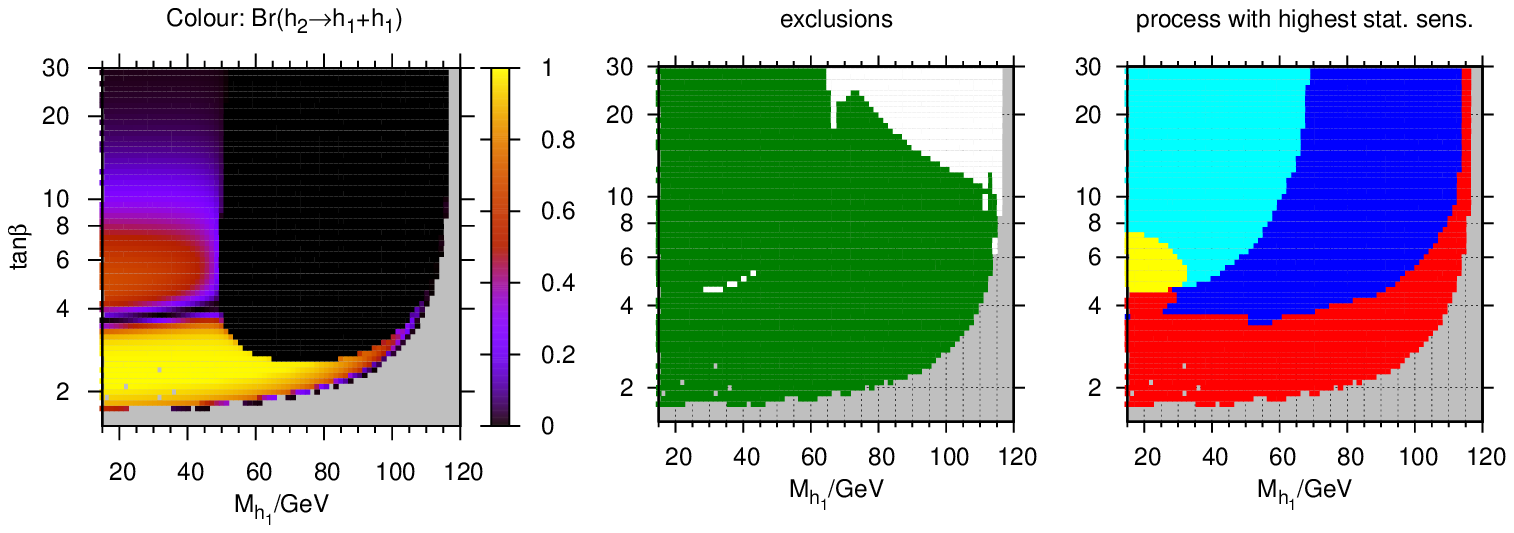}
}
\resizebox{\textwidth}{!}{%
\includegraphics{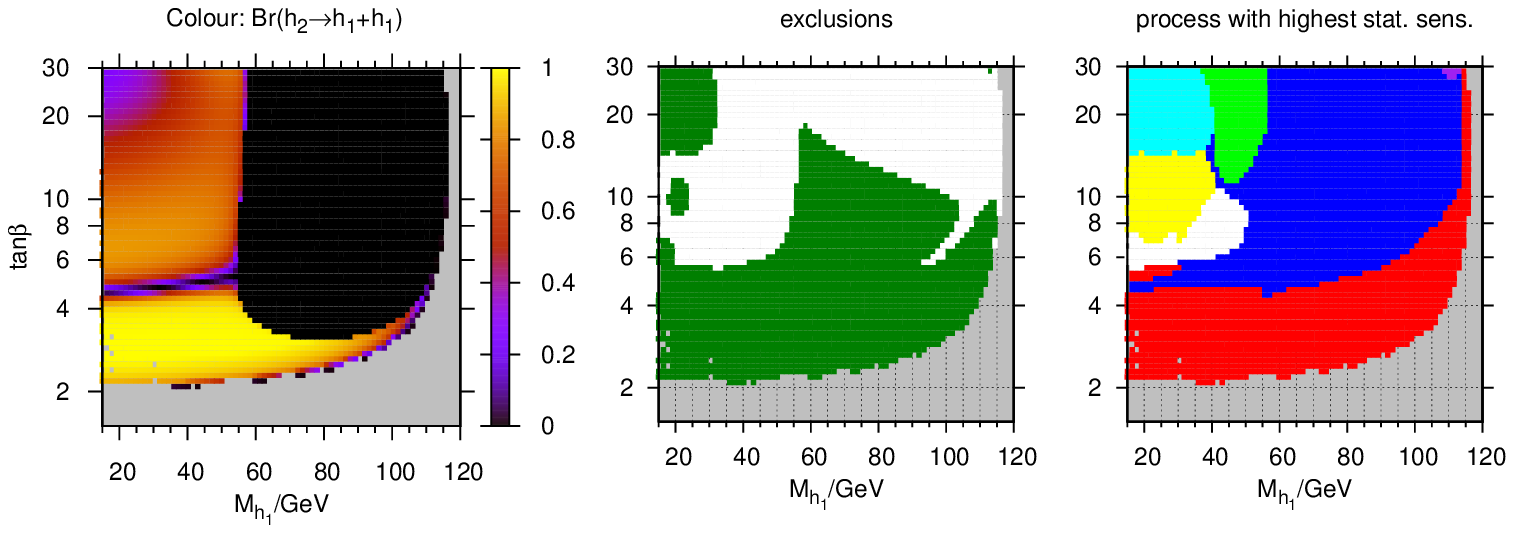}
}
\end{center}
\caption{The $h_2\to h_1 h_1$ branching ratio, LEP exclusions and channels with the highest statistical sensitivity plotted in the $M_{h_1}-\tan \beta$ plane. Upper graphs show $\mu=0.9 \times 2000 \gev$, lower graphs show $\mu=1.1 \times 2000 \gev$, other parameters taken from CPX scenario. See the captions of \reffi{CPXreg} and \reffi{CPXexcl} for the colour codes of the plots in the second and third columns.\label{MHpTBLEPmuebmuec}}
\end{figure}

It is also interesting to consider the effect of varying the Higgsino mass parameter $\mu$. The branching ratios shown in \reffi{MHpTBLEPmuebmuec} are qualitatively very similar to those in \reffi{MHpTBLEPcapBcapC}. As $\mu$ increases, $|g^{{\rm eff}}_{h_2ZZ}|^2$ is enhanced at the expense of $|g^{{\rm eff}}_{h_2h_1Z}|^2$, and this determines the relative sizes of the regions involving these couplings. The plot with $\mu=1.1\times 2000 \gev$ in \reffi{MHpTBLEPmuebmuec} has a large region in which the channel $\HZtohhZtobbbbZ$ has the highest statistical sensitivity. The unexcluded regions have increased in size substantially as $\mu$ increases and they have joined up.

\begin{figure}
\begin{center}
\resizebox{\textwidth}{!}{%
\includegraphics{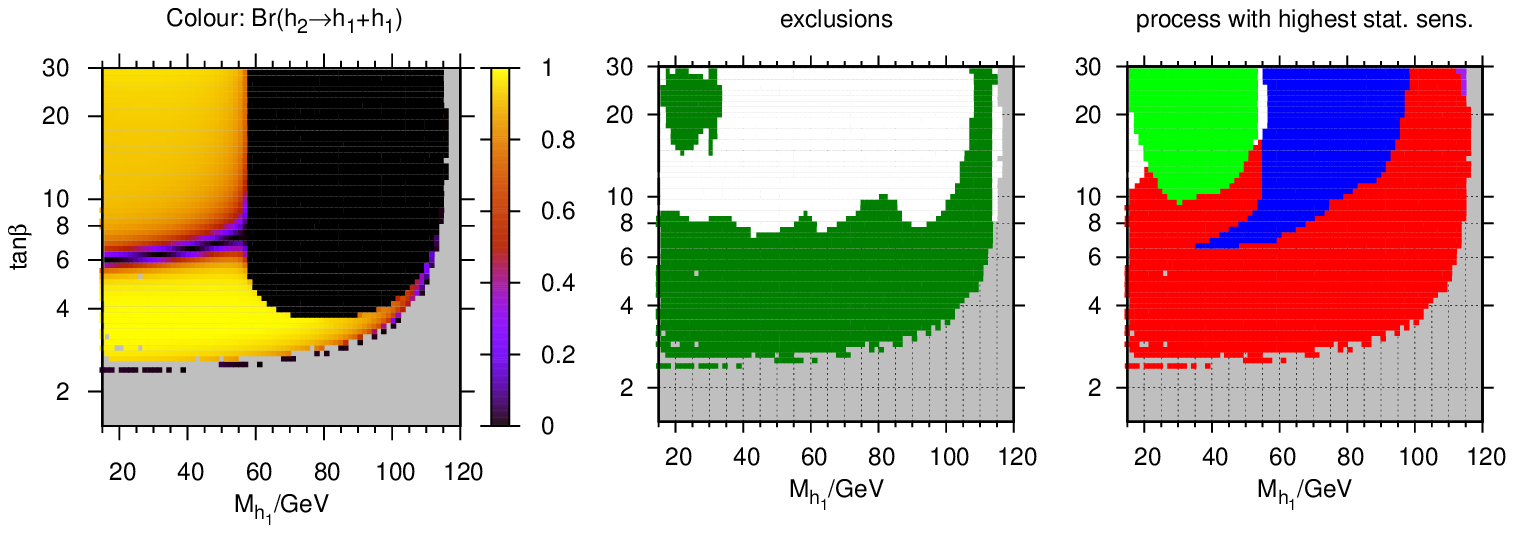}
}
\resizebox{\textwidth}{!}{%
\includegraphics{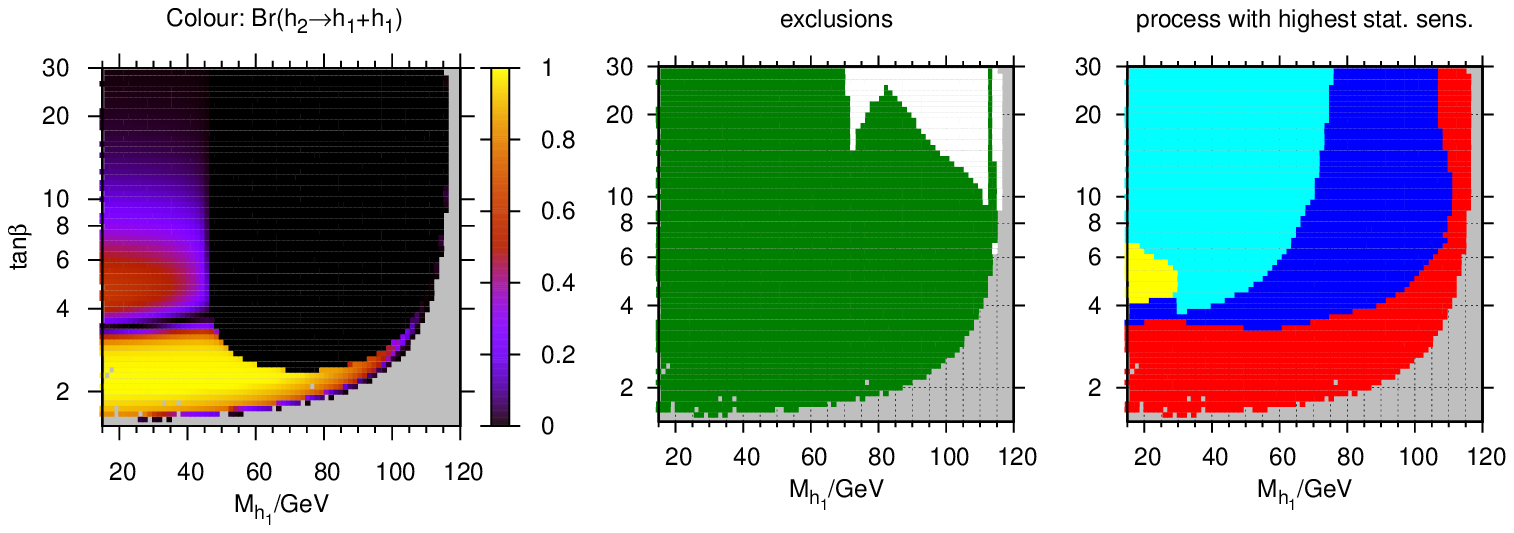}
}
\end{center}
\caption{The $h_2\to h_1 h_1$ branching ratio, LEP exclusions and channels with the highest statistical sensitivity plotted in the $M_{h_1}-\tan \beta$ plane. Upper graphs show $M_{\rm SUSY}=0.9 \times 500 \gev$, lower graphs show $M_{\rm SUSY}=1.1 \times 500 \gev$, other parameters taken from CPX scenario. See the captions of \reffi{CPXreg} and \reffi{CPXexcl} for the colour codes of the plots in the second and third columns.\label{MHpTBLEPMStQbMStQc}}
\end{figure}

Similarly, the effect of decreasing the soft-breaking term $M_{\rm
SUSY}$ by 10\%, as shown in \reffi{MHpTBLEPMStQbMStQc}, can be explained
by an enhancement of $|g^{{\rm eff}}_{h_2ZZ}|^2$ at the expense of
$|g^{{\rm eff}}_{h_2h_1Z}|^2$ and a suppression of BR($h_2\to b
\bar{b}$) as $M_{\rm SUSY}$ decreases. This results in a very large
unexcluded region, covering the majority of the parameter space above
$\tb\sim 9$. In contrast, increasing the value of $M_{\rm SUSY}$ to $M_{\rm SUSY}=1.1 \times 500 \gev$ greatly reduces the amount of unexcluded parameter space. In addition, note that the parameter space is populated at lower values of $\tb$ at this value of $M_{\rm SUSY}$.


\subsection{Comparison with {\cpsh}}
\label{sec:comp_cpsh}

\begin{figure}[h]
\includegraphics[width=\linewidth,angle=0]{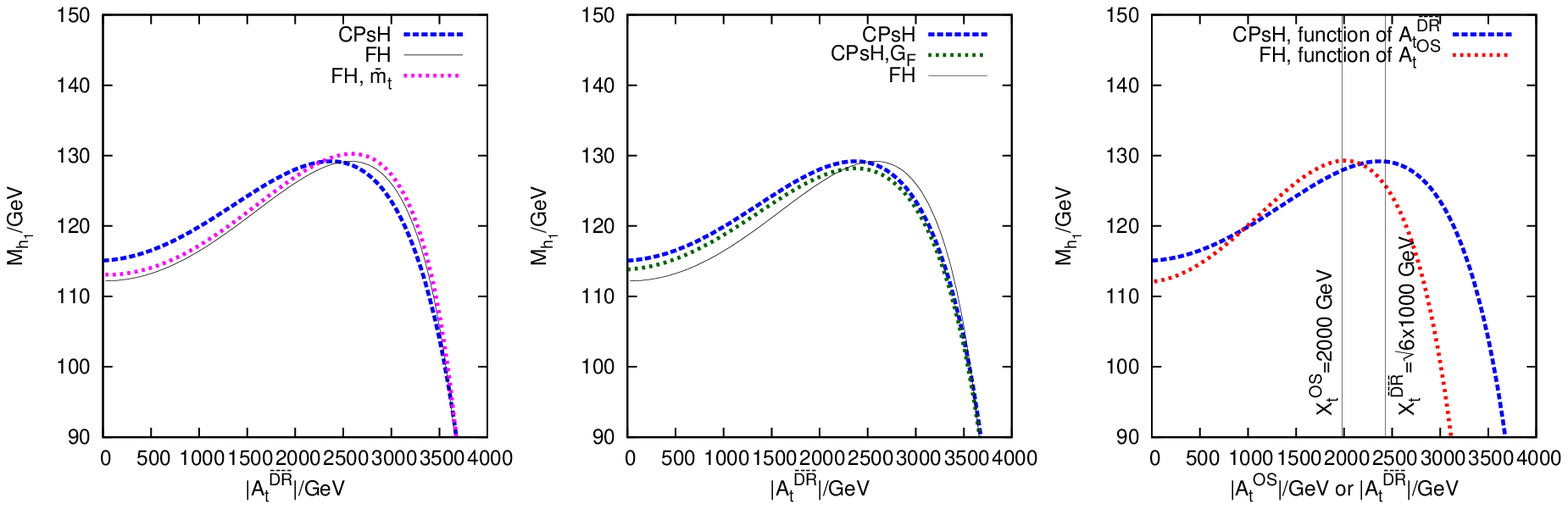}
\begin{center}(a)\hspace{5cm}(b)\hspace{5cm}(c)\end{center}
\caption{Lightest neutral Higgs mass as a function of $A^{\drbarm}_t$ in the  {\mhmax} scenario, with $\tb=10$, $\MHp=1000\gev$. All {\fh} results shown here use the full Higgs self-energies, and the $\drbarm$ to on-shell parameter conversion involves terms at both $\mathcal{O}(\alpha_s)$ and $\mathcal{O}(\alpha_t)$, and the calculation is parametrised in terms of $m^{OS}_t$ unless otherwise stated. \newline
(a) blue dashed: {\cpsh}, black solid: {\fh}, magenta dotted: {\fh} with code edited such that calculation is parametrised in terms of $\overline{m}_t$.\newline
(b) blue dashed: {\cpsh}, green dotted: {\cpsh} but with the result
parametrised in terms of the Fermi constant instead of $\alpha(M_Z)$,
black solid: {\fh}. \newline
(c) blue dashed: {\cpsh} as a function of $A_t^{\drbarm}$ as before, red
dotted: {\fh} as a function of $A_t^{OS}$.
\label{fig:compmhmax}}
\end{figure}

\begin{figure}[h]
\begin{center}
\includegraphics[width=0.7\linewidth,angle=0]{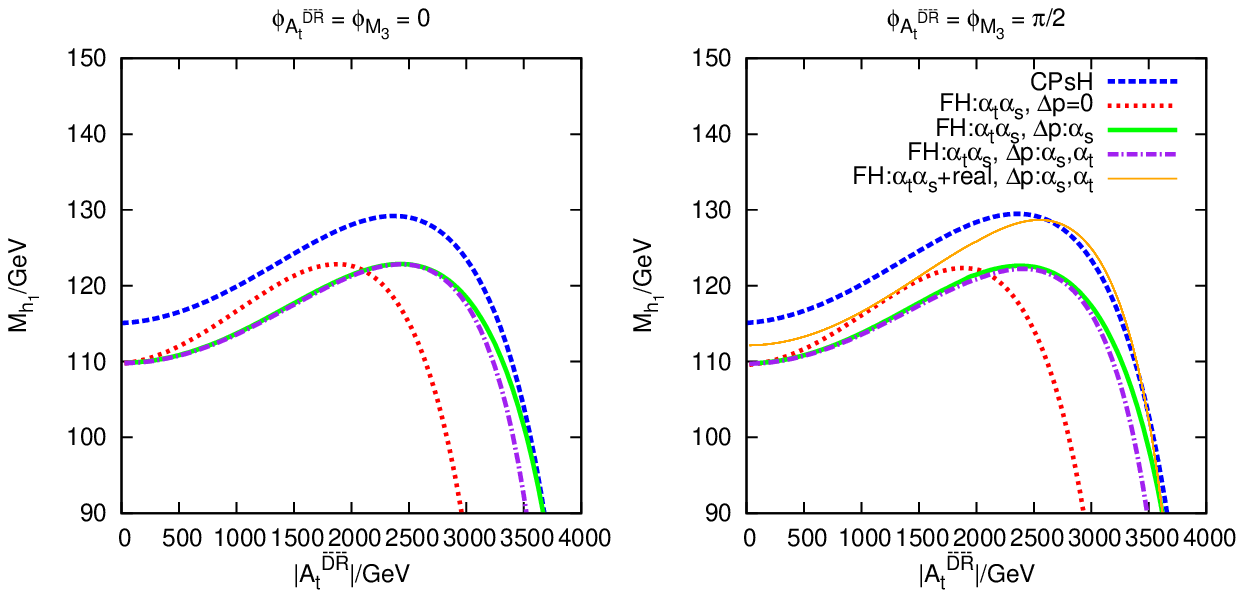}
\end{center}
\caption{Lightest neutral Higgs mass as a function of $A_t^{\drbarm}$ in the {\mhmax} scenario (left) and a CP-violating {\mhmax}-like scenario (right), where $\phi_{A_t^{\drbarm}}=\phi_{M_3}=\pi/2$. In both plots, $\tb=10$ and $\MHp=1000\gev$. All {\fh} results use Higgs self-energies up to $\mathcal{O}(\alpha_t\alpha_s)$ only (with the exception of the orange (thin solid) line) and the calculation is parametrised in terms of $m^{OS}_t$.
Blue dashed: {\cpsh} as in \reffi{fig:compmhmax}, red dotted: {\fh}
result without parameter shifts for the input parameters i.e.\  $\Delta
p=0$, green solid (thick): {\fh} with $\Delta p$ at
$\mathcal{O}(\alpha_s)$, purple dot-dashed: {\fh} result with $\Delta p$
involving contributions at $\mathcal{O}(\alpha_s)$ and
$\mathcal{O}(\alpha_t)$, orange solid (thin): as for purple, except
subleading 2-loop corrections are included in the Higgs self-energies,
calculated at all possible combinations of $\phi_{X_t}=0,\pi$,
$\phi_{X_b}=0,\pi$ and $\phi_{M_3}=0,\pi$. Therefore, there are eight
orange lines in total, but they lie too close to each other to be distinguishable by eye.
\label{fig:compmhmax2}}
\end{figure}

As we discussed in \refse{sec:convren}, when comparing our calculation with the results from {\cpsh}, we need to make sure that we account for the difference between the definitions of the input parameters. 

When comparing LEP exclusion plots, we shall investigate both the $\CPXdrbar$ and $\CPX$ scenarios, as defined in \refse{section:cpx} (recall that, in the $\CPXdrbar$ scenario, the input parameters $A_t$, $M_L^2$ and $M_{\tilde{t}R}^2$ are defined in the $\drbarm$ scheme, while in the $\CPX$ scenario, they are defined in the on-shell scheme). The $\CPXdrbar$ scenario is used most frequently in the literature as an example of an interesting CP-violating MSSM scenario and closely resembles the CP-violating scenario used by the LEP Higgs Working Group \cite{Schael:2006cr}. The $\CPX$ scenario is a more natural scenario in which to perform a Feynman-diagrammatic calculation and therefore has been used throughout the majority of this paper, up to this section.


However, we shall begin the discussion by an investigation of two
scenarios with less extreme parameter values than the $\CPXdrbar$ and
$\CPX$ scenarios, i.e.\ where the higher-order effects are expected to
be smaller. We will use the (CP-conserving) {\mhmax} scenario, where we
interpret the stop sector parameters $A_t$, $M_L$ and $M_{\tilde{t}R}$
as $\drbarm$ parameters ($\tb=10$,
$M_{H^{\pm}}=M_L^{\drbarm}(M_S)=M^{\drbarm}_{\tilde{t}R}(M_S)=1000
\gev$, $\mu^{\drbarm}(M_S)=200\gev$, $M_3=800\gev$). We also consider an
`{\mhmax}-like' scenario with CP violation (as for {\mhmax}, but
$\phi^{\drbarm}_{A_t}=\phi^{\drbarm}_{M_3}=\pi/2$). Since we have
defined $A_t$, $M_L^2$, $M_{\tilde{t}R}^2$ and $\mu$ in the $\drbarm$ renormalisation scheme at $M_S$, we must make use of the parameter conversions described in \refse{sec:convren}, in order to obtain $A_t^{\rm on-shell}$, $M_L^{\rm on-shell}$, $M_{\tilde{t}R}^{\rm on-shell}$ and $\mu^{\drbarm}(m_t)$, which are the input parameters that need to be given to {\fh}.

In \reffi{fig:compmhmax} (a), we display the result from {\cpsh} (blue dashed) and the result from {\fh} using $\mathcal{O}(\alpha_s)$ and $\mathcal{O}(\alpha_t)$ contributions to the parameter shifts (black solid) (as discussed in \refse{sec:convren}) in the {\mhmax} scenario. We can see that the maximum value of $M_{h_1}$ agrees to within $0.1\gev$ between the two codes, although the maximum occurs at a value of $A_t^{\drbarm}$ which is $\sim 200\gev$ higher for the {\fh} result than the {\cpsh} result. Note also that there is a somewhat larger discrepancy between the two codes at small values of $A_t^{\drbarm}$, where {\cpsh} predicts a mass $\sim 3\gev$ higher than {\fh}. Since {\fh} contains the full $\mathcal{O}(\alpha_t\alpha_s)$, $\mathcal{O}(\alpha_t^2)$ and the complete 1-loop corrections to the neutral Higgs self-energies in the MSSM with real parameters, while {\cpsh} only contains the leading logarithmic contributions, we should not expect our $\mathcal{O}(\alpha_s)$ and $\mathcal{O}(\alpha_t)$ parameter shifts to convert perfectly between the two calculations.

Also shown in \reffi{fig:compmhmax} (a) is the result where the {\fh} calculation is parametrised in terms of the $\msbarm$ top mass (magenta dotted), as discussed in \refse{sec:choiceofmtinFH}, instead of the on-shell top mass (black solid). This changes the Higgs mass by less than $2\gev$ (for $A_t^{\drbarm}<3600 \gev$), which is appropriately small since this is formally a higher order effect. 

Another contribution to the remaining discrepancy between the results
for the Higgs mass predictions in the two codes is the difference in the
way that the electric charge is parameterised at lowest order in the two
codes. We investigate the magnitude of this effect in the following way. The input to {\cpsh} is supposed to be $\alpha(M_Z)$ (blue dashed), but we can investigate the effect of changing it to $\alpha \to G_F 2\sqrt{2}M_W^2s_W^2/(4 \pi)$ (green dotted), which should only affect the calculation at the level of the unknown higher order contributions. In \reffi{fig:compmhmax} (b), we can see that this changes the lightest Higgs mass by less than $1.3\gev$ for $A_t^{\drbarm}<3600 \gev$. 




A meaningful comparison between the two Higgs mass calculations can also
be carried out in the {\mhmax} scenario by neglecting the parameter
shifts in $M_L$ and $M_{\tilde{t}R}$ and using the values of $X_t^{OS}$
and $X_t^{\drbarm}$, which maximise the lightest Higgs masses from {\fh}
and {\cpsh}, respectively (corresponding to independently defining a
{\mhmax} scenario in the on-shell scheme and a {\mhmax} scenario in the
$\drbarm$ scheme). Using the leading 1-loop and 2-loop corrections, it
is possible to predict that this will occur at approximately
$X_t^{OS}=2000\gev$ and $X_t^{\drbarm}=\sqrt{6}\times 1000 \gev$, see
\citere{hep-ph/0001002}. In \reffi{fig:compmhmax} (c), we plot the {\cpsh} 
result for $M_{h_1}$ against $A_t^{\drbarm}$ (blue dashed, as before), together with the {\fh} result for $M_{h_1}$ against $A_t^{OS}$ (red dotted). The maxima in $M_{h_1}$ occur at the points $(X_t^{OS},M_{h_1})=(1990 \gev,129.3\gev)$ and $(X_t^{\drbarm},M_{h_1})=(2350\gev,129.2\gev)$, i.e.\  this definition of the {\mhmax} scenario does indeed give remarkable agreement between the two codes for the value of $M_{h_1}$. The two vertical lines show $X_t^{OS}=2000\gev$ and $X_t^{\drbarm}=\sqrt{6}\times 1000 \gev$ (we can see that using these approximate values of $X_t^{OS}$ and $X_t^{\drbarm}$ predicts lightest Higgs masses which are within $0.1\gev$ of the true values at the maxima).

It is interesting to compare the corresponding shift in $A_t$ i.e.\  $\widetilde{\Delta} A_t=\widetilde{\Delta} X_t= X_t^{\drbarm}-X_t^{OS}=2350-1990=360\gev$, with the shift $\Delta A_t$ which we obtain from performing the parameter conversion. At $X_t^{\drbarm}=\sqrt{6}\times 1000 \gev$, $\Delta A_t=510\gev$ when just $\mathcal{O}(\alpha_s)$ corrections are included in the parameter conversion and $\Delta A_t=600\gev$ when both $\mathcal{O}(\alpha_s)$ and $\mathcal{O}(\alpha_t)$ corrections are included. 
Comparison with the analysis in \citere{hep-ph/0001002}, which was performed for the leading 1-loop and 2-loop contributions, shows a similar behaviour if only the $\mathcal{O}(\alpha_s)$ correction to the parameter conversion is taken into account. Incorporating also the $\mathcal{O}(\alpha_t)$ correction turns out to give rise to a slightly larger discrepancy between the shifts $\widetilde{\Delta} A_t$ and $\Delta A_t$. The numerical effect on the lightest Higgs mass, however, is small over this range of $A_t$.

As discussed in \refse{section:Higgsmasses}, the only 2-loop pieces
contained within {\fh} which have been calculated with explicit phase
dependence are of $\mathcal{O}(\alpha_t\alpha_s)$. Before we start to
discuss the CP-violating MSSM, it is instructive to compare the results
from {\cpsh} with the results given by {\fh} when the only 2-loop pieces
which are included are $\mathcal{O}(\alpha_t\alpha_s)$. This is shown in
\reffi{fig:compmhmax2} (left). Also shown are the {\fh} results when the
parameter shifts are neglected (red dotted), in order to give an idea of
the size of the parameter shifts. We have a choice for the parameter
shifts: we can either use the shifts at both $\mathcal{O}(\alpha_s)$ and
$\mathcal{O}(\alpha_t)$ (purple dot-dashed) as before, or restrict to
$\mathcal{O}(\alpha_s)$ pieces (green solid (thick)). Theoretically,
restricting to $\mathcal{O}(\alpha_s)$ pieces (green solid (thick)) is
preferable, since neither Higgs mass calculation now involves the full
set of $\mathcal{O}(\alpha_t^2)$ corrections. It can be seen that this
choice makes very little difference in this scenario. We can see that,
when we restrict to 2-loop pieces at $\mathcal{O}(\alpha_t\alpha_s)$ in
{\fh}, the resulting Higgs mass at the peak in the plot is approximately
$7\gev$ lighter than when all available corrections in {\fh} are used
(and therefore also significantly lighter than the mass given by {\cpsh}
at the peak). In \reffi{fig:compmhmax2} (right), we introduce CP
violation: this figure has the same parameters as \reffi{fig:compmhmax2}
(left) ($\tb=10$,
$M_{H^{\pm}}=M_L^{\drbarm}(M_S)=M^{\drbarm}_{\tilde{t}R}(M_S)=1000
\gev$, $\mu^{\drbarm}(M_S)=200\gev$, $|M_3|=800\gev$), except that $A_t$, $A_b$ and $M_3$ are now complex: $\phi_{A_t^{\drbarm}}=\phi_{A_b}=\phi_{M_3}=\pi/2$.
Note that both the $\mathcal{O}(\alpha_s)$ and $\mathcal{O}(\alpha_t)$
parameter shifts include the full phase dependence, as described in
\refse{sec:convren}. We can see that, in this scenario, the introduction
of complex phases does not significantly affect the lightest Higgs mass.
Therefore, we can deduce that the dominant source of the discrepancy
between the masses in this CP-violating example are the neglected 2-loop
corrections. It is instructive to consider whether the calculation of
these 2-loop corrections to the neutral Higgs self-energies for the MSSM
with real parameters can be used to give an estimate of these
corrections in the complex MSSM. It is possible to evaluate these
additional corrections at a phase of $0$ and $\pi$ for each complex
parameter and then perform an interpolation between these results (this
is an optional feature in {\fh}). In \reffi{fig:compmhmax2} (right), we
show predictions for the lightest Higgs mass (orange solid (thin)) in
the $\CPX$ scenario when the neutral Higgs self-energies are calculated
up to $\mathcal{O}(\alpha_t \alpha_s)$ with full phase dependence (as
previously) and the subleading 2-loop corrections are calculated at
combinations of $\phi_{X_t}=0,\pi$, $\phi_{X_b}=0,\pi$ and
$\phi_{M_3}=0,\pi$. These eight combinations give very similar results,
and their curves can not be distinguished in the plot, and, obviously,
an interpolation between these results would also lie on this same line.
It is, however, difficult to evaluate the accuracy of this Higgs mass
prediction. On the one hand, it provides better agreement with the
{\cpsh} result, and the fact that the eight CP-conserving limits give
very similar results could be interpreted as an indication that a fully 
phase-dependent calculation of these corrections may have only a mild 
dependence on the phase. However, a confirmation of this interpretation
would require the actual result of a fully phase-dependent calculation.
In an `extreme' scenario like the $\CPX$ scenario, which is close to
unstable regions of parameter space, the evaluation of the combinations
of real parameters needed as input for the interpolation can be
problematic. In particular, higher-order contributions in the
CP-conserving limit $\phi_{X_t}=0$ turn out to be unphysically large in the
$\CPX$ scenario and cause the mass calculation to fail (to obtain a real
and positive mass value, the limit $\phi_{A_t}=0$ must be used instead).
Since we will now turn our attention to such `extreme' examples, 
exhibiting very large higher order corrections, we shall restrict in
these cases the Higgs self-energies calculation in the complex MSSM to
$\mathcal {O}(\alpha_t \alpha_s)$, as we have done throughout the rest of the
paper.

\begin{sidewaysfigure}[p]
\includegraphics[width=\linewidth,angle=0]{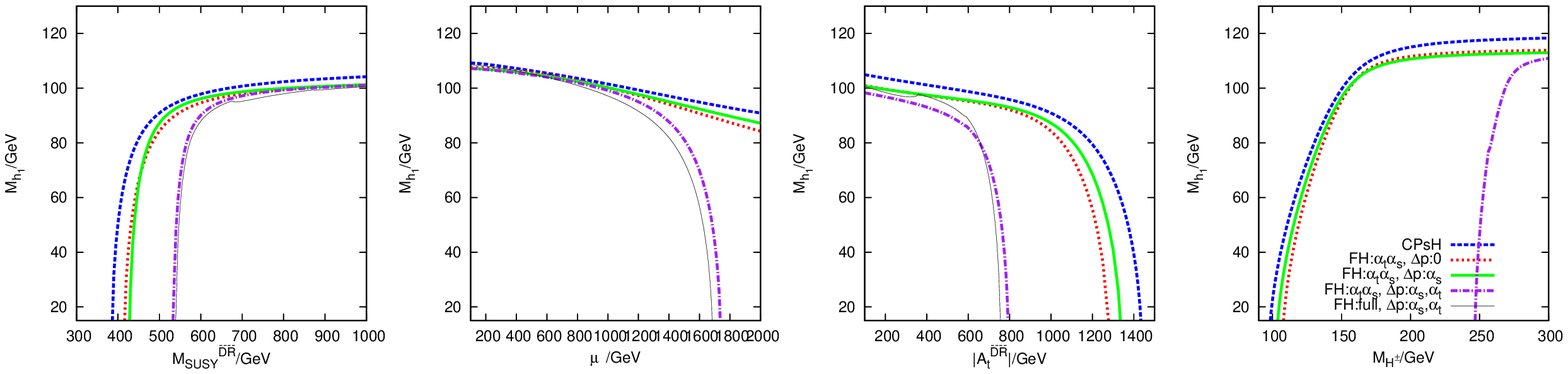}
\caption{Lightest neutral Higgs mass in a CP-conserving version of the $\CPXdrbar$ scenario at $\tb=10,M_{H^{\pm}}=140\gev$, as function of (left to right) $M_{SUSY}^{\drbarm}(M_S)$, $\mu^{\drbarm}(M_S)$, $|A_t^{\drbarm}(M_S)|$ and $M_{H^{\pm}}$. 
Blue dashed: {\cpsh} result, red dotted: {\fh} with Higgs self-energies up to $\mathcal{O}(\alpha_t\alpha_s)$, with $\Delta p=0$,
green solid (thick): {\fh} with Higgs self-energies up to $\mathcal{O}(\alpha_t\alpha_s)$ corrections, with $\Delta p$ at $\mathcal{O}(\alpha_s)$, 
purple dot-dashed: {\fh} with Higgs self-energies up to $\mathcal{O}(\alpha_t\alpha_s)$ corrections, with $\Delta p$ at $\mathcal{O}(\alpha_s)$ and $\mathcal{O}(\alpha_t)$, 
black solid (thin): {\fh} with full Higgs self-energies, with $\Delta p$ at $\mathcal{O}(\alpha_s)$ and $\mathcal{O}(\alpha_t)$.
\label{fig:compcpxreal} 
}
\includegraphics[width=\linewidth,angle=0]{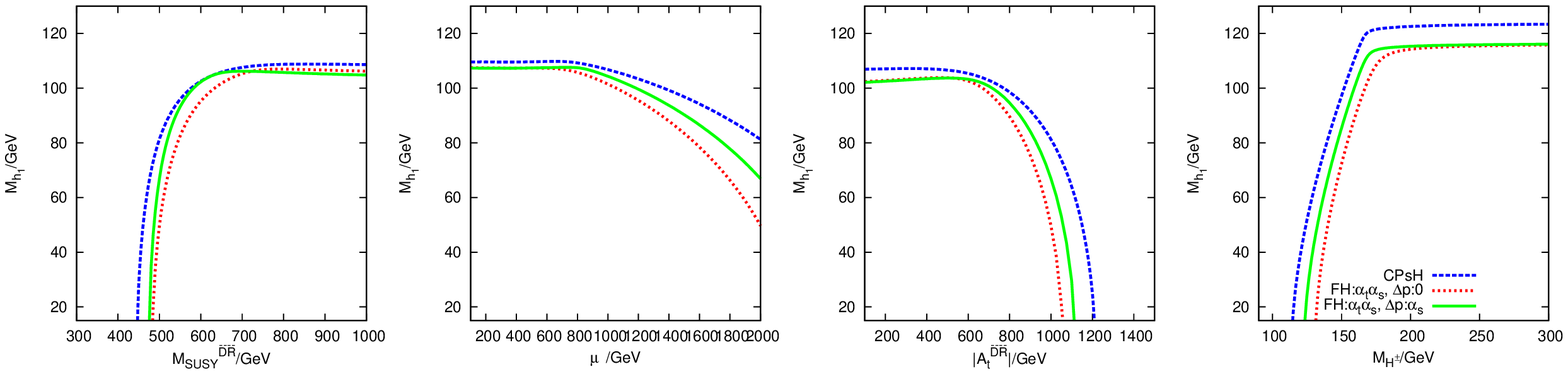}
\caption{Lightest neutral Higgs mass in the $\CPXdrbar$ scenario at $\tb=10,M_{H^{\pm}}=140\gev$, as function of (left to right) $M_{SUSY}^{\drbarm}(M_S)$, $\mu$, $|A_t^{\drbarm}(M_S)|$ and $M_{H^{\pm}}$. (See \reffi{fig:compcpxreal} for the definitions of the abbreviations labelling the different curves.)
\label{fig:compcpxcompl} 
}
\end{sidewaysfigure}

 \afterpage{\clearpage}

We shall first consider a CP-conserving version of the $\CPXdrbar$
scenario i.e. the $\CPXdrbar$ scenario with all phases set to zero (and
$\mu^{\drbarm}(M_S)=2000\gev$), as this will exhibit the main
characteristics of the $\CPXdrbar$ scenario (and, since we have more
available corrections in the MSSM with real parameters than the MSSM
with complex parameters, it will be easier for us to investigate). In
\reffi{fig:compcpxreal}, we compare the {\cpsh} result (blue dashed) to
the full {\fh} result with $\Delta p$ at $\mathcal{O}(\alpha_s)$ and
$\mathcal{O}(\alpha_t)$ (black solid (thin)), the {\fh} result with
Higgs self-energies only up to $\mathcal{O}(\alpha_t\alpha_s)$ and
$\Delta p$ at $\mathcal{O}(\alpha_s)$ and $\mathcal{O}(\alpha_t)$
(purple dot-dashed), the {\fh} result with Higgs self-energies up to
$\mathcal{O}(\alpha_t\alpha_s)$ and $\Delta p$ only at
$\mathcal{O}(\alpha_s)$ (green solid (thick)). Note that, when the
$\mathcal{O}(\alpha_t)$ contributions 
to $\Delta p$ are included, this incorporates the
conversion of $\mu^{\drbarm}(M_S)$ to $\mu^{\drbarm}(m_t)$, using
\refeq{eq:mueshift}. Also illustrated is the {\fh} result when the
parameter shifts are neglected (red dotted), in order to demonstrate the
numerical effect of the parameter shifts. In each plot in
\reffi{fig:compcpxreal}, we can see that the effect of the parameter 
shifts at $\mathcal{O}(\alpha_t)$ on the mass calculation becomes very 
large once the nominal values of this CP-conserving $\CPXdrbar$-like 
scenario are approached ($M_{SUSY}^{\drbarm}=500\gev$, 
$\mu^{\drbarm}=2000\gev$,
$A_t^{\drbarm}=1000\gev$, $M_{H^{\pm}}=140\gev$). Indeed, for the
nominal values of this scenario, the lightest on-shell stop mass is
driven to imaginary values when both the $\mathcal{O}(\alpha_s)$ and
$\mathcal{O}(\alpha_t)$ parameter shifts are used (purple dot-dashed and
black solid (thin)). These unphysically large effects are an indication
that this scenario is close to an unstable region, which causes
instabilities in the parameter conversion (this could imply that it
would be appropriate to explicitly decouple the heavier particles in the
conversion to the $\drbarm$ scheme). In order to avoid these
instabilities, we will, in the following, restrict the parameter
conversion to $\mathcal{O}(\alpha_s)$. Application of this parameter
conversion, in conjunction with 2-loop mass corrections at
$\mathcal{O}(\alpha_t\alpha_s)$ (green solid (thick)), yields a
relatively good agreement between the {\fh} and {\cpsh} results.

We observe the same quantitative behaviour when we switch on the CP-violating phases in this scenario (\reffi{fig:compcpxcompl}), and therefore we can use the knowledge gained from the CP-conserving version to help interpret these characteristics. Once again, we find that the parameter conversions at $\mathcal{O}(\alpha_s)$ and $\mathcal{O}(\alpha_t)$ result in unphysically large corrections to the lightest Higgs mass and therefore restrict the parameter conversion to $\mathcal{O}(\alpha_s)$ (green solid), which compares relatively well to the {\cpsh} result (blue dashed).

\begin{sidewaysfigure}[p]
\resizebox{\textwidth}{!}{%
  \includegraphics{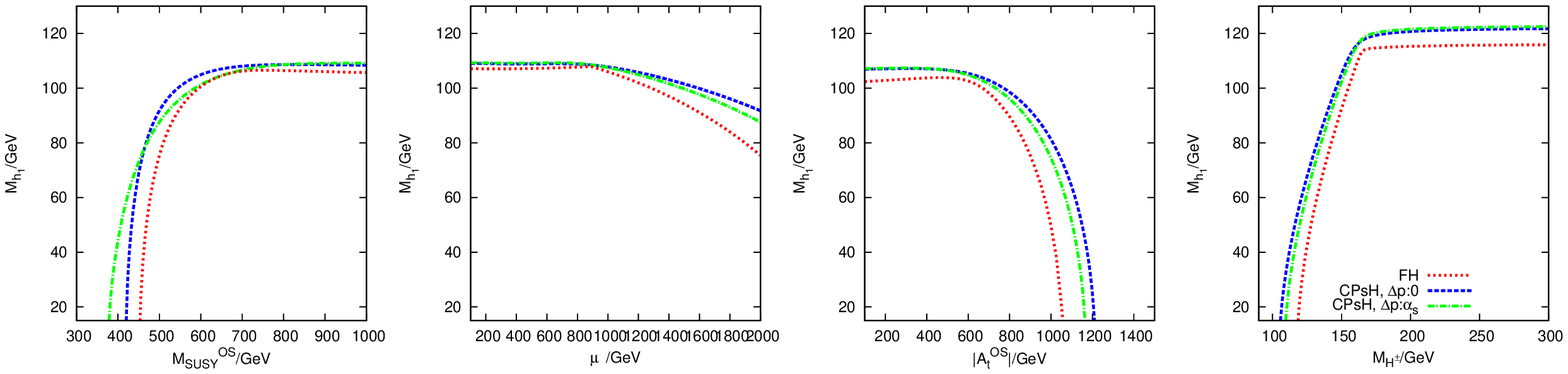}
}
\caption{Lightest neutral Higgs mass in the $\CPX$ scenario at $\tb=10,M_{H^{\pm}}=140\gev$, as function of (left to right) $M_{SUSY}^{OS}$, $\mu$, $|A_t^{OS}|$ and $M_{H^{\pm}}$. 
Red dotted: {\fh} with Higgs self-energies up to $\mathcal{O}(\alpha_t\alpha_s)$ corrections,
blue dashed: {\cpsh} with $\Delta p=0$,
green dot-dashed: {\cpsh} with $\Delta p$ at $\mathcal{O}(\alpha_s)$,
}
\label{fig:cpxagain}       
\end{sidewaysfigure}

 \afterpage{\clearpage}

We will now carry out a complementary analysis in the $\CPX$ scenario,
which is defined in terms of on-shell stop sector parameters. Therefore,
the parameter shifts are now used to convert from on-shell to $\drbarm$
parameters, which can then be given as input to {\cpsh}. Results for the
lightest Higgs mass are shown in \reffi{fig:cpxagain} for {\cpsh} with
no parameter conversion (blue dashed) and {\cpsh} with parameter
conversion at $\mathcal{O}(\alpha_s)$ (green dot-dashed). Also shown is
the {\fh} result where the masses are calculated at
$\mathcal{O}(\alpha_t\alpha_s)$ (red dotted). Once again, we can see
that this scenario lies near the border of stable parameter space, which
is necessary if we would like to investigate very low lightest Higgs
masses. We find that, when the parameter conversion is performed in the
on-shell to $\drbarm$ direction, the effect of the parameter shifts
tends to push the results into more stable regions of parameter space
(rather than into more unstable regions, as it happens in the case of 
the $\CPXdrbar$ scenario). However, we have confirmed that this parameter conversion still gives rise to large stop mass corrections at $\mathcal{O}(\alpha_t)$.

Before starting a comparison of the LEP exclusions in $\CPXdrbar$ or $\CPX$ parameter space, it is useful to describe the method that was used by the LEP Higgs Working Group to tackle this issue of consistently comparing results calculated in using different renormalisation schemes. The LHWG analysis was carried out in the $\CPXdrbar$ scenario and both {\fh} version 2.0 \cite{Heinemeyer:1998yj} and {\cph} \cite{Carena:2000ks} (a predecessor of the program {\cpsuperh}) were considered. The relation
\BEA
\tilde{X}_t^{\rm CPH}&=&\tilde{X}_t+\frac{\alpha_s}{3\pi}M_{\rm SUSY}\left[8+\frac{4\tilde{X}_t}{M_{\rm SUSY}}-\frac{\pi\tilde{X}^{\rm FH}_t}{M_{\rm SUSY}}\log\left(\frac{m_t^{{\rm OS,}2}}{M^2_{\rm SUSY}}\right)\right]
\label{eq:LEPapprox}
\EEA
was used \cite{emailfromPhilip} to convert between different definitions for $|A_t|$, with $\alpha_s=0.108$ and $\tilde{X}_t=|A_t|-\mu/\tb$. \footnote{This is analogous to the expression for on-shell to $\msbarm$ conversion in the MSSM with real parameters given by \cite{hep-ph/0001002}: $X_t^{\msbarm}=X_t^{\rm OS} + \frac{\alpha_s}{3\pi}M_S\left[8+\frac{4X_t}{M_S}-\frac{X_t^2}{M_S^2}-\frac{3X_t}{M_S}\log\left(\frac{m_t^2}{M_S^2}\right)\right]$, where $M_S=\sqrt{M_{\rm SUSY}^2+m_t^2}$, which was obtained for $m_{\tilde{g}}=M_{\rm SUSY}$, $\mu_{\rm ren}=M_S$ and the assumptions $m_t/M_S\ll 1$ and $m_t X_t/M_S^2\ll 1$.} 
The shifts in $\phi_{A_t}$ and $M_{\rm SUSY}(= M_L=M_{\tilde{t}_R})$
were neglected, and $A_b$ was set to be the same as $A_t$. 

Near the region $M_{h_1}\sim 45 \gev $, $\tb \sim 7$, this approximation
gives a larger shift in $|A_t|$ than would be given from the full
$\mathcal{O}(\alpha_s)$ shifts, and therefore increases the agreement
between the two Higgs mass codes. The LHWG then carried out two parallel
analyses, one using Higgs sector predictions from {\cph} and one using
Higgs sector predictions from {\fh} (apart from the triple Higgs
coupling, as discussed in \refse{sec:LHWGresults}). Since they found
significant differences between the two results, they considered a point in $\CPXdrbar$ parameter space to be excluded only if it could be excluded in both the analysis based on the {\cph} results and the analysis based on the {\fh} results.

\begin{figure}[tb]
\begin{center}
\resizebox{\textwidth}{!}{%
\includegraphics{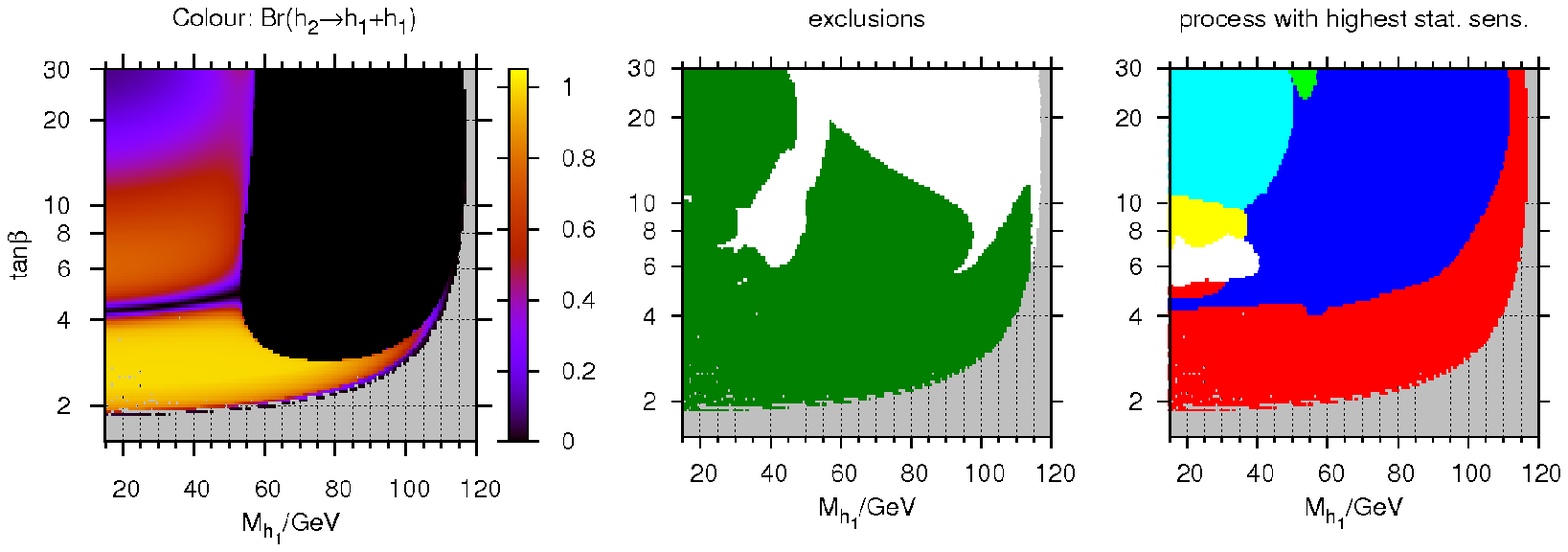}
}
 \resizebox{\textwidth}{!}{%
\includegraphics{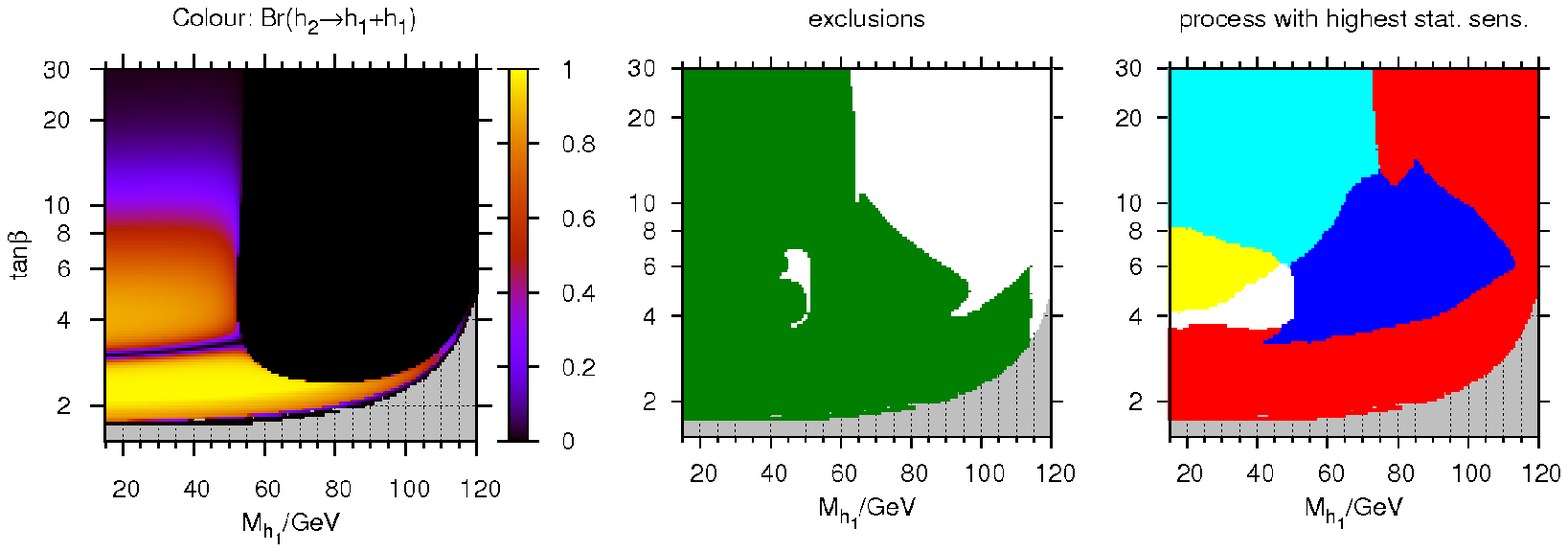}
 }

\end{center}
\caption{
The $h_2\to h_1 h_1$ branching ratio, LEP exclusions and channels with the highest statistical sensitivity for the $\CPXdrbar$ scenario, plotted in the $M_{h_1}-\tan \beta$ plane. 
Top row: Using the calculations presented in this paper, parameter
conversions at $\mathcal{O}(\alpha_s)$ as described by \refeqs{eq:convren1}--(\ref{eq:convren2}), and neutral Higgs self-energies from {\fh} up to $\mathcal{O}(\alpha_t \alpha_s)$.  
Bottom row: Using Higgs masses, couplings and branching ratios calculated by {\cpsuperh}. See the captions of \reffi{CPXreg} and \reffi{CPXexcl} for the colour codes of the plots in the second and third columns.
\label{MHpTBLEPconv}}
\end{figure}

 \afterpage{\clearpage}

In \reffi{MHpTBLEPconv} (top row), we use the topological cross section
limits from LEP in conjunction with updated results for the Higgs
masses, couplings and branching ratios, in the $\CPXdrbar$ scenario
using the parameter conversions at $\mathcal{O}(\alpha_s)$. In
particular, we use our complete 1-loop diagrammatic calculation for the
$h_i\to h_j h_k$ decay processes with full phase dependence as described
in \refse{sec:hihjhk} and we use renormalised neutral Higgs
self-energies obtained from {\fh} (which includes corrections at
$\mathcal{O}(\alpha_t \alpha_s)$ with full phase dependence). We compare
this to the results using the Higgs masses, couplings and branching
ratios from the program {\cpsh}, shown in \reffi{MHpTBLEPconv} (bottom
row). The {\cpsh} prediction for the $h_2\to h_1h_1$ branching ratio is qualitatively similar to our result, although the thin `knife-edge' region is at $\tb\sim 3$ rather than $\tb \sim 4.5$ and the peak at moderate $\tb$ is higher but narrower, in particular, around $M_{h_1}\sim 45\gev$. Also, different processes have the highest statistical sensitivity at different parts of the parameter space (\reffi{MHpTBLEPconv} right column). Most significantly, the area where the process $\HZtobbZ$ has the highest statistical sensitivity is larger when using the Higgs sector predictions outlined in this paper, rather than the predictions given by {\cpsh}. As a result, the unexcluded part of parameter space is much larger for $M_{h_1}<60\gev$ when using our calculation and the unexcluded regions join together. However, for $M_{h_1}>60\gev$, the unexcluded area is larger when the {\cpsh} results are used.

\begin{figure}[tb]
\resizebox{\textwidth}{!}{%
  \includegraphics{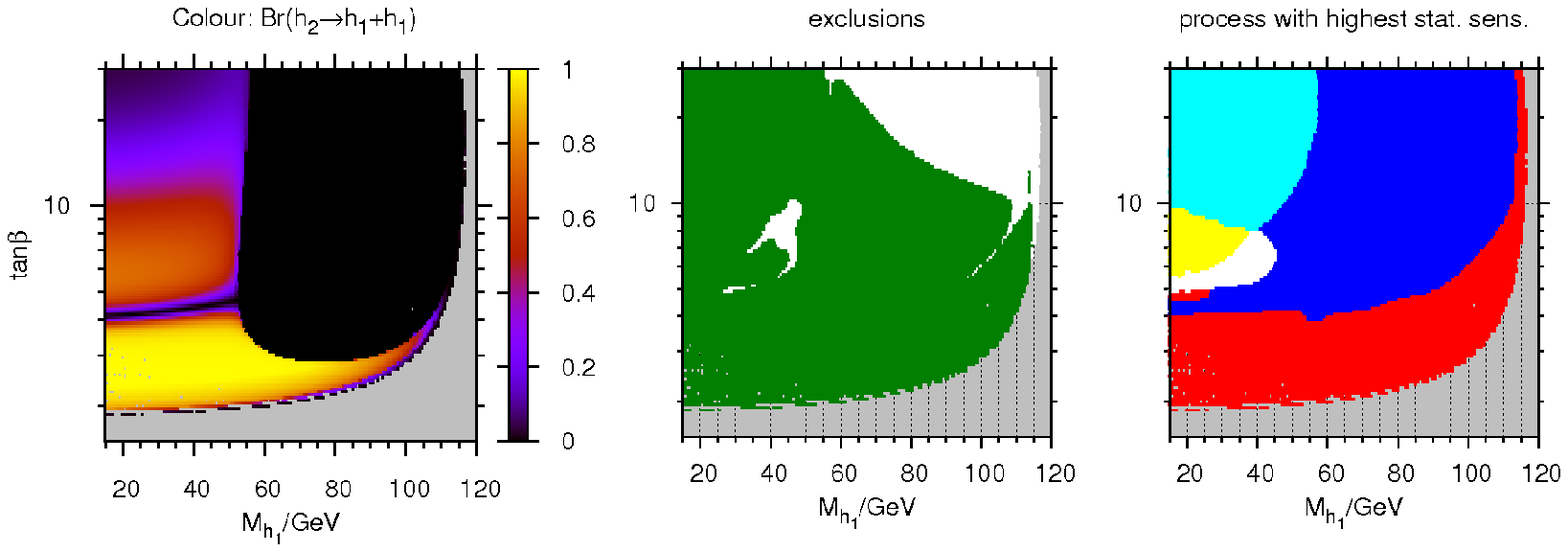}
}
\resizebox{\textwidth}{!}{%
  \includegraphics{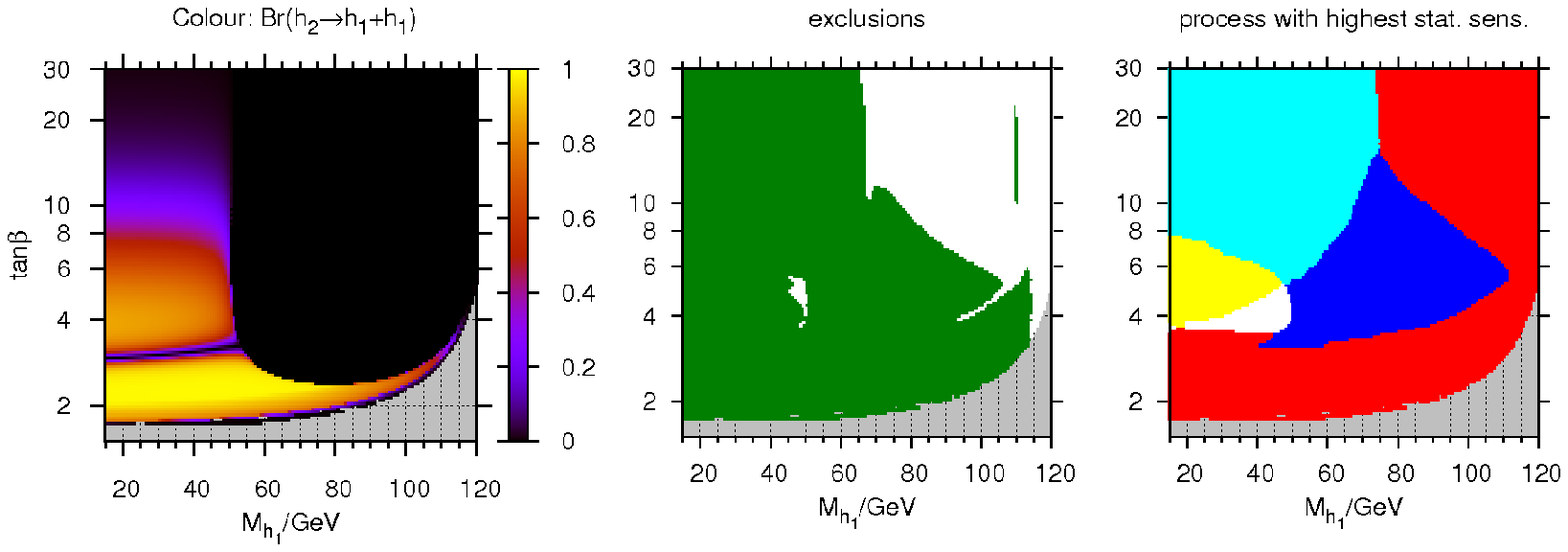}
}
\caption{
The $h_2\to h_1 h_1$ branching ratio, LEP exclusions and channels with the highest statistical sensitivity for the $\CPX$ scenario, plotted in the $M_{h_1}-\tan \beta$ plane. 
Top row: Using the calculations presented in this paper and neutral Higgs self-energies from {\fh} up to $\mathcal{O}(\alpha_t \alpha_s)$.  
Bottom row: Using Higgs masses, couplings and branching ratios
calculated by {\cpsuperh} and parameter conversions at
$\mathcal{O}(\alpha_s)$ as described by \refeqs{eq:convren1}--(\ref{eq:convren2}). See the captions of \reffi{CPXreg} and \reffi{CPXexcl} for the colour codes of the plots in the second and third columns.
}
\label{fig:cpxagain2}       
\end{figure}

 \afterpage{\clearpage}

Finally, in \reffi{fig:cpxagain2}, we compare the results from {\cpsh}
and our calculation (as shown previously in \reffi{Br2} (left),
\reffi{CPXreg} (right) and \reffi{CPXexcl} (right)) for the Higgs
cascade decay width and LEP exclusions in the $\CPX$ scenario (defined
in terms of on-shell stop sector parameters). Again, the {\cpsh}
prediction for the $h_2\to h_1h_1$ branching ratio is qualitatively similar to our result, although, once again, the thin `knife-edge' region is at lower $\tb$ and the maximum at moderate $\tb$ is higher. As we saw in \reffi{MHpTBLEPconv}, the {\cpsh} result has a large region where $\hZtobbZ$ has the highest statistical sensitivity at medium-to-high $\tb$ and $M_{h_1}$, which is not seen in our result. Once again, the area where processes involving $h_1h_2$ pair-production (${\color{cyan}\blacksquare}$ and ${\color{yellow}\blacksquare}$) have the highest statistical sensitivity is larger in the {\cpsh} result than in ours. As a result, the region where $\HZtobbZ$ has the highest statistical sensitivity does not significantly overlap with the region with large $h_2\to h_1h_1$ branching ratio, and therefore the region $M_{h_1}<65\gev$ is almost entirely excluded. However, once again, the {\cpsh} result has a larger unexcluded region at medium-to-high $\tb$ and $M_{h_1}$.

\begin{figure}[tb]
\resizebox{\textwidth}{!}{%
  \includegraphics{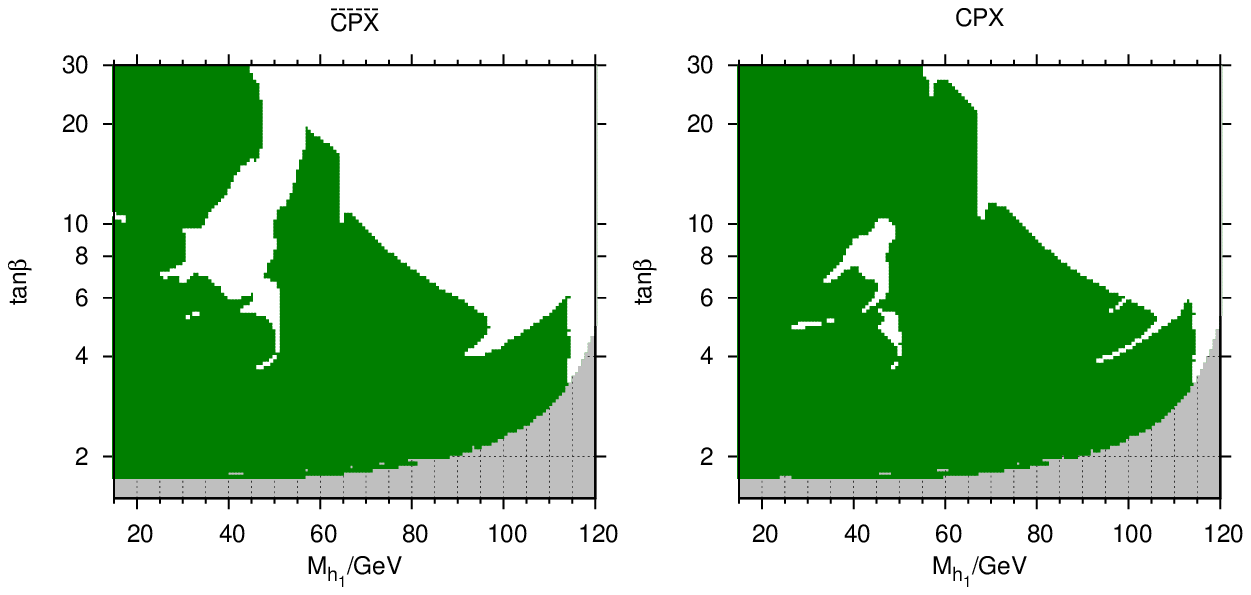}
}
\caption{Combining results for the LEP exclusions in the $\CPXdrbar$ (left, using \reffi{MHpTBLEPconv}) and $\CPX$ (right, using \reffi{fig:cpxagain2}) parameter space. Green (darker grey): point in parameter space can be excluded using the topological LEP limits and the Higgs sector predictions from both our calculation (which uses neutral Higgs self-energies from {\fh}) and the calculation implemented into {\cpsh}, white: is excluded by only one or by neither of the calculations.
}
\label{fig:cpxcomb}       
\end{figure}

In summary, we have compared the results presented in this paper with
those given by the program {\cpsh}. Since these results are very
sensitive to the Higgs masses, we started with a comparison between the
masses given by {\fh} and {\cpsh} (including appropriate parameter
transformations, which we have extended to the complex MSSM) in the
{\mhmax} scenario, in a {\mhmax}-like scenario with added CP violation,
the $\CPXdrbar$ scenario (in which the stop sector parameters have
$\drbarm$ definitions), the $\CPXdrbar$ scenario with vanishing phases,
and the $\CPX$ scenario (in which the stop sector parameters have
on-shell definitions). We find good agreement between the mass
calculations in the  {\mhmax} and {\mhmax}-like scenarios once the
appropriate parameter shifts are taken into account. The $\CPXdrbar$
scenario has been used by the LEP Higgs Working Group in their analysis
and is the most frequently used CP-violating scenario in the literature.
We find that the $\mathcal{O}(\alpha_t)$ corrections in the parameter
transformations for the stop masses are very large in this scenario and
drive the $\tilde{t}_1$ mass to unphysical values. We interpret this as
an indication that this scenario is close to an unstable region of
parameter space (i.e.\  a region with relatively large higher-order
corrections), where we must be careful to avoid unphysically large
corrections in the parameter shifts (which may require to explicitly 
decouple heavy particles in the conversion). When restricting to the
$\mathcal{O}(\alpha_s)$ parameter conversions, we find rough agreement
between the mass predictions, and we observed similar behaviour when we considered the $\CPX$ scenario with appropriate parameter conversions at $\mathcal{O}(\alpha_s)$. The variation in the Higgs sector predictions given by our calculation (using neutral Higgs self-energies from {\fh}) and {\cpsh} are reflected in significant differences in the size and shape of the regions of parameter space that could not be excluded by the LEP results. When the topological cross section limits from LEP are confronted with the predictions of our calculation, significantly less parameter space can be excluded at low $M_{h_1}$. 
However, there is qualitative agreement about the existence of an unexcluded region at $M_{h_1}\sim 45\gev$. 
Using either our calculation or the {\cpsh} calculation results in large unexcluded regions at medium-to-high $\tb$ and $M_{h_1}$. 
In general, more of the $\CPX$ parameter space can be excluded by LEP, as compared to the $\CPXdrbar$ parameter space. Once again, both calculations are consistent with an unexcluded region at $M_{h_1}\sim 45 \gev$ in the $\CPX$ scenario, although, when {\cpsh} Higgs sector predictions are used, this region is very small, and might be covered by a dedicated analysis of this scenario which could make use of a combination of different LEP Higgs search channels (as discussed in \refse{sec:exclplots}). However, both calculations once again yield large unexcluded regions of parameter space at medium-to-high $\tb$ and $M_{h_1}$. Since the shape, size and position of the unexcluded regions vary significantly depending on which method was used to calculate the Higgs sector predictions, we adopt the LEP Higgs Working Group philosophy and only consider a point in parameter space excluded if it can be excluded using both methods independently. We therefore show these `combined' exclusion plots in \reffi{fig:cpxcomb} for the $\CPXdrbar$ and $\CPX$ scenarios. We conclude that there is still a sizeable area of parameter space in these scenarios below the kinematical limit at LEP which the LEP Higgs topological cross section limits are unable to exclude (in agreement with the LEP Higgs Working Group analysis) and therefore these scenarios with a very light neutral MSSM Higgs boson remain phenomenologically very interesting.

 \afterpage{\clearpage}

\section{Conclusions}
\label{sec:concl}

In this paper we have presented theoretical predictions concerning the Higgs
sector in the MSSM with complex parameters, which are especially relevant
when considering MSSM scenarios that are particularly challenging for
the past and present Higgs searches at LEP, the Tevatron and the LHC.

We have calculated the full 1-loop vertex corrections within the
Feynman-diagrammatic approach for the process $h_a\to h_b h_c$, taking
into account the dependence on all complex phases of the supersymmetric
parameters and the full momentum dependence.  We have
included the full propagator corrections, using neutral Higgs
self-energies as provided by {\feynhiggs}, and we have consistently included
1-loop mixing with the $Z$ boson and the unphysical Goldstone-boson
degree of freedom. Our results are currently the most precise
predictions for the $h_a\to h_b h_c$ decay width. These results will be
included in the publicly available program {\fh}.

We have found that the genuine vertex corrections to the triple Higgs
vertex are numerically very important. This holds both for the
cases of CP conservation and CP violation. The inclusion of the genuine 
vertex corrections changes the
predictions for the decay widths drastically as compared to an
approximation based solely on propagator-type corrections. Using the
leading Yukawa contributions
 or the full fermion/sfermion vertex corrections at
zero incoming momentum yields a prediction for the decay width
which is closer to the full result, but we still find deviations of
$\sim 30\%$ and $\sim 15\%$, respectively, in the example of the CPX parameter 
space at $\tb \sim 9$, 
$M_{h_1}\sim 40\gev$. Yet, these approximations can be useful in terms of 
effective couplings. The Yukawa approximation including all leading corrections 
can be expressed in a very compact form, thus providing a convenient
way to go beyond the tree level vertex contributions. The effective
coupling created from the full fermion/sfermion contributions at 
vanishing external momentum is a more sophisticated effective coupling
approximation. These effective couplings can be used for theoretical
predictions of cross sections for processes such as $e^+e^- \to h_1 Z \to h_1
h_1 Z$ at a future Linear Collider and the corresponding process at the LHC, 
which are of great interest in order to directly access the Higgs
self-couplings and thus investigate a crucial element of the Higgs
mechanism. We find that the genuine vertex corrections enhance the
triple-Higgs squared coupling of the light CP-even Higgs boson of the MSSM by 
20--60\% in the phenomenologically relevant region as compared to the 
lowest order coupling in the Standard Model for the same value of the
Higgs mass. This could have interesting implications for the sensitivity 
of searches at the LHC and a future Linear Collider to effects of the 
triple-Higgs coupling.

We also have presented the full 1-loop electroweak vertex corrections to
the $h_a\to f\bar{f}$ decay width in the complex MSSM, including full
phase dependence. We have supplemented these contributions with 1-loop
QED and SM QCD corrections, resummed SUSY QCD contributions, propagator
corrections calculated using neutral Higgs self-energies from {\fh}, and
1-loop propagator mixing with Goldstone bosons and Z bosons. Our calculation
has already been incorporated into the program {\fh}. 

We performed on-shell to $\drbarm$ parameter conversions at 
$\mathcal{O}(\alpha_s)$  and $\mathcal{O}(\alpha_t)$ for the general
case of complex supersymmetric parameters, to be used when comparing 
results of calculations based on different renormalisation schemes.
We also present a simple approximation to the $\mathcal{O}(\alpha_s)$ contribution to these parameter shifts, which is valid in the complex MSSM. 

We investigated the numerical impact of parametrising the neutral Higgs self-energies (in the Feynman-diagrammatic approach) in terms of the $\msbarm$ top mass, rather than the on-shell top mass, which is formally a 3-loop effect. This will be incorporated as a possible option in {\fh}.

Furthermore, the investigated $t,\tilde{t},b,\tilde{b}$ genuine vertex
corrections to the LEP Higgsstrahlung and LEP pair production processes in the CP-violating MSSM were found to have a small numerical impact in the CPX parameter space. 

Using our new theoretical predictions in conjunction with the topological cross
section limits from the LEP Higgs searches (there exist no relevant
limits for this parameter region from the Tevatron), we were able to investigate
the effect of the new contributions on the exclusion regions in the CPX
benchmark scenario. Of particular interest is the region $30 \gev\lsim
M_{h_1} \lsim 50 \gev$, $3\lsim\tb\lsim 10$, which previous analyses had
not been able to exclude, despite the relatively low values of $M_{h_1}$
involved. Since the coupling of the lightest Higgs boson to gauge bosons is
suppressed in much of this region, processes involving the second
heaviest or the heaviest neutral Higgs boson are important. Across the majority
of this region, the $h_2\to h_1h_1$ decay is dominant and therefore a
precise theoretical prediction for this decay width is crucial for
investigating whether an unexcluded region actually exists and for
mapping out its extent. Our results confirm the existence of an
unexcluded region at $M_{h_1}\sim 45\gev$. Furthermore, we find that
there is a rather large unexcluded region in the CPX-type scenarios 
below the LEP limit for a SM-like Higgs, occurring at medium to high
values of $\tb$ and extending down to $M_{h_1}$ values of about
$60\gev$. We have performed a detailed analysis of the dependence of the
unexcluded parameter regions on variations of the MSSM parameters.

After discussing in detail the issues involved in such a comparison, we 
have compared our results for the impact of the existing Higgs search
results on the parameter region of the CPX scenario with the results 
obtained using the publicly available program {\cpsh}. While both 
analyses confirm the existence of unexcluded regions in similar parts of 
the CPX parameter space, sizable differences occur in the detailed 
extent of these regions. 

Our results should serve as a motivation to focus Higgs searches at
the LHC not only on the mass range above about $114\gev$,
corresponding to the allowed region for a SM-like Higgs, but to perform
dedicated searches also for much lighter Higgs states.

\section{Acknowledgements}

We would like to thank Oliver Brein, Alison Fowler, Howard Haber, Thomas Hahn, Sven Heinemeyer, Wolfgang Hollik, David Lopez Val, Sophy Palmer and Pietro Slavich for valuable discussions and advice. Work supported in part by the Helmholtz Alliance HA-101 'Physics at the Terascale', the European Community's Marie-Curie Research
Training Network under contract MRTN-CT-2006-035505 `Tools and Precision
Calculations for Physics Discoveries at Colliders' (HEPTOOLS) and
MRTN-CT-2006-035657 `Understanding the Electroweak Symmetry Breaking
and the Origin of Mass using the First Data of ATLAS' (ARTEMIS).

\appendix

\section{Relations between $\Sigma_{h_iG}$ and $\Sigma_{h_iZ}$}
\label{section:gzrelations}

Using a Feynman-diagrammatic calculation, we find that the following relations hold at 1-loop order in the $R_{\xi}$-gauge:
\BEA
M_Z\Sigma_{\phi_2G}(p^2)+i p^2 \Sigma_{\phi_2Z}(p^2)&=&-\frac{e}{2c_Ws_W}c_{\beta}t_A\non,\\
M_Z\Sigma_{\phi_1G}(p^2)+i p^2 \Sigma_{\phi_1Z}(p^2)&=&\frac{e}{2c_Ws_W}s_{\beta}t_A\non,\\
M_Z\Sigma_{AG}(p^2)+i p^2 \Sigma_{AZ}(p^2)+M_Z(p^2-m_A^2)f_0(p^2)&=&\frac{e}{2c_Ws_W}\left(c_{\beta}t_{\phi_2}-s_{\beta}t_{\phi_1}\right)\non,\\
M_Z\Sigma_{GG}(p^2)+2i p^2 \Sigma_{GZ}(p^2)-\frac{p^2}{M_Z}\Sigma^L_{ZZ}(p^2)&=&\frac{e}{2c_Ws_W}\left(s_{\beta}t_{\phi_2}+c_{\beta}t_{\phi_1}\right)\non,\\
M_W\Sigma_{H^-G^+}(p^2)- p^2 \Sigma_{H^-W^+}(p^2)+M_W(p^2-\mHpm^2)f_{\pm}(p^2)&=&\frac{e}{2s_W}\left(c_{\beta}t_{\phi_2}-s_{\beta}t_{\phi_1}+it_A\right)\non,\\
M_W\Sigma_{G^-G^+}(p^2)- 2p^2 \Sigma_{G^-W^+}(p^2)-\frac{p^2}{M_W}\Sigma^L_{WW}(p^2)&=&\frac{e}{2s_W}\left(s_{\beta}t_{\phi_2}+c_{\beta}t_{\phi_1}\right),
\EEA
where
\BEA
f_0(p^2)&=&-\frac{\alpha}{16\pi s_W^2M_W^2}s_{\beta-\alpha}c_{\beta-\alpha}M_Z^2\xi_Z\left[B_0\left(p^2,m_h^2,M_Z^2\xi_Z\right)-B_0\left(p^2,m_H^2,M_Z^2\xi_Z\right)\right],\label{eq:fzero}\\
f_{\pm}(p^2)&=&-\frac{\alpha}{16\pi s_W^2M_W^2}s_{\beta-\alpha}c_{\beta-\alpha}M_W^2\xi_W\left[B_0\left(p^2,m_h^2,M_W^2\xi_W\right)-B_0\left(p^2,m_H^2,M_W^2\xi_W\right)\right].\non\\
\EEA
$f_0,f_{\pm}$ are finite and do not contribute if the particle is on-shell.

In terms of renormalised quantities, the relations are given by
\BEA
\hat{\Sigma}_{hG}(p^2)+\frac{ip^2}{M_Z}\hat{\Sigma}_{hZ}(p^2)&=&0,\label{eq:hGhZ}\\
\hat{\Sigma}_{HG}(p^2)+\frac{ip^2}{M_Z}\hat{\Sigma}_{HZ}(p^2)&=&0,\\
\hat{\Sigma}_{AG}(p^2)+\frac{ip^2}{M_Z}\hat{\Sigma}_{AZ}(p^2)+(p^2-m_A^2)f_0(p^2)&=&0,\label{eq:AGAZ}\\
\hat{\Sigma}_{GG}(p^2)+\frac{2ip^2}{M_Z}\hat{\Sigma}_{GZ}(p^2)-\frac{p^2}{M_Z^2}\hat{\Sigma}^L_{ZZ}(p^2)&=&0,\\ 
\hat{\Sigma}_{H^-G^+}(p^2)-\frac{p^2}{M_W}\hat{\Sigma}_{H^-W^+}(p^2)+(p^2-\mHpm^2)f_{\pm}(p^2)&=&0,\\
\hat{\Sigma}_{G^-G^+}(p^2)-\frac{2p^2}{M_W}\hat{\Sigma}_{G^-W^+}(p^2)-\frac{p^2}{M_W^2}\hat{\Sigma}^L_{WW}(p^2)&=&0.
\EEA

These relations have been checked algebraically using the $t,\tilde{t},b,\tilde{b}$ sector and the gauge and Higgs boson sector. These relations have also been checked numerically for the entire MSSM. These expressions first appeared in \citere{mythesis} and were derived using a BRST transformation in \citere{Baro:2008bg}. 

\section{Explicit form of counterterms}
\label{sec:apprencon}
In this section, we briefly summarise the counterterms used in this paper.

We renormalise the gauge bosons on-shell:

\BEA
\delta M_Z^2&=&{\rm Re}\Sigma^T_{ZZ}(M_Z^2),\\
\delta M_W^2&=&{\rm Re}\Sigma^T_{WW}(M_W^2),\\
\delta Z_{VV}&=&-{\rm Re}\Sigma_{VV}^{\prime},\\
\delta Z_{\gamma Z}&=&-2\frac{{\rm Re}\Sigma^{T}_{\gamma Z}(M_Z^2)}{M_Z^2},\\
\delta Z_{Z\gamma}&=&2\frac{\Sigma^{T}_{\gamma Z}(0)}{M_Z^2},
\label{eq:gaugefieldren}
\EEA
where `T' denotes the transverse part of the self-energy,
and as before the prime indicates the derivative
w.r.t.\ the external momentum squared.
We use a combination of on-shell and $\drbarm$ renormalisation for the Higgs sector, as in \citere{Frank:2006yh}:

\BEA
\delta M_{H^{\pm}}^2&=&{\rm Re}\Sigma_{H^{\pm}}(M_{H^{\pm}}^2),\\
\delta Z_{\cHe}&=&-\left[{\rm Re}\Sigma_{\phi_1\phi_1}^{\prime}\right]^{\rm div},\\
\delta Z_{\cHz}&=&-\left[{\rm Re}\Sigma_{\phi_2\phi_2}^{\prime}\right]^{\rm div}.
\EEA
`${\rm div}$' indicates that we have just kept the terms proportional to the divergent piece $\frac{2}{4-D}-\gamma_E+\log(4\pi)$. We define the $\tb$ counterterm using
\BEA
\tanb &\rightarrow& \tanb (1+\dtanb)=\tanb(1+\frac{1}{2}\left(\delta Z_{\cHz}-\delta Z_{\cHe}\right)). 
\EEA
This choice has been shown 
in \citeres{Brignole:1992uf,Frank:2002qa,Freitas:2002um}
to yield numerically stable results. We also need to fix the renormalisation scale for $\delta Z_{\cHe},\delta Z_{\cHz}$, which we choose to be $\mu_{\rm ren}=m_t$, as in \citere{Frank:2006yh}. 
We set 
\BEA
\delta T_{h_i}&=&-T_{h_i},
\EEA
to ensure that the renormalised tadpoles vanish.

In all of our calculations involving the full complex MSSM, we choose to
parametrise the result in terms of $\alpha(M_Z^2)$, where 
$\alpha(M_Z^2)=\alpha(0)/\left(1-\De\al\right)$, and $\Delta \alpha$ is a finite quantity that can be split into the contribution from the $e,\mu,\tau$ leptons and the contribution from the light quarks (i.e.\ all quarks except $t$),  $\De\al=\De\al_{\textup{lept}}+\De\al^{(5)}_{\textup{had}}$. $\De\al_{\textup{lept}}$ has been calculated to 3-loop order \cite{Steinhauser:1998rq} as
\BEA 
\De\al_{\textup{lept}}&=&0.031497687,
\EEA
while $\De\al^{(5)}_{\textup{had}}$ has been determined using experimental data via a dispersion relation \cite{Hagiwara:2003da} as
\BEA 
\De\al^{(5)}_{\textup{had}}&=&0.02755\pm0.0023.
\EEA

We therefore use the charge counterterm:
\BEA
\delta Z_e^{e(M_Z^2)}
&=&
 \frac{1}{2} \frac{\partial}{\partial q^2}\left.\Sigma^{\rm all\,loops}_{\gamma\gamma}(q^2)\right|_{q^2=0}
-\frac{1}{2} \frac{\partial}{\partial q^2}\left.\Sigma^{{{\rm light}\,f}{\rm in \,loops}}_{\gamma\gamma}(q^2)\right|_{q^2=0}
\non\\
 &+&\frac{1}{2M_Z^2}{\rm Re}\Sigma^{{{\rm light}\,f}{\rm in \,loops}}_{\gamma\gamma}(M_Z^2)
 + \frac{\sw}{\cw} \frac{\Si^{\rm T}_{\gamma Z}(0)}{\MZ^2},\label{eq:de}
\EEA 
Note that the second term in the right-hand side of \refeq{eq:de}
cancels the light fermion contributions in the first term, which involve
large contributions proportional to $\alpha\log\left(\frac{m^2_{{\rm
light}\,f}}{\mu_{\rm ren}^2}\right)$, arising from the running of $\alpha$ from $q^2=0$ to a higher energy scale, which are problematic because the masses of the light quarks are not well defined.

For calculations which only involve Standard Model fermions and their superpartners in loops, we choose to parametrise the electric charge in terms of the Fermi constant $G_F$ (as used, for example, in \citere{Hahn:2002gm}), yielding the charge renormalisation counterterm
\BEA
\delta Z_e^{G_F}
&=&\frac{\delta s_W}{s_W}-\frac{1}{2 M_W^2}\left(\Si^{\rm T}_{WW}(0)-\delta M_W^2\right).
 \label{eq:debyeGF}
\EEA


We renormalise the quark sector on-shell using the
conditions\footnote{Note
that the quark field renormalisation factors given in \refeq{dZL} and
\refeq{dZR} are sufficient for renormalisation purposes but they do not
necessarily take into account the full wave function normalisations 
that are required in the most general case to ensure the correct
on-shell properties of external quark fields, see \citere{AlisonPhD}.
As the quark masses of the final-state fermions are small in the Higgs
decays that we are considering (i.e., we do not consider Higgs decays
into top quarks),
in our calculations $\delta Z^L$ and $\delta Z^R$
account for the whole wave function corrections.}

\BEA
\delta m&=&\frac{1}{2}{\rm Re}\left[ m\left(\Sigma^L(m^2)+\Sigma^R(m^2)\right)+\Sigma^l(m^2)+\Sigma^r(m^2)\right],\\
\delta Z^L&=&\widetilde{\rm Re}\left[-\Sigma^L(m^2)+\frac{1}{2m}\left(\Sigma^l(m^2)-\Sigma^r(m^2)\right)\right.\non\\
&&\left.-m^2\left(\Sigma^{L'}(m^2)+\Sigma^{R'}(m^2)\right)-m\left(\Sigma^{l'}(m^2)+\Sigma^{r'}(m^2)\right)\right], \label{dZL} \\
\delta Z^R&=&\widetilde{\rm Re}\left[-\Sigma^R(m^2)+\frac{1}{2m}\left(\Sigma^r(m^2)-\Sigma^l(m^2)\right)\right.\non\\
&&\left.-m^2\left(\Sigma^{L'}(m^2)+\Sigma^{R'}(m^2)\right)-m\left(\Sigma^{l'}(m^2)+\Sigma^{r'}(m^2)\right)\right].  \label{dZR}
\EEA

In order to perform the $\drbarm$ to on-shell parameter conversion in \refse{sec:convren}, we apply the renormalisation conditions which were used in the calculation of the leading 2-loop corrections to the Higgs self-energies \cite{Heinemeyer:2007aq}, which were incorporated into the program {\feynhiggs}:
\BEA
\matr{\left(U_{\tilde{t}} \delta M_{\tilde{t}} U_{\tilde{t}}^{\dagger}\right)_{11}}&=\delta m_{\tilde{t}_1}^2&={\rm Re}\Sigma_{\tilde{t}_{11}}(m^2_{\tilde{t}_1}),\\
\matr{\left(U_{\tilde{t}} \delta M_{\tilde{t}} U_{\tilde{t}}^{\dagger}\right)_{22}}&=\delta m_{\tilde{t}_2}^2&={\rm Re}\Sigma_{\tilde{t}_{22}}(m^2_{\tilde{t}_2}),\\
\matr{\left(U_{\tilde{t}} \delta M_{\tilde{t}} U_{\tilde{t}}^{\dagger}\right)_{12}}&=\delta Y_{\tilde{t}}&=\frac{1}{2}\left(\widetilde{\rm Re}\Sigma_{\tilde{t}_{12}}(m^2_{\tilde{t}_1})+\widetilde{\rm Re}\Sigma_{\tilde{t}_{12}}(m^2_{\tilde{t}_2})\right),\\
\matr{\left(U_{\tilde{t}} \delta M_{\tilde{t}} U_{\tilde{t}}^{\dagger}\right)_{21}}&=\delta Y_{\tilde{t}}^*&=\frac{1}{2}\left(\widetilde{\rm Re}\Sigma_{\tilde{t}_{21}}(m^2_{\tilde{t}_1})+\widetilde{\rm Re}\Sigma_{\tilde{t}_{21}}(m^2_{\tilde{t}_2})\right).
\EEA

\section{Triple Higgs vertex corrections and Higgs self-energies in the Yukawa approximation for the special case where the stop masses are equal}
\label{app:yukequalsquarkmasses}

In \refse{section:massesYukCPV}, we investigated the Yukawa corrections
to the triple Higgs vertex and Higgs self-energies. For completeness, we
give these quantities here in the limit where the stop masses in the
Yukawa approximation are equal.

The triple Higgs vertex corrections in this limit are:
\BEA 
\Delta \lambda^{\rm Yuk}_{\phi_1\phi_1\phi_1}&=&
\Delta \lambda^{\rm Yuk}_{\phi_1\phi_1A}=
\Delta \lambda^{\rm Yuk}_{\phi_1\phi_2A}=
\Delta \lambda^{\rm Yuk}_{\phi_2\phi_2A}=
\Delta \lambda^{\rm Yuk}_{\phi_1AA}=
\Delta \lambda^{\rm Yuk}_{AAA}=
0,\\
\Delta \lambda^{\rm Yuk}_{\phi_1\phi_1\phi_2}&=&-\frac{3 e^3 m_t^4}{32 \pi ^2 M_W^3 s_W^3 s_{\beta }^3}  \left\{\frac{|\mu|^2}{m_{\tilde{t}_1}^2}\right\},\\
\Delta \lambda^{\rm Yuk}_{\phi_1\phi_2\phi_2}&=&\frac{3 e^3 m_t^4}{16 \pi ^2 M_W^3 s_W^3 s_{\beta }^5}
\left\{s_{\beta } c_{\beta }   \frac{|\mu|^2} {m_{\tilde{t}_1}^2}
\right\}, \\
\Delta \lambda^{\rm Yuk}_{\phi_2\phi_2\phi_2}&=&\frac{3 e^3 m_t^4}{16 \pi ^2 M_W^3 s_W^3 s_{\beta }^3}\left\{
2-\frac{2m_t^2}{m_{\tilde{t}_1}^2}
-3\log{\frac{m_{\tilde{t}_1}^2}{m_t^2}}
-\frac{3|A_t|^2}{2m_{\tilde{t}_1}^2}
\right\} ,\\
\Delta \lambda^{\rm Yuk}_{\phi_2AA}&=&\frac{3 e^3 m_t^4}{32 \pi ^2 M_W^3 s_W^3 s_{\beta }^5}
\left\{
-\frac{|\mu|^2}{m_{\tilde{t}_1}^2}
-2s_{\beta }^2 c_{\beta }^2\log{\frac{m_{\tilde{t}_1}^2}{m_t^2}}
   \right\},
\EEA  
while the Higgs self-energies in this limit are:
\BEA
\hat{\Sigma}^{(1)\phi_1\phi_1}_{\rm Yuk}&=&\frac{3 e^2  m_t^4 }{16 \pi^2  M_W^2 s_W^2 s_{\beta }^2}\left(|\mu|^2
   \frac{\CfuncLeq}{2} \right),\\
\hat{\Sigma}^{(1)\phi_1\phi_2}_{\rm Yuk}&=&-\frac{3 e^2  m_t^4 }{16 \pi^2  M_W^2 s_W^2 s_{\beta }^2}\left(\frac{ |\mu|^2}{t_{\beta }}
   \frac{\CfuncLeq }{2}\right),\\
\hat{\Sigma}^{(1)\phi_2\phi_2}_{\rm Yuk}&=&-\frac{3 e^2  m_t^4 }{16 \pi^2  M_W^2 s_W^2 s_{\beta }^2}\left( 2\logMSteMStzbyMTsq-\frac{ |\mu|^2
    }{t_{\beta }^2}\frac{\CfuncLeq}{2}\right),\\
\hat{\Sigma}^{(1)\phi_1A}_{\rm Yuk}&=&\hat{\Sigma}^{(1)\phi_2A}_{\rm Yuk}=0,\\
\hat{\Sigma}^{(1)AA}_{\rm Yuk}&=&\frac{3 e^2  m_t^4 }{16 \pi^2  M_W^2 s_W^2 s_{\beta }^4}\left( |\mu|^2
  \frac{\CfuncLeq}{2} \right),
\EEA
where
\BEA
\CfuncLeq&=&\text{C}_0\left(0,0,0,m_{\tilde{t}_1}^2,m_{\tilde{t}_1}^2,M_L^2\right).
\EEA

In this limit, the Yukawa corrections to both the triple Higgs vertex and the Higgs self-energies are independent of the CP-violating MSSM phases.

\bibliographystyle{h-elsevier3-newarxivid-leftjust-moreauthors}

\bibliography{main}

\end{document}